\documentclass[sigconf]{acmart}

\usepackage{multirow}
\usepackage{makecell}
\usepackage{ulem}
\usepackage{array,ragged2e}
\usepackage{tabularx}
\usepackage{booktabs}

\newlength{\cuaindent}
\usepackage{array,longtable,booktabs}
\newcolumntype{P}[1]{>{\raggedright\arraybackslash}p{#1}}
\usepackage{makecell}

\AtBeginDocument{%
  }

\newcommand{\system}{A11y-CUA}

\newcommand\revision[1]{\textcolor{black}{#1}}
\newcommand\revtwo[1]{\textcolor{black}{#1}}

\begin{document}

%%
%% The "title" command has an optional parameter,
%% allowing the author to define a "short title" to be used in page headers.
\title{\system{} Dataset: Characterizing the Accessibility Gap in Computer Use Agents}

\author{Ananya Gubbi Mohanbabu}
\authornote{Equal contribution.}
\affiliation{%
  \institution{University of California, Berkeley}
  \city{Berkeley, CA}
  \country{USA}}
\email{ananyagm@berkeley.edu}

\author{Rosiana Natalie}
\authornotemark[1]
\affiliation{%
  \institution{University of Michigan}
  \city{Ann Arbor, MI}
  \country{USA}}
\email{rosianan@umich.edu}

\author{Brandon Kim}
\affiliation{%
  \institution{University of Michigan}
  \city{Ann Arbor, MI}
  \country{USA}}
\email{kimbd@umich.edu}

\author{Anhong Guo}
\authornote{Equal supervision.}
\affiliation{%
  \institution{University of Michigan}
  \city{Ann Arbor, MI}
  \country{USA}}
\email{anhong@umich.edu}

\author{Amy Pavel}
\authornotemark[2]
\affiliation{%
  \institution{University of California, Berkeley}
  \city{Berkeley, CA}
  \country{USA}}
\email{amypavel@berkeley.edu}

\begin{abstract}

Computer Use Agents (CUAs) operate interfaces by pointing, clicking, and typing---mirroring interactions of sighted users (SUs) who can thus monitor CUAs and share control. CUAs do not reflect interactions by blind and low-vision users (BLVUs) who use assistive technology (AT). BLVUs thus cannot easily collaborate with CUAs. To characterize the accessibility gap of CUAs, we present \system{}, a dataset of BLVUs and SUs performing 60 everyday tasks with \revision{40.4 hours} and \revision{158,325 events}.
\revision{Our dataset analysis reveals that our collected interaction traces quantitatively confirm distinct interaction styles between SU and BLVU groups (mouse- vs. keyboard-dominant) and demonstrate interaction diversity within each group (sequential vs. shortcut navigation for BLVUs).} We then compare collected traces to state-of-the-art CUAs under default and AT conditions (keyboard-only, magnifier). The default CUA executed \revision{78.3\%} of tasks successfully. But with the AT conditions, CUA's performance dropped \revision{to 41.67\% and 28.3\% with keyboard-only and magnifier conditions respectively,} and did not reflect nuances of real AT use. 
\revision{With our open \system{} dataset}, we aim to promote collaborative and accessible CUAs for everyone.

\end{abstract}

\copyrightyear{2026}
\acmYear{2026}
\setcopyright{cc}
\setcctype{by}
\acmConference[CHI '26]{Proceedings of the 2026 CHI Conference on Human Factors in Computing Systems}{April 13--17, 2026}{Barcelona, Spain}
\acmBooktitle{Proceedings of the 2026 CHI Conference on Human Factors in Computing Systems (CHI '26), April 13--17, 2026, Barcelona, Spain}
\acmPrice{}
\acmDOI{10.1145/3772318.3791896}
\acmISBN{979-8-4007-2278-3/2026/04}




% \received{20 February 2007}
% \received[revised]{12 March 2009}
% \received[accepted]{15 January 2026}

\begin{abstract}

Computer Use Agents (CUAs) operate interfaces by pointing, clicking, and typing---mirroring interactions of sighted users (SUs) who can thus monitor CUAs and share control. CUAs do not reflect interactions by blind and low-vision users (BLVUs) who use assistive technology (AT). BLVUs thus cannot easily collaborate with CUAs. To characterize the accessibility gap of CUAs, we present \system{}, a dataset of BLVUs and SUs performing 60 everyday tasks with \revision{40.4 hours} and \revision{158,325 events}.
\revision{Our dataset analysis reveals that our collected interaction traces quantitatively confirm distinct interaction styles between SU and BLVU groups (mouse- vs. keyboard-dominant) and demonstrate interaction diversity within each group (sequential vs. shortcut navigation for BLVUs).} We then compare collected traces to state-of-the-art CUAs under default and AT conditions (keyboard-only, magnifier). The default CUA executed \revision{78.3\%} of tasks successfully. But with the AT conditions, CUA's performance dropped \revision{to 41.67\% and 28.3\% with keyboard-only and magnifier conditions respectively,} and did not reflect nuances of real AT use. 
\revision{With our open \system{} dataset}, we aim to promote collaborative and accessible CUAs for everyone.

\end{abstract}

\maketitle
\section{Introduction}
Recent advances in multimodal large language models have enabled the development of computer use agents (CUAs) that interact with computer applications to complete everyday tasks. CUAs such as OpenAI’s Operator~\cite{Introducing_Operator}, Anthropic’s Computer Use Tool ~\cite{Computer_use_tool}, Microsoft Copilot~\cite{sadadow}, and Google DeepMind’s Project Astra~\cite{Project_Astra} have demonstrated this emerging capability.  Given natural language instructions (e.g., \textit{``change the brightness of my computer to 50\%''} or \textit{``book the cheapest round trip flight from Austin to New York from September 2 to 4 on Expedia website''}), CUAs interpret screen pixels from screenshots ~\cite{zhou2023webarena, zhang2024ufo} and structural context such as DOM or accessibility trees ~\cite{xie2024osworld}, in order to plan actions. They then perform low-level operations such as clicking, scrolling, and typing--mirroring the computer use behavior of sighted users (SUs) ~\cite{ComputerUsingAgent}. SUs can thus easily follow CUA interactions and flexibly trade control when needed~\cite{huq2025cowpilot}. 

Current CUAs however, overlook interaction practices of blind and low-vision users (BLVUs) who use assistive technologies (AT) such as screen readers, magnifiers, or high contrast settings for computer use. 
\revision{While prioritizing CUA interaction efficiency over AT use can be beneficial when fully automating tasks, current CUAs still frequently encounter errors~\cite{OSWorld-Website,huq2025cowpilot} and make choices that conflict with user preferences~\cite{peng2025morae} such that users need to observe and understand CUA state~\cite{huq2025cowpilot,peng2025morae}.} 
However, BLVUs receive minimal accessible
feedback from CUAs and thus cannot easily observe or collaborate with CUAs in the same way as SUs. 
But, with the rise of CUAs, \revtwo{it is important to evaluate and improve their compatibility} with computer use for everyone--for example, screen reader users should also be able to easily monitor agent interactions to prevent potential errors or privacy risks, and magnification users should be able to hand off control efficiently even while they’re viewing a magnified screen. 
\revision{Further, we envision CUAs may be used for a broad range of downstream applications including personalized task demonstrations, in-situ guidance, synthetic user testing~\cite{hamalainen2023evaluating}, or improving automated accessibility testing~\cite{taeb2024axnav}, all of which} \revtwo{would benefit from CUAs that can operate under AT constraints.}
Existing benchmarks ~\cite{xie2024osworld, he2024webvoyagerbuildingendtoendweb, zhou2023webarena} and datasets ~\cite{deka2017rico, deng2023mind2web, pan2024webcanvas, wei2025browsecomp} overwhelmingly reflect SU interaction behavior and are primarily web-based, leaving a gap in understanding how CUAs perform in AT–mediated desktop and cross-application environments. Identifying this gap requires real-world data on BLVU computer use—yet, to our knowledge, no such dataset currently exists.

To characterize the accessibility gap of CUAs, we introduce \system{}, a multimodal dataset of BLVUs and SUs completing 60 everyday computer tasks across desktop and web applications. The dataset comprises \revision{40.4 hours} and \revision{158,325 events} collected from 16 participants (8 BLVUs, and 8 SUs). Our \revision{open-source} computer use recorder captures a dense, time-synchronized interaction history: screen video, system audio, OS-level input (keystrokes, mouse, scroll), window/element context, accessibility settings, and periodic \revision{UI Automation (UIA)} snapshots (UI elements' roles, names, states). For the web, it additionally logs per-tab DOM, accessibility tree, webpage metadata (URL, title), yielding replayable traces for analysis and simulation. 

\revision{Our dataset analysis reveals that A11y-CUA includes and quantifies distinct interaction \revtwo{differences} between SUs and BLVUs (i.e., SUs are mouse-dominant and BLVUs are keyboard-dominant) that has been qualitatively established ~\cite{ borodin2010more, jordan2024information, webaim2025screen} by prior studies in fewer task domains \cite{bigham2007webinsitu}.} \revision{Our analysis also demonstrates that A11y-CUA contains extensive within-group variations and recurring strategies: within SU and BLVU groups, participants approach the same task differently (e.g., using context-menu vs. shortcuts vs. drag-and-drop for text editing in SUs; using sequential vs. shortcut navigation in BLVUs).}
Both SUs and BLVUs try alternate interaction methods to complete a task if the prior approach failed to work. 

\revtwo{Our goal is to understand whether current CUAs, when constrained under AT conditions have the potential to assist BLVUs for the tasks we envision. So,} we evaluate how \revision{two general purpose} state-of-the-art models: \revision{a closed model (Claude Sonnet 4.5) and an open model (Qwen3-VL-32B-Instruct)} perform under default and two AT conditions: (1) keyboard-only to \revtwo{approximate input constraints in the workflow of BLVUs using screen readers} and (2) magnified viewport to \revtwo{approximate low vision users' magnified viewing constraints.} 
\revision{Sonnet 4.5's} CUA under default condition successfully completed \revision{78.33\%} of the tasks in \system{}, whereas its performance drops to \revision{41.67\%} when forced to use only keyboard and to \revision{28.33\%} when made to operate even with minimal magnifications at 150\% magnified viewport. \revision{Qwen3-VL model on the other hand, successfully completed 20\% of the tasks under default condition but its performance sharply dropped to 0\% under both AT conditions.} We characterize the accessibility gap in CUAs into perception, cognitive, and action gaps. \revision{We release the interaction traces of SUs, BLVUs, and CUAs along with our computer use recorder for supporting future research\footnote{huggingface.co/datasets/berkeley-hci/A11y-CUA}. Our work provides a foundation towards instruction, testing, and collaborative uses of CUAs.}
In summary, our work makes the following contributions:
\begin{itemize}
    \item A dataset of SUs, BLVUs, \revision{and CUAs} performing 60 real-world everyday computing tasks across desktop and web applications.

    \item A computer use recorder that captures and creates replayable traces of real computer use.

    \item A comparative analysis of SU and BLVU interaction styles, highlighting between-group and within-group variability.
    
    \item An evaluation of state-of-the-art CUAs under default and AT (keyboard-only, magnified viewport) conditions, revealing accessibility gaps in current CUAs.
\end{itemize}
\begin{figure*}
    \centering
    \includegraphics[width=\linewidth]{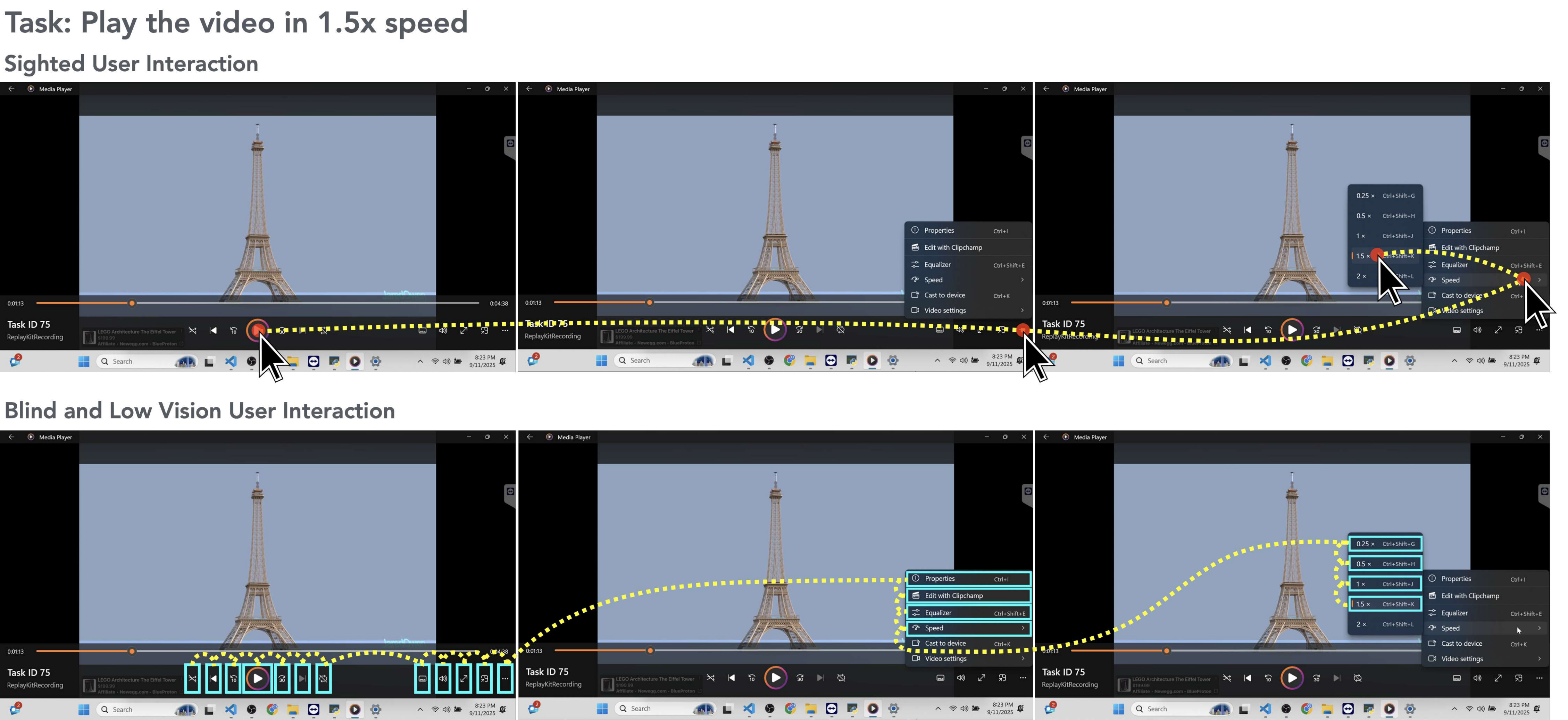}
    \caption{Illustrative example of interaction traces from SUs and BLVUs performing a task in the A11y-CUA dataset. SUs complete the task primarily through mouse interactions, resulting in fewer steps to complete the tasks. In contrast, BLVUs use keyboard navigation and screen reader feedback, which generally leads to longer interaction sequences.}
    \Description[]{The figure shows the comparison of sighted versus blind user interaction with a computer on a video player interface. The video player is showing a picture of the Eiffel Tower. The task of the interaction is "Play the video in 1.5x speed". The figure has two rows. The top row (Sighted Interaction): A sighted user moves the mouse pointer along a yellow dotted path, clicking small icons and navigating context menus. The sequence shows the cursor traveling from the play button to a menu icon and then selecting an option from a dropdown menu. The bottom row (Blind and Low Vision Interaction): A blind or low vision user interacts using a screen reader and focuses on a different button in the interface. The yellow dotted path connects to large, highlighted screen reader navigation boxes. The sequence demonstrates how the screen reader cycles through interface options, reading them out and presenting enlarged blue selection boxes around each element (buttons and menus) before selecting an action.}
    \label{fig:interaction_sample}
\end{figure*}

\section{Related Work}
\subsection{Computer Use Agents and Accessibility Gaps}
Computer Use Agents (CUAs), like Operator~\cite{Introducing_Operator}, Anthropic~\cite{Computer_use_tool}, Copilot~\cite{sadadow}, and Project Astra~\cite{Project_Astra} operate interfaces by pointing, clicking, and typing, similar to how SUs interact with computer. For example, in OpenAI Operator, users give natural language prompts, the agent then takes a screenshot of the screen and determines which actions to take based on the instruction. Because CUAs operate visually, they are inherently "sighted", which allows sighted users to easily collaborate with and oversee agent interactions. However, this reliance on visual perception and mouse-based interactions leaves a notable gap: CUAs have not been thoroughly evaluated in accessibility contexts, such as navigation using screen readers, where interaction is primarily keyboard-driven and when providing visuals is not feasible to the BLVUs.

CUA performance has steadily improved on established benchmarks ~\cite{xie2024osworld, zhou2023webarena, he2024webvoyagerbuildingendtoendweb}. According to OSWorld benchmark results~\cite{OSWorld-Website}, state-of-the-art CUAs such as Claude Sonnet 4.5 by Anthropic ~\cite{Computer_use_tool} have achieved up to 62.9\% task success in general computer use. These findings highlight the rapid progress of CUAs and promising direction for computer use automation. Yet, benchmarks to date focus exclusively on SU interactions style and do not capture the practices of people using assistive technologies. As a result, the actual performance of CUAs in accessibility-mediated environments remains unknown.

At present, no dataset exists that captures real-world interactions of BLVUs to support benchmarking in accessibility contexts. We build upon the foundational work of the existing datasets such as Rico~\cite{deka2017rico}, the first large-scale mobile GUI video dataset; Android in the Wild~\cite{rawles2023androidinthewild}, which contains 715k episodes of sequential GUI images; and other efforts spanning multi-platform analysis~\cite{chen2024gui}, mobile interactions~\cite{sun2022meta,li2022spotlight,tian2024mmina,zhang2024mobile}, web-based GUIs~\cite{lu2024weblinx,zhou2023webarena, yao2022webshop,koh2024visualwebarena,liu2018reinforcement, deng2023mind2web}, and desktop GUIs~\cite{zheng2024agentstudio, kapoor2024omniactdatasetbenchmarkenabling, mialon2023gaia,zheng2024gpt,liu2023webglm,niu2024screenagent}. Our work fills this critical gap by introducing a dataset that captures both BLVU and SU performance across 60 everyday computer tasks, comprising \revision{40.4 hours} of interaction traces and over \revision{158,325 actions}. 

Our dataset also extends prior approaches to evaluating agents by grounding the evaluation in real-world data. For example, Park et al.~\cite{park2024generative} assess generative agents using data from 1,000 real individuals. Natalie et al.~\cite{natalie2025not} construct a dataset with 40 low-vision individuals to evaluate VLM-based agents that simulate vision perception under low-vision conditions. Proxona~\cite{choi2024proxona} generates agent personas derived from real-world comments of content creators. Similarly, Hämäläinen et al.~\cite{hamalainen2023evaluating} validate their LLM-generated synthetic data by comparing it with real-world survey responses from video game players. Our dataset comprises the interaction from 8 BLVUs performing the tasks in our controlled Windows environment with our custom recorder.

\subsection{Computer Use of Blind and Low Vision Users}
Studies of BLVUs interacting with computers have examined several specific interaction methods, including text entry~\cite{anu2017performance,zhang2019text, shi2019vipboard,komninos2023review,zhang2024accessible,fakrudeen2017finger}, as well as pointing, touching and selection~\cite{10.1145/3447526.3472022, fakrudeen2017finger, 10.1145/3677846.3677867}. A substantial body of work also focuses on screen reader use ~\cite{godfrey2013statistical,kodandaram2023detecting, 10.1145/3649223,10.1145/3733155.3737910, 10.1145/3441852.3471202,10.1145/3493612.3520454,10.1145/3411764.3445242, shirogane2008accessibility,10.1145/3131785.3131837}, investigating aspects such as how screen readers operate~\cite{10.1145/3441852.3471202,10.1145/3493612.3520454,10.1145/3411764.3445242} and its integration with braille displays~\cite{shirogane2008accessibility} and other modalities (e.g., voice~\cite{10.1145/3131785.3131837}). While valuable, these studies typically isolate one modality rather than examining the full range of interactions required to complete everyday tasks. They also provide a limited insight into distinct interaction styles between SUs and BLVUs, as well as variability within user groups.

Other works focus on task-based evaluations of web and application use ~\cite{bigham2007webinsitu, savva2017understanding, Lazar01052007, godfrey2013statistical}. For instance, Lazar et al.~\cite{Lazar01052007} documented strategies BLVUs adopt to cope with slower input rates. While these studies offer comprehensive evaluations of accessibility barriers, they generally lack direct comparisons with SUs and focus primarily on success rates, rather than unpacking the different strategies and approaches that BLVUs use when operating computers.

Moving beyond specific interactions and task-based evaluation, emerging work has begun to address a more higher-level evaluation of computer interaction in accessibility contexts~\cite{kodandaram2024enabling, zhang2024ufo}, . For example, Savant~\cite{kodandaram2024enabling} leverages LLMs to unify and simplify screen reader interactions so that the computer can perform a single action. While Savant aims to automate parts of screen reader usage, it is limited to single actions rather than handling continuous and complex tasks like CUA in general. Despite prior work focusing on computer interactions of BLVUs, no large-scale dataset has yet been provided to capture their interaction practices and serve as a benchmark to characterize the accessibility gap of CUAs.

Datasets have also been developed to study input performance in accessibility research~\cite{findlater2020input}. For example, Input Accessibility~\cite{findlater2020input} analyzed mouse and touchscreen interactions from over 700 participants across diverse ages and motor abilities. However, this dataset is limited to four low-level interaction tasks, which are pointing, dragging, crossing, and steering, rather than capturing a more high-level task (e.g., "\textit{copy my speaking notes from MS Word to sticky notes app.}"). To the best of our knowledge, no existing dataset systematically captures the high-level  task interactions of BLVUs who rely on assistive technologies such as screen readers.

Our work complements and extends these efforts by compiling a dataset of both BLVU and SU interactions across real-world computing tasks (e.g., browsing, document editing, system operations, workflows, and media use). To support the multimodal requirements of CUA evaluation, our dataset is designed to be inherently multimodal, comprising synchronized screen recordings, system audio, OS window and element context, web DOM and accessibility tree information, input events (e.g., keystrokes, mouse interactions, hotkeys), UI automation snapshots, and accessibility settings. This study further contrasts interaction styles between SU and BLVU groups and highlights variability within each group.
\section{Designing the \system{} Dataset}

\begin{table*}[t]
\small
\centering
\begin{tabular}{@{} P{0.1\textwidth} P{0.23\textwidth} P{0.27\textwidth} P{0.1\textwidth} P{0.2\textwidth} @{}}
\toprule
\textbf{Category} & \textbf{Task Context} & \textbf{Task Instruction} & \textbf{Applications} & \textbf{End State} \\
\midrule
Browsing and Web &
I've been getting into cooking recently, so I've decided to upgrade my frying pan. &
Visit \texttt{walmart.com} on Chrome. Search for a frying pan, then click on the first product with a pan size of exactly 10 inches and add it to cart. &
Chrome &
The webpage is showing product page of the pan with 10 inches and the cart button has an additional item.\\

Media &
I want to make a new playlist to organize my music better. I want to add "Baby Shark" video in my playlist.  &
Create the playlist "Fun with Kids" on Media Player and add Task ID 76.mp3 in Documents > Task 76 folder to the playlist. &
File Explorer, Windows Media Player &
A new playlist called "Fun with Kids" is created. Baby Shark song is added to the newly created playlist. \\

Workflow &
I want to make a recorded demonstration of a simple calculation using BODMAS rule for my school presentation. &
Turn on screen recording using the Snipping Tool.
Open calculator app and perform the calculation 26-12/2.
Now clear the calculator results and calculate (26-12)/2.
Stop the recording on Snipping Tool. 
Insert this video in Task ID 61.pptx in Documents > Task 61 folder. &
Calculator, Snipping Tool, MS PowerPoint &
The recording of BODMAS calculation is inserted in MS PowerPoint. \\

Document Editing &
I am tracking all the books I've read in an excel sheet. I have columns for "Book Title", "\# Days to Finish", and "Rating (1-10)". I want to make a bar chart for how many days I took to read all the books. &
Open Task ID 43.xlsx in Documents > Task 43 folder, create a bar chart where the x-axis is "Book Title" and the y-axis is the "Days to Finish". Change the title of the chart to "Summer Reading". &
File Explorer, MS Excel &
A "Bar Chart" is added with the x-axis titled "Book Title", y-axis titled "Days to Finish", and chart titled "Summer Reading". \\

System Operations &
I'm browsing on Chrome but it is frozen. &
Open Chrome and close the application from Task Manager &
Chrome, Task Manager &
Chrome is closed and Task Manager does not show active Chrome application running. \\

\bottomrule
\end{tabular}
\caption{Sample tasks from our \system{} dataset, which covers five categories of tasks including Browsing \& Web, Media, Workflow, Document Editing, and System Operations. Every task comes with the Task Context and Instruction that situate the user and specify the required actions. Our tasks have a wide range of application scope, \revision{26 tasks involve a single application, and the remaining 34 tasks require two or more applications.} To determine successful completion, each task is associated with one or more End States that define the expected outcome.}
\label{tab:ten-tasks-fixed}
\end{table*}

Prior benchmarks of computer use evaluate agents and have advanced planning and UI grounding ~\cite{xie2024osworld, deng2023mind2web, zhou2023webarena,liu2018reinforcement, mialon2023gaia}. Our dataset also targets everyday tasks with verifiable end states, but differs in two ways: (1) it includes traces from both SUs and BLVUs, and (2) it logs accessibility interactions (e.g., screen-reader announcements, window/element context, magnification/zoom changes) alongside standard UI actions (keystrokes, mouse events, DOM events). These additions enable us to analyze how BLVUs use AT, identify breakdowns, and benchmark agents on real tasks. 

Prior work has evaluated CUAs on their ability to complete everyday computing work ~\cite{xie2024osworld, deng2023mind2web, zhou2023webarena}. To support comparable evaluation, we include routine activities that \textbf{reflect real needs} such as browsing the web, editing documents, managing files, and adjusting system settings. Each task is framed with context, and supporting resources, to capture the reasoning behind a user’s path. 
We also cover tasks \textbf{supporting different interaction types} such as pointing and clicking, typing and shortcuts, search and navigation (headings/ landmarks), drag-and-drop, context menus, clipboard actions, text editing, and timeline/media controls. This lets us compare strategies across modalities (mouse, keyboard, screen reader commands, magnifier) without forcing a single “right” path. Prior work shows variation in screen reader strategies \revision{qualitatively}, motivating this breadth \cite{jordan2024information} 
We also cover tasks of \textbf{varying complexity} by including both short, \revision{single-application but multi-step tasks (e.g., create and rename a file) and longer, multi-application tasks} (e.g., record a demonstration of BODMAS rule on Calculator using Snipping Tool and add it to PowerPoint). This supports understanding planning, error recovery, and long-horizon control seen in prior agent benchmarks \cite{deng2023mind2web, zhou2023webarena}.
Each task has a concrete, \textbf{verifiable success criteria} (e.g., “file created and renamed to \texttt{X}”, “Word document set to double-spaced”) to enable consistent human annotation on success or failure and support execution-based evaluation, similar to prior work ~\cite{xie2024osworld}.
Tasks are stated as \textbf{high-level instructions} rather than as step-by-step instructions. This design supports natural variation in strategies (e.g. menu navigation vs. keyboard shortcuts; visual search vs. in-page search), allowing us to capture different paths to task completion both within and across SUs and BLVUs groups for maximum coverage of interaction patterns. See Fig.~\ref{fig:interaction_sample} for the differences in interaction styles between SUs and BLVUs.

\begin{table*}[t]
\centering
\small
\renewcommand{\arraystretch}{1.2}
\begin{tabularx}{\textwidth}{l l l X}
\toprule
\textbf{Event Type} & \textbf{Source} & \textbf{Triggers} & \textbf{Fields Captured} \\
\midrule
\texttt{mouse\_click}    & Desktop     & Mouse button pressed                        & cursor, hovered element (UIA), application/window, timestamp \\
\texttt{mouse\_up}       & Desktop     & Mouse button released                       & cursor, hovered element (UIA), application/window, timestamp \\
\texttt{mouse\_move}     & Desktop     & Mouse moves (throttled to $\sim$2Hz)       & delta, cursor position \\
\texttt{scroll}          & Desktop+Web & Mouse scroll event                          & delta, direction \\
\texttt{drag\_drop}      & Desktop     & If cursor moves $\geq$6px while button held & from/to apps + UIA targets, duration, slider value change (if applicable) \\
\texttt{key\_press}      & Desktop+Web & Key pressed (non-modifiers only)            & normalized key name, modifiers, context, target element (UIA) \\
\texttt{hotkey}          & Desktop     & Modifiers + non-modifier                    & combo string (e.g., Ctrl+C), classified intent (e.g., copy, paste, etc.) \\
\texttt{input}           & Web         & User types into an input field              & selector, element metadata \\
\texttt{focus} / \texttt{blur} & Web  & Focus changes                               & target element selector + metadata \\
\texttt{page}            & Web         & On page load or URL change                  & DOM + accessibility tree snapshot, tab ID \\
\bottomrule
\end{tabularx}
\caption{List of event types, their sources, triggers, and fields captured by the computer use recorder.}
\label{tab:events}
\end{table*}

\subsection{Data Collection Procedure}
\subsubsection{Tasks}
To ground the tasks in real user needs, we drew tasks from public “how-to” sources (e.g., help articles, tutorials) and community forums (e.g., discussions similar to Reddit or Stack Overflow) and adapted those tasks so that each has a clear, verifiable end state. We collected a total of 60 tasks that belong to five categories: web \& browsing, workflow, media, document editing, and system operations, with 12 tasks in each category. To align these tasks with prior work ~\cite{xie2024osworld}, we also integrated \revision{15} OSWorld tasks (primarily for Ubuntu OS) and adapted them for Windows. We deliberately excluded programming tasks in OSWorld to keep the dataset representative of everyday computer use. We include multi-step, cross-application workflow tasks to observe planning, inter-app hand-offs (e.g., file saves, clipboard transfers), and small-error recovery capabilities that stress both human performance and agent control (see A.4 for a full list of tasks in \system{} dataset). 
Windows is the most commonly used computer operating system for BLVUs  as they commonly use the AT applications it supports (NVDA, JAWS, Narrator, ZoomText)~\cite{webaim2025screen}.
It also exposes stable OS-level APIs (Win32 for input hooks and process/window introspection, and UI Automation (UIA) for control trees, roles, names, states, bounding boxes, and focus/selection events), letting us record high-resolution, timestamped streams for mouse/keyboard, window focus, and UI state. Our dataset uses widely used Windows applications: system utilities (File Explorer, Settings, Media Player) and third-party tools (Chrome, Word, Excel). Tasks in the \system{} dataset span different complexity levels: 26 tasks require the use of one application, 27 require two, 6 require three, and 1 requires four. \revision{All of the tasks consist of more than one step action.} For every task, we provide a short description for task context, task instruction, and end-state checks. We also provide any supporting resources required to complete the task (e.g., documents, media assets, URLs, login credentials) in the task instruction. We validated that every task is feasible in the target environment using both mouse-driven and keyboard-centric workflows, and we verified that our recorder consistently captured the corresponding interactions when these tools are used. \revision{The tasks chosen were complex enough to sufficiently show the accessibility gap in the CUAs under AT conditions.  }

\subsubsection{Participants}
We recruited a total of 16 participants (8 SUs, 8 BLVUs) through our mailing list of previous study contacts, local organizations, and snowball sampling. Participants ranged in age from 20 to 60 years \revision{(SUs: $\mu$ = 32.62, $\sigma$ = 11.83; BLVUs: $\mu$ = 31.0, $\sigma$ = 7.50).} 
On average, SUs reported 20.25 years ($\sigma$ = 5.23) of experience using computers, whereas BLVUs reported 18.75 years of experience using computers ($\sigma$ = 6.92). BLVUs reported using screen readers for an average of 14 years ($\sigma$ = 5.78). Refer to Table~\ref{tab:blvu-demo-at} for the full list of \revision{all participants' demographics}. 

In terms of everyday computer use, all participants engaged in web browsing (e.g., Google Chrome, Edge), document editing (e.g., Microsoft Office, Google Workspace, Notepad), and messaging (e.g., WhatsApp). We also recruited one participant who used both a magnifier and a screen reader. To simplify the analysis of BLVUs who primarily use screen readers, we excluded this participant’s data from the BLVU interaction analysis. However, their participation motivated us to also evaluate CUA performance with a magnifier.

\begin{table*}[t]
\small
\centering
\setlength{\tabcolsep}{2.7pt}
\renewcommand{\arraystretch}{1.05}

\begin{tabularx}{\textwidth}{
  % --- BLVU (8 cols) ---
  l c c c
  >{\raggedright\arraybackslash}p{2.35cm}  
  >{\raggedright\arraybackslash}l    
  c
  >{\raggedright\arraybackslash}p{3cm} 
  % --- divider between groups (tighter) ---
  !{\hspace{0.25em}\vrule\hspace{0.25em}}
  % --- SU (4 cols) ---
  l c c c
}
\toprule
\textbf{PID} & \textbf{Age} & \textbf{Gender} &
\makecell{\textbf{\# Years of}\\\textbf{Computer Use}} &
\makecell{\textbf{Vision}\\\textbf{Impairment}} &
\textbf{Onset} &
\makecell{\textbf{\# Years}\\\textbf{AT Use}} &
\makecell{\textbf{AT Used}}
&
\textbf{PID} & \textbf{Age} & \textbf{Gender} &
\makecell{\textbf{\# Years of}\\\textbf{Computer Use}} \\
\midrule

BLVU1 & 28 & M & 25 & Congenital Glaucoma & Since birth & 10 & JAWS, NVDA
& \revision{SU1} & \revision{32} & \revision{M} & \revision{22} \\

BLVU2 & 20 & M & 10 & Completely blind & Since birth & 10 & NVDA, VoiceOver
& \revision{SU2} & \revision{28} & \revision{M} & \revision{23} \\

BLVU3 & 24 & M & 11 & Completely blind & Since birth & 12 & NVDA, JAWS
& \revision{SU3} & \revision{60} & \revision{M} & \revision{28} \\

BLVU4 & 34 & M & 19 & Completely blind & Since birth & 8 & Braille Display, NVDA
& \revision{SU4} & \revision{31} & \revision{F} & \revision{20} \\

BLVU5 & 31 & M & 25 & Completely blind (left eye), Glaucoma (right eye) & Since birth & 12 & NVDA, JAWS, Braille Display
& \revision{SU5} & \revision{35} & \revision{F} & \revision{13} \\

BLVU6 & 37 & F & 20 & Completely blind & Since birth & 25 & JAWS, VoiceOver, NVDA, Narrator, Braille Display
& \revision{SU6} & \revision{22} & \revision{M} & \revision{15} \\

BLVU7 & 44 & M & 25 & Retinitis Pigmentosa & 9 years old & 20 & JAWS, NVDA, VoiceOver, Android TalkBack
& \revision{SU7} & \revision{24} & \revision{F} & \revision{16} \\

BLVU8 & 30 & F & 15 & Pathologic Myopia & 2 years old & 15 & JAWS, Braille Display, ZoomText
& \revision{SU8} & \revision{29} & \revision{M} & \revision{25} \\

\bottomrule
\end{tabularx}
\caption{\revision{Demographics of SUs and} BLVUs (with assistive technology usage).}
\label{tab:blvu-demo-at}
\end{table*}

\subsubsection{Method}

We conducted data recording sessions using both in-person and remote computer access. All BLVUs accessed the research team's laptops remotely through Team Viewer ($N$=6) or JAWS Tandem ($N$=2). The laptop was equipped with JAWS and NVDA screen readers. 
3 SUs used TeamViewer to participate in data recording sessions and the remaining 5 SUs completed them in-person. All participants each completed all 60 tasks. At the start of the session, \revision{SUs were given time to familiarize themselves with the computer use recorder interface and the setup of the laptop that they were accessing remotely or using in-person. Similarly, BLVUs} were given time to customize the assistive technologies on the research team's laptops until they felt comfortable with the setup. Prior to adopting this hybrid setup, the research team verified that potential disruptions (e.g., latency) introduced by remote access did not meaningfully affect the user experience. \revision{We asked the participants to perform tasks on our laptops instead of their personal systems} to ensure that (1) participants’ privacy was protected and (2) all required applications and resources (e.g., pre-prepared documents for document editing tasks) were available in a clean, default environment. As the participants were working on the tasks, our computer use recorder ran in the background to capture interaction data during task execution. \revision{We kept task categories in a fixed sequence for all participants as they introduced a difficulty progression, but we randomized the task order within each category for every participant.} Participants could use search engines to look up how to perform a task if needed. \revision{Throughout the study, the researcher observed both groups of participants and recorded task execution. Tasks were concluded when participants reached the pre-defined end state or after approximately eight minutes.} As recording all 60 tasks took more than two hours, participants were allowed to complete the study over multiple sessions to reduce fatigue and maintain consistent performance. Our study procedure has been approved by our university's IRB and all participant received USD 30 per hour for their participation.

\begin{figure*}
    \centering
    \includegraphics[width=\linewidth]{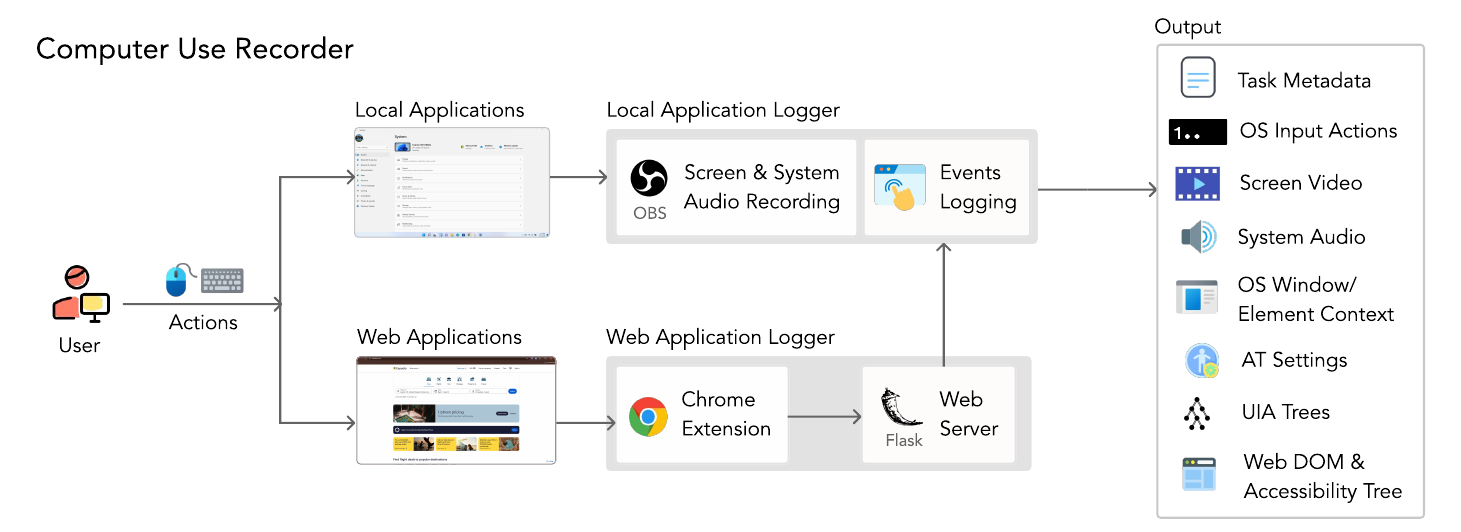}
    \caption{Computer Use Recorder. The Local Application Logger runs the pipeline: it presents tasks, resets the environment, records screen video and system audio, logs desktop actions, and receives web actions from a Chrome Extension through a Flask web logger. All streams are aligned on a single timeline to produce synchronized outputs: task metadata, OS input actions, screen video, system audio, OS window/element context, AT settings, UIA trees, and web DOM and accessibility tree events.}
    \Description[]{A flow diagram titled Computer Use Recorder showing how user actions are logged from local and web applications.
    On the left, a user performs actions. Local Applications feed into a Local Application Logger, which includes OBS screen and system audio recording and events logging. Web Applications feed into a Web Application Logger, which includes a Chrome extension sending data to a Flask web server. Both loggers connect to outputs on the right. Outputs include: task metadata, OS input actions, screen video, system audio, OS window/element context, assistive technology settings, UI Automation (UIA) trees, and web DOM & accessibility tree. This diagram illustrates how computer use is captured across applications and translated into multiple detailed logging outputs.}
    \label{fig:recorder-pipeline}
\end{figure*}

\subsection{Computer Use Recorder}
To make real-world interaction data easier to collect and share, we built a lightweight computer use recorder suitable for adoption in other studies. It captures screen video, system audio, OS-level input (keystrokes, mouse, scroll), window/element context, accessibility settings, and periodic UIA snapshots (roles, names, states). For the web, it additionally logs per-tab DOM, accessibility tree, webpage metadata (URL, title). This combination yields human-interaction traces well-suited to accessibility analysis and simulation, going beyond OSWorld’s environment, which exposes screenshots and optional accessibility trees to agents~\cite{xie2024osworld}.
At a high level, the system has two parts: (1) a local application logger that monitors desktop interactions, accessibility settings, and UIA snapshots and coordinates recording; and (2) a web logger server that receives structured events from a Chrome extension, giving DOM-level and accessibility updates not visible at the OS layer. Together, these components capture cross-application behavior across native Windows applications and the browser in a single, replayable log. \revision{Our computer use recorder is open-source to support future work and can be easily adapted to record similar traces for other AT setups, user groups, and tasks, to enable broader accessibility-focused CUA studies.}

\subsubsection{Implementation}
The local application logger is implemented in Python \texttt{tkinter} library for GUI components and uses \texttt{json} for loading tasks. The interface is built using \texttt{tkinter} and allows users to select tasks from a pre-loaded JSON task list. Each task is displayed in a modal dialog containing both a high-level context and a concise instruction.

When the task is started, the local application logger first enforces a clean environment by closing all applications with visible windows, except for a small whitelist to ensure the computer use recorder is running as wanted (i.e., Python itself, the OBS recorder, Windows Search Host, and File Explorer). This process is managed using \texttt{psutil} for process inspection and \texttt{pywin32} APIs for window management. This ensures minimal background interference and consistent logging conditions. In contrast, OSWorld typically resets VM state via snapshots in its \texttt{DesktopEnv}. Our process/window–level cleanup achieves a similar “clean slate” while running on a fixed Windows host.

\textbf{Capturing Video \& Audio.} The local application logger then initiates multimodal recording with OBS Studio~\cite{OBS}. Using \texttt{obsws\_python} (v5), the system connects to OBS Studio via its WebSocket API to begin synchronized capture of both the display and system audio. The output is subsequently split using \texttt{ffmpeg} into three separate streams: a full-resolution screen recording, an isolated track of the system’s internal audio, and an optional microphone channel. Microphone recordings, however, are excluded from the dataset release in order to protect participants' privacy. Along with audio-visual recording, the logger continuously captures fine-grained action events from local applications, incoming web events from the Chrome extension, periodic accessibility settings snapshots, and UIA accessibility tree dumps. Actions from local applications are captured via the \texttt{pynput} library for mouse and keyboard hooks, and \texttt{uiautomation} for accessibility tree queries. The local application logger generates a metadata file for each task that links together these heterogeneous logs, summarizing the applications launched, the number and type of events recorded, and pointers to the corresponding accessibility trees and replayable resources.

\textbf{Recording Desktop Actions.} The local application logger captures a detailed set of desktop interactions. Mouse actions include \texttt{mouse\_click} on button press, \texttt{mouse\_up} on release, \texttt{scroll} with direction and delta values, and \texttt{drag\_drop}, which is synthesized at mouse release if the pointer has moved $\ge$ 6 pixels while a button was held. Drag-and-drop actions store detailed contextual fields such as source and destination application, window, and element snapshots, along with pixel distance moved and interaction duration in milliseconds. Keyboard events are logged as \texttt{key\_press} for all non-pure modifier keys. Normalization is handled with Python’s built-in \texttt{keyboard} utilities and mapped consistently using \texttt{pynput}’s key constants. Key names are normalized (e.g., letters lowercased, modifier keys standardized) and mapped across common classes such as Ctrl/Alt/Shift, Windows keys, function keys, and arrows. When a modifier is combined with a non-modifier, a \texttt{hotkey} event is emitted, storing both the compact key combination (e.g., Ctrl+C) and a classified intent where possible (e.g., copy, paste, save, open new tab). 
Each event record includes a timestamp, process name, window title, cursor position, focused element (via UIA name and type), element under cursor (including bounding rectangle), window handle, process ID, window bounds, and window state (normal, minimized, maximized). Refer to Table ~\ref{tab:events} for the list of all events the computer use recorder captures. 
\revision{We selected these event types with the goal to ensure that the logs were sufficient to fully reconstruct a user’s interaction timeline, including timing, cursor trajectory, on-screen state focus changes, and navigation between applications.  We then verified that this event set allows us to accurately replay the recorded sessions. We omit interactions, like key releases, short drags (<6 px) or unintentional micro-movements that provide no additional information about a user's interaction. \texttt{mouse\_move} events are still recorded at a throttled rate for trajectory reconstruction, but are excluded from the \texttt{mouse\_actions} count as they would can inflate the totals.}

\textbf{Recording Web Actions.} For web-based tasks, a Chrome extension injects a content script into each page and relays events to the web application logger via the local \texttt{flask} server, since the DOM and its accessibility tree run inside the browser’s sandboxed renderer and OS-level APIs see only top-level UI (e.g., tabs, address bar), missing in-page changes.
Events such as \texttt{input}, \texttt{focus}, \texttt{blur}, and \texttt{page} are recorded along with DOM metadata and accessibility tree snapshots. Page events are captured both at initial load and upon each URL change, including a base64 snapshot of the rendered DOM with inline styles. Each browser tab is assigned a unique session ID and order index to maintain consistent logging across multiple simultaneous tabs. 

\textbf{Capturing Accessibility Trees.} Accessibility tree snapshots are collected according to specific triggers  using the \texttt{uiautomation} package for UIA traversal and serialized into JSON with Python’s built-in \texttt{json} module. A new tree is logged when focus shifts to a different window surface, or when the existing tree has changed after a 15-second cooldown. These trees are stored as JSON files in the task directory and referenced from the metadata file. System-wide accessibility settings, such as whether a screen reader is running, whether high contrast mode is enabled, and whether a magnifier is active, are polled and diffed every second to capture state changes over the course of task execution. This provides ground truth for how accessibility features were configured at task onset and how participants adapted them while working.

\textbf{Synchronizing Streams.} All data for a task is stored within a dedicated folder structure under the participant’s identifier. This includes audio-visual recordings, per-application JSON, web interaction logs, accessibility tree files, and the consolidated metadata summary. Metadata files include task-level attributes (task ID, instruction, context, familiarity and difficulty ratings before and after, task success and failure reasons), session-level attributes (start and end time, applications by order, output folder paths), and per-application statistics (number of events, event breakdowns by type, associated accessibility tree files). By aligning all data streams, our computer use recorder produces synchronized human-interaction traces that better support accessibility analysis, simulation, and agent benchmarking than conventional agent-based logs.

\section{Analysis of Human Performance and Behavior Across Sighted and Blind and Low-Vision User Groups} We analyze the \system{} dataset and summarize the dataset and the behavioral characteristics it captures. We compare sighted users (SUs) and blind and low-vision users (BLVUs) on both outcomes and process: task success, completion time, and interaction mix (mouse vs. keyboard, hotkeys, screen-reader navigation, magnifier behavior). We also examine within-group diversity, quantifying variability and recurring strategies (e.g., Tab/Arrow “walkers”, hotkey “jumpers”, mouse-dominant flows) using event-type measures.

\begin{figure*}
    \centering
    \includegraphics[width=\linewidth]{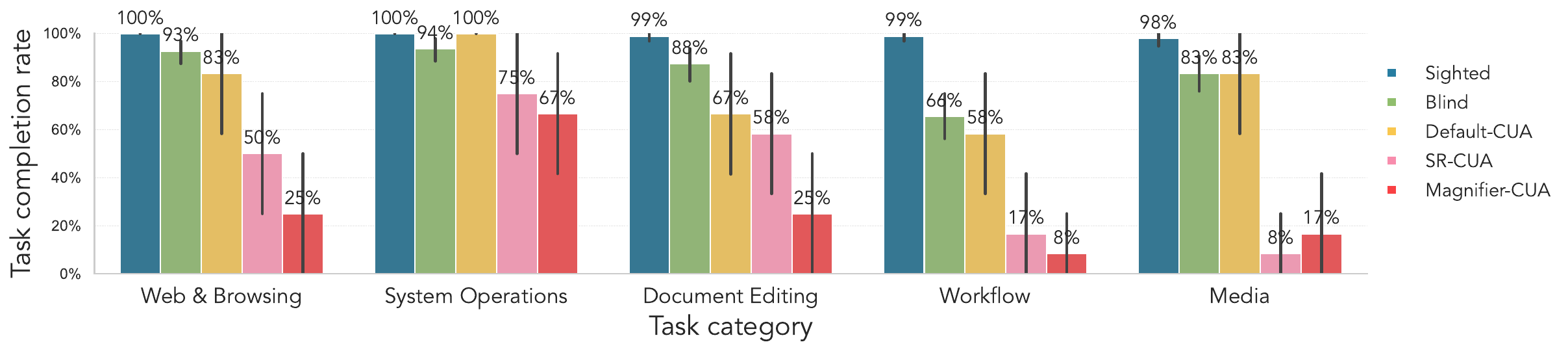}
    \caption{\revision{Task completion rates by task category for SUs, BLVUs, and three CUA configurations (Default-CUA, SR-CUA, Magnifier-CUA) with Claude Sonnet 4.5 model. SUs complete nearly all tasks across categories, while BLVUs show slightly lower success rates: especially for workflow tasks. Default-CUA reaches moderate performance, approaching BLVUs for web \& browsing, system operations, and media, but falls further behind on document editing and workflow tasks. SR-CUA and Magnifier-CUA perform substantially worse overall, with particularly low completion rates on workflow and media tasks.}}
    \Description[]{A bar chart showing task completion rates across five task categories for different user groups: Sighted, Blind, Default-CUA, SR-CUA, and Magnifier-CUA. Web & Browsing: Sighted 100\%, Blind 93\%, Default-CUA 92\%, SR-CUA 33\%, Magnifier-CUA 0\%. System Operations: Sighted 100\%, Blind 94\%, Default-CUA 84\%, SR-CUA 21\%, Magnifier-CUA 0\%. Document Editing: Sighted 99\%, Blind 83\%, Default-CUA 83\%, SR-CUA 0\%, Magnifier-CUA 0\%. Workflow: Sighted 99\%, Blind 64\%, Default-CUA 33\%, SR-CUA 0\%, Magnifier-CUA 8\%. Media: Sighted 98\%, Blind 83\%, Default-CUA 42\%, SR-CUA 0\%, Magnifier-CUA 0\%. ›Sighted users have the highest completion rates across all categories, followed by blind users with assistive technologies. CUA groups (Default, Screen Reader, Magnifier) show significantly lower performance, especially SR-CUA and Magnifier-CUA, which often score near zero.}
    \label{fig:task-success-rate-by-category}
\end{figure*}

\subsection{Dataset Overview}
\system{} dataset contains a total of \revision{158,325 events (44,409 from SUs and 113,916 from BLVUs), including 24,927 mouse actions and 96,253 keyboard actions}. Across all sessions, we collected 40.439 hours 
of recorded interactions from 16 participants (8 SUs, 8 BLVUs).
Overall, SUs completed tasks with an average of 92.35 seconds ($\sigma= 78.05$ seconds) and a success rate of 99.16\%. In contrast, BLVUs took an average of 211.18 seconds per task ($\sigma= 154.99$ seconds) with a success rate of 85\%.
In addition to capturing interactions (i.e., mouse and keyboard actions), events include app/tab focus changes, window/element context, web navigation and DOM/accessibility tree updates and accessibility-settings changes.
We record participants performing various types of actions such as mouse clicks, drag-and-drops, hotkey presses, scrolls while completing the tasks. We confirm that SUs show mouse-dominant interaction behaviors (51.93 mouse actions per task compared to 21.06 keyboard actions per task) in contrast to BLVUs, who use keyboard-only actions (no mouse actions) for navigation and information finding (179.45 keyboard actions per task).
Across task categories, success rates for SUs range from 97.9\% (Media) to 100\% (Web \& Browsing, System Operations), while for BLVUs, they range from 65.62\% (Workflow) to 93.75\% (System Operations).

\begin{figure*}
    \centering
    \includegraphics[width=\linewidth]{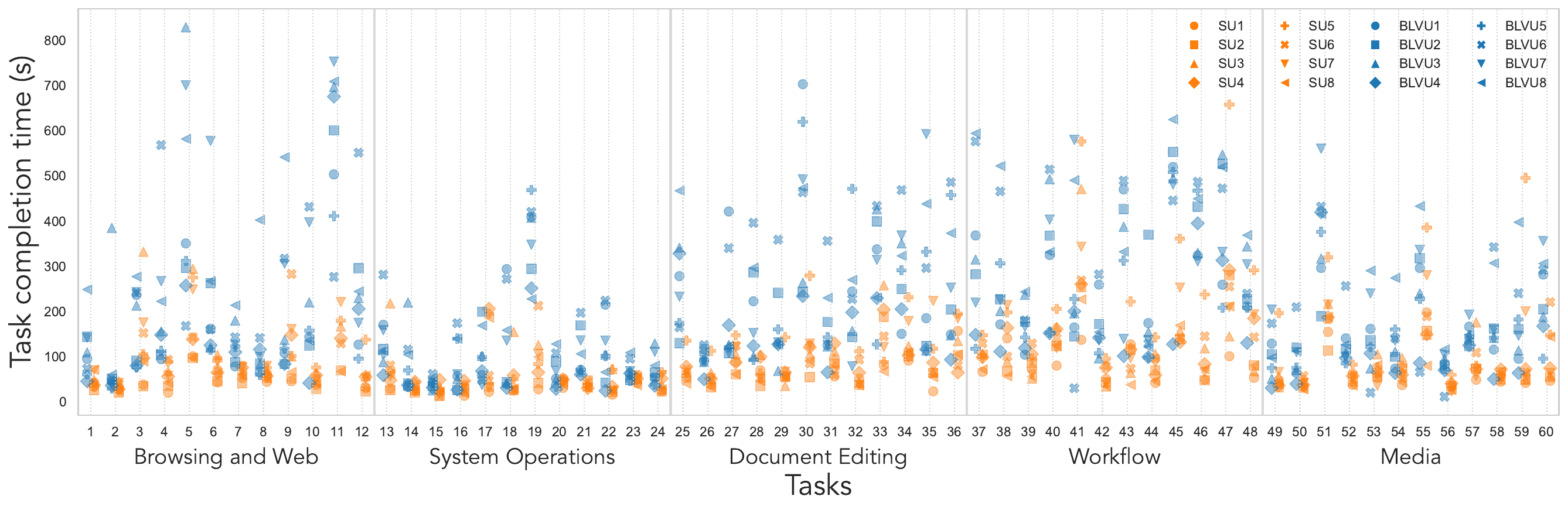}
    \caption{Per-task completion times across 60 tasks. Each marker is one participant×task; orange symbols denote sighted users (SU1–SU8) and blue symbols denote blind and low-vision users (BLVU1–BLVU8). Tasks are grouped by category along the x-axis (vertical dividers). SUs complete most tasks quickly with a tighter spread (often <150 s), whereas BLVUs show higher median and greater variance, especially in Document Editing, Workflow and Media categories. The wide dispersion within both groups shows substantial within-group strategy differences.}
    \Description[]{A scatter plot showing task completion times (in seconds) for 60 tasks across five categories: Browsing and Web, System Operations, Document Editing, Workflow, and Media. The x-axis lists tasks grouped by category. The y-axis shows task completion time from 0 to 800 seconds. Orange markers represent sighted users (SU1–SU4), while blue markers represent blind and low vision users (BLVU1–BLVU8). Sighted users cluster near the lower time range (below 200 seconds), while blind and low vision users show much higher variability, with many data points exceeding 400 seconds and some approaching 800 seconds. Across all categories, blind and low vision users consistently take longer to complete tasks than sighted users.}
    \label{fig:per_user_tct}
\end{figure*}

\subsection{Contrasting Sighted and Blind and Low-Vision Users' Interactions}

\system{} reveals contrasts in how SUs and BLVUs' approach common computer tasks. To understand the distinctness in their interaction styles, we highlight the differences in SUs and BLVUs approaches, speed, success, error recovery and time for task completion within and across both groups.

\subsubsection{Task Completion Time \& Speed of Action Execution} 
SUs completed tasks on an average of 92.35 seconds ($\sigma = 78.05$ seconds) by adopting point-and-click strategy. BLVUs on the other hand, navigated interfaces sequentially using a screen reader resulting in 2.28x longer task completion time ($\mu = 211.18, \sigma = 154.99$ seconds). \revision{We conducted Mann-Whitney U test for significance testing and obtained a statistically significant result ($p < 0.0001, Z = 15.08$).} BLVUs executed substantially more number of keyboard actions per task ($\mu = 179.45, \sigma = 163.41$) compared to SUs' keyboard and mouse actions ($\mu = 72.99$) combined per task, \revision{and this was a statistically significant result as well ($p < 0.0001, Z =13.84$)}. Even though BLVUs executed a much larger number of actions per task, their speed (0.85 actions per second) was only slightly higher compared to SUs (0.79 actions per second) as BLVUs often paused to let the screen reader's audio output end to determine which window or UI element was in focus before proceeding. Both SUs and BLVUs completed System Operations tasks the fastest (SUs: $\mu= 52.48, \sigma= 42.85$ seconds; BLVUs: $\mu = 113.92, \sigma = 97.24$ seconds). These tasks typically involved a single application with straightforward navigation. In contrast, Workflow tasks took the longest for both groups (SUs: $\mu= 142.51, \sigma= 107.48 $ seconds; BLVUs: $\mu= 296.86 , \sigma = 152.91$ seconds), as they required using two or more applications simultaneously or in sequence and included steps that are not visually obvious and are nested. We observed that switching apps often reset the focused element and navigation landmarks, so BLVUs had to rebuild their context before continuing. Accordingly, BLVUs show a wider spread in task-completion time (Fig. ~\ref{fig:per_user_tct}): variability is small for System Operations ($\sigma = 97.24$) and Media ($\sigma = 109.02$), but much larger for Document Editing ($\sigma = 141.9$) and Workflow ($\sigma = 152.9$) categories, indicating greater coordination overhead. 

\subsubsection{Task Success Rates}
Across the five categories, SUs succeeded on 99.16\% of tasks on average, but success rate dropped to 84.6\% for BLVUs \revision{and this was statistically significant (p < 0.0001, Fisher’s test).} The rare SU failures occurred when the required step was buried in a non-obvious location. For example, SU2 knew PowerPoint supported changing a slide’s orientation but could not locate the control: SU2 looked for a slide-layout control under ``Layout'', even an app-wide \texttt{Ctrl+F} didn’t help because it was nested under ``Design'' → ``Slide Size'' → ``Custom Slide Size''.
This reasoning extends to BLVUs, with lower success particularly in Document Editing ($\mu=87.5\%$) and Workflow ($\mu=65.6\%$) tasks (Figure ~\ref{fig:task-success-rate-by-category}). All 8 BLVU participants did not complete the Workflow task which involved selecting and recording a particular screen area (presentation-demo scenario) and required using complex keyboard shortcuts, but these were unfamiliar to them. However, all 8 SUs were able to successfully complete the same task by using drag-and-drop-action. Thus, even though tasks in the \system{} dataset are accessible and achievable, these tasks demand substantially more time and effort from BLVUs. From manual review of BLVUs’ recordings, we saw a consistent \textbf{verify-before-commit} routine: in addition to waiting for the screen reader to finish, they re-performed the target action to confirm the outcome. This slowed progress but reduced premature or incorrect actions. SUs, by contrast, rarely repeated actions for confirmation and relied on immediate visual feedback. We also observed that when one command path failed, BLVUs tried workarounds to continue task execution. For example, in a task involving Bookmarking a webpage, BLVUs cycled focus with Tab/arrow or opened the context menu (\texttt{Shift+F10}) if toolbar buttons weren’t reachable. Sometimes when the hotkeys they attempted did not work as expected, they tried alternate shortcut sequences. This pattern also appeared in SUs: when a familiar/easy route didn’t work, they tried a different path (toolbar vs. right-click context menu), scanned neighboring icons in the applications to locate the target control.

\subsubsection{Input Modality and Navigation}
We confirm that the interaction traces from SUs are mouse-dominant. SUs performed an average of 51.93 mouse actions ($\sigma = 66.29$), 13.52 scrolls ($\sigma = 53.85$), 1.26 drag-and-drops ($\sigma = 2.56$) and 21.06 key presses ($\sigma = 32.71$). BLVUs traces, in contrast, are keyboard-dominant. BLVUs interactions averaged 179.45 key presses ($\sigma = 163.41$) and  13.63 hotkey activations ($\sigma = 25.41$) per task with no mouse interactions. Arrow keys (including \texttt{Tab} presses) accounted for a large fraction of their keystroke activity (72.23\%), indicating that BLVUs spent a large share of their time navigating and moving focus to the target UI element than acting on controls. Repeated keypresses of Tab or arrow keys inflated action counts, since navigation required serial exploration, retries, and backtracking across multiple UI elements. In addition, BLVUs also made certain workflows efficient via heavy hotkey use (7.5\% of total keyboard actions on average), particularly in document editing (10.7\%) and workflow tasks (11.3\%) as compared to other (web: 5.1\%, system operations: 3.1\%, media: 2.9\%) categories. Modifier-specific hotkeys expose distinct functionalities. BLVUs use more \texttt{Ctrl/Alt/Win/Shift} hotkeys across categories: \texttt{Ctrl} to jump or edit text (e.g., \texttt{Ctrl+L/F/C/V}), \texttt{Alt} to navigate announced menus/ribbons or switch tasks, \texttt{Win} to open OS functions and recover focus (\texttt{Win+I/E/R, Win+1–9}), and \texttt{Shift} for reverse-tabbing and precise range selection. SUs rarely used \texttt{Alt/Win/Shift} (0.18\% of total keyboard actions). SUs could visually “skip” irrelevant items and move directly to the target with a single mouse click. Additionally, SUs tend to use keystrokes primarily tasks involving web browsing (0.6:1 keyboard to mouse actions ratio) or editing a document (0.7:1 keyboard to mouse actions ratio) as they naturally involve typing (e.g., form-filling, URLs, search queries, drafting a document) compared to tasks in system operations (0.13:1 keyboard to mouse actions ratio), media (0.25:1 keyboard to mouse actions ratio) or workflow (0.29:1 keyboard to mouse actions ratio) categories. 

\begin{figure*}
    \centering
    \includegraphics[width=\linewidth]{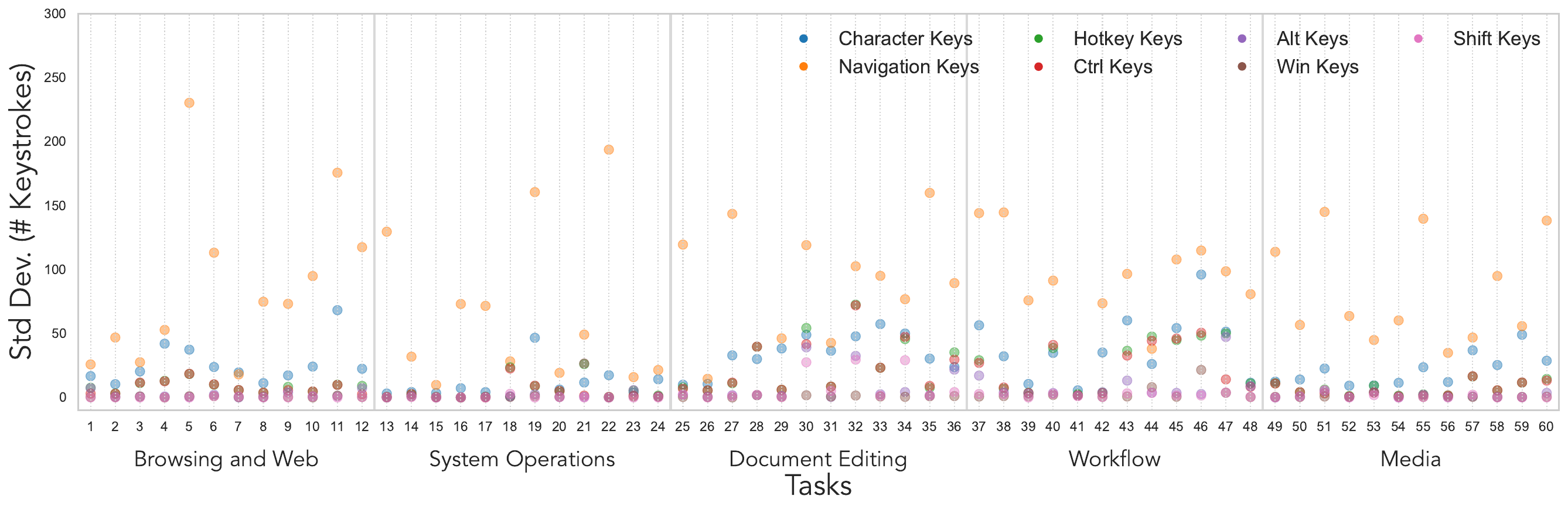}
    \caption{Cross-BLVU standard deviation in keystrokes by task and keystroke type (lower is more consistent). Each dot reports, for a given task, the variance across BLVUs in one keystroke category: character input, navigation (Tab/Arrow), hotkeys (Ctrl, Alt, Win, Shift). Variability is generally small for browsing \& web and system operations, but rises sharply for document editing (largest spikes, driven by character and navigation counts) and remains elevated in Workflow and media. Hotkeys categories are typically low-variance, with occasional Ctrl spikes during editing. The high dispersion indicates diverse strategies (typing vs. navigation vs. shortcuts) for the same task. }
    \Description[]{A scatter plot showing the standard deviation of keystrokes across 60 tasks in five categories: Browsing and Web, System Operations, Document Editing, Workflow, and Media. The x-axis represents tasks grouped by category. The y-axis shows the standard deviation in number of keystrokes, ranging from 0 to 300. Different colors represent key types: blue (Character Keys), orange (Navigation Keys), green (Hotkey Keys), red (Ctrl Keys), brown (Win Keys), purple (Alt Keys), and pink (Shift Keys). Navigation Keys (orange) consistently have the highest variability, with many points exceeding 100 and some above 200. Character Keys (blue) show moderate variability, while all modifier keys (Ctrl, Alt, Win, Shift, Hotkeys) generally remain clustered near the lower end (0–40). The trend suggests that navigation input contributes the most variability in keystroke usage across tasks.}
    \label{fig:per_user_tct}
\end{figure*}

\begin{figure*}
    \centering
    \includegraphics[width=\linewidth]{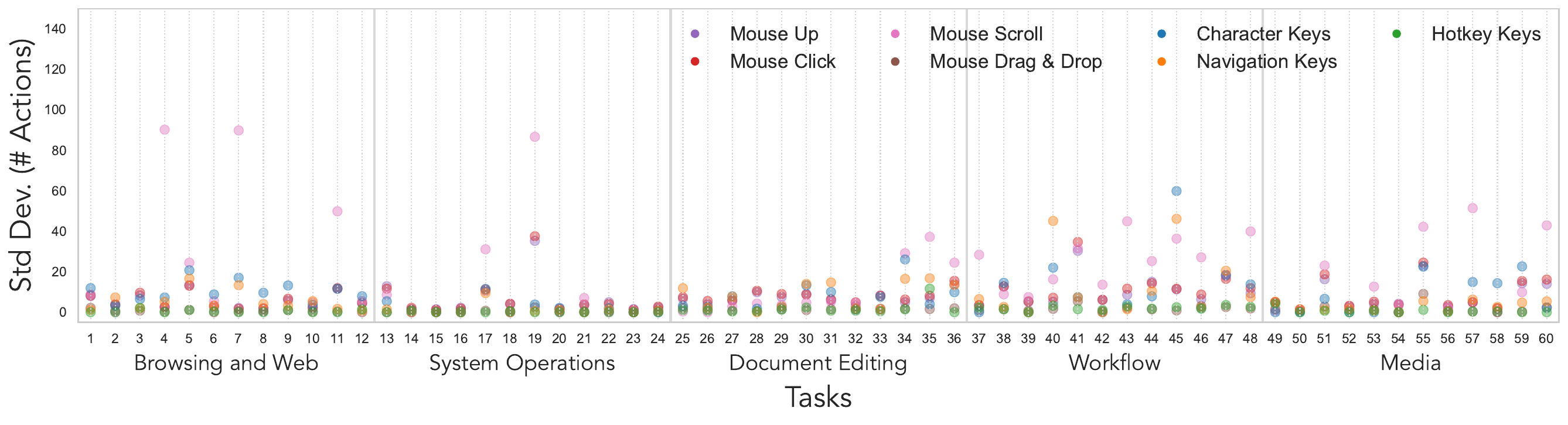}
    \caption{Cross-SU standard deviation in keyboard and mouse actions by task and action type (lower is more consistent). Each dot shows the standard deviation of action counts for a given task (1–60) and actions. Larger values indicate less predictable behavior. Mouse scroll exhibits the highest dispersion with occasional large spikes. Mouse clicks and navigation keys are also variable, especially in workflow and media. Hotkeys remain near zero across tasks, and drag-and-drop only spikes on tasks that require it.}
    \Description[]{A scatter plot showing the standard deviation of user actions across 60 tasks in five categories: Browsing and Web, System Operations, Document Editing, Workflow, and Media. The x-axis represents tasks grouped by category. The y-axis shows the standard deviation in number of actions, ranging from 0 to 140. Different colors represent action types: purple (Mouse Up), pink (Mouse Scroll), red (Mouse Click), brown (Mouse Drag & Drop), blue (Character Keys), orange (Navigation Keys), and green (Hotkey Keys). Mouse Scroll (pink) and Mouse Up (purple) have the highest variability, often exceeding 40 and peaking above 120. Other actions, such as Mouse Clicks, Character Keys, Navigation Keys, and Hotkeys, generally remain in the lower range (0–30). The trend suggests mouse-based interactions (especially scrolling and button release) contribute the most variability in task execution.}
    \label{fig:per_user_tct}
\end{figure*}

\subsection{Interaction Diversity within Sighted and Blind and Low-Vision User Groups} 

We observed that participants in the same group often completed the same task in different ways. In this section we examine within-group variation: how SUs and BLVUs differ from one another within their own groups. We define ``interaction method'' along three dimensions (i) action mix (e.g., mouse input, character input, navigation via tabbing, and hotkeys), (ii) tempo (time per action), and (iii) navigation structure (e.g., point-and-click, shortcuts, menu traversal, or search-first strategies). The following subsections ground these patterns with per-user, per-task examples drawn from our logs in \system{} dataset.

\subsubsection{Sighted Users (SUs)} SUs complete tasks with a mixture of pointing-clicking and typing. Their main within-group difference is mouse-first versus shortcut-friendly behavior, with a few users mixing both. 
\revision{For instance, the same shopping task (task 4: adding a 10 inch frying pan to cart on \texttt{walmart.com}) shows three distinct styles. SU2 uses typed targeting with a few precise clicks (35.4 seconds, 40 actions, 83\% keyboard, 18\% mouse, tempo $\approx$ 0.89 seconds/action) minimizing steps by jumping directly to search and filters. SU6 uses browsing and scrolling (94.0 s, 291 actions, 6\% keyboard, 94\% mouse, tempo $\approx$ 0.32 seconds/action) executing each action quickly but accumulating many more of them. SU3 uses a hybrid approach (68.8 seconds, 61 actions, 56\% keyboard, 44\% mouse, $\approx$ 1.13 seconds/action).} The within-SU group variation is more evident when menus are involved for task completion, users either navigate visually with the mouse or stitch in shortcuts.
We see a clear contrast between drag-and-drop and a non-drag menu path in a Systems Operation task of moving a file from one folder to another. SU3 completes the move with a single drag (26.0 seconds, 10 actions, 0\% keyboard, 100\% mouse, tempo $\approx$ 2.60 seconds/action) and SU8 does similarly (26.0 seconds, 14 actions, 0\% keyboard, 100\% mouse, tempo $\approx$ 1.86 seconds/action). By comparison, SU2 skips dragging and uses a sequence of clicks/menus (21.3 seconds, 14 actions, 14\% keyboard, 86\% mouse, 1.52 seconds/action) beating the drag-and-drop times.
In the Workflow task requiring users to copy text from Word to the Notes section in PowerPoint (task 37) shows mouse-dominant flows with a few hotkeys: SU2 (104.9 seconds, 10 keyboard + 91 mouse actions), SU6 (130.0 seconds, 3 keyboard + 62 mouse actions), SU8 (72.7 seconds, 11 keyboard + 50 mouse actions). Across applications, SUs mainly point and click, using small bursts of hotkeys (Ctrl+C/V, Alt+Tab) when convenient.
Tasks like “Change the video speed” (task 49) and “Fast-forward a video by 30 seconds” (task 48) are uniformly pointer-centric (e.g., SU6: 57.9 seconds, 23 mouse actions; SU2: 31.7 seconds, 4 keyboard + 20 mouse actions). Variation exists but is smaller than in Browsing and Workflow tasks because the UI exposes direct, visible controls.

\subsubsection{Blind and Low-Vision Users (BLVUs)}\label{BLVU-interaction}
BLVUs complete the same goals with different keyboard-based methods as they pair screen-reader feedback with distinct navigation methods they’ve learned over time. We describe three recurring methods: walking (\texttt{Tab}/Arrow step-by-step), chunk-jumping (\texttt{Ctrl/Shift} shortcuts that act on larger units or jump to targets), and ribbon/OS routes (\texttt{Alt/Win} menus and system dialogs). We demonstrate the within-group differences in interaction for BLVUs using Web \& Browsing, Document Editing, and Workflow tasks where we observe maximum variance in the input keyboard actions. For Web \& Browsing tasks, the BLVU group splits into search-first versus navigation-first behaviors. For example, for a task requiring participants to find a specific information on a webpage (task 4), BLVU1 uses a compact search pattern using \texttt{Ctrl+L/F} (237.13 seconds, 64 actions tempo $\approx$ 3.70 seconds/action, longer confirmation per step but very few steps). In contrast, BLVU6 walks in smaller increments (76.86 seconds, 77 actions, tempo $\approx$  0.99 seconds/action), trading more steps for continuous feedback. The same split scales up on multi-step flows. For a web task which needed participants to log into \texttt{target.com} website and modify a user profile (task 5), BLVU8 walks extensively (700.72 seconds, 479 actions; 80\% navigation), while BLVU2 proceeded to login page quickly (296.70 seconds, 298 actions; 50\% characters keys, 36\% navigation keys, 26\% hotkeys).
For tasks in Document Editing category such as creating a chart in Excel (task 30), BLVU1 relied on walking with periodic hotkeys use (703.4 seconds, 598 actions; 408 navigation steps, 134 hotkeys), while BLVU8 uses ribbon/selection sequences (473.7 seconds, 245 actions; 113 hotkeys with prominent \texttt{Alt/Shift} usage). Word document formatting task (task 36) shows a clean chunk-jumping example of B1 using many \texttt{Ctrl} sequences (244.7 seconds, 448 actions; 211 hotkeys, Ctrl=210 actions), whereas BLVU7 succeeds with very few decisive steps (79.3 seconds, 39 actions). Thus, when tasks involve menus, ribbons, or other structured content, the method a user chooses (walking, chunk-jumping, or ribbon routes) largely determines how many operations are needed and, in turn, drives the wide variation in completion times.

Workflow tasks needed users to coordinate across applications.
A task on extracting text from an image (task 46), surfaced three BLVU routes. BLVU2 worked with Win-key launches/toggles (432.5 seconds, 179 actions; Win=63), BLVU7 runs Ctrl-centric cycles for selection/transfer (311.3 seconds, 315 actions; Ctrl=142) but is unsuccessful, and BLVU8 primarily walks the UI (450.0 s, 475 actions; 95\% navigation) successfully. Creating a desktop shortcut shows similar diversity: BLVU5 completed the task with a short, direct path (208.8 seconds, 192 actions), while BLVU3 used Alt/ribbon menus across apps (547.4 seconds, 337 actions; Alt=139), indicating that when workflows span tools, \texttt{Alt/Win} use rises for launching applications and navigating menus. Modifier patterns make these methods visible in logs (\texttt{Ctrl/Shift} for chunk-jumping, \texttt{Alt/Win} for menus/OS,  \texttt{Tab/Arrow keys} for ``walking''), providing a clear and comparable way to describe within-group diversity.

\begin{table*}[]
\small
\setlength{\tabcolsep}{3pt}
\begin{tabular}{llcccc|*{3}{cc}|*{3}{cc}}
\toprule
\multicolumn{2}{l}{} &
\multicolumn{4}{c|}{\textbf{Participants}} &
\multicolumn{6}{c|}{\textbf{Claude Sonnet 4.5}} &
\multicolumn{6}{c}{\textbf{Qwen3-VL-32B-Instruct}} \\
\multicolumn{2}{l}{} &
\multicolumn{2}{c}{SUs} &
\multicolumn{2}{c|}{BLVUs} &
\multicolumn{2}{c}{Default-CUA} &
\multicolumn{2}{c}{SR-CUA} &
\multicolumn{2}{c|}{Magnifier-CUA} &
\multicolumn{2}{c}{Default-CUA} &
\multicolumn{2}{c}{SR-CUA} &
\multicolumn{2}{c}{Magnifier-CUA} \\
\multicolumn{2}{l}{} &
$\mu$ & $\sigma$ &
$\mu$ & $\sigma$ &
$\mu$ & $\sigma$ &
$\mu$ & $\sigma$ &
$\mu$ & $\sigma$ &
$\mu$ & $\sigma$ &
$\mu$ & $\sigma$ &
$\mu$ & $\sigma$ \\
\midrule
\multicolumn{2}{l}{Task Success  Rate (\%)} & 99.1\% &   & 84.6\% &  & \revision{78.3\%} &  & \revision{41.6\%} &  & \revision{28.3\%} & &\revision{20.0\%} &\revision{} &\revision{0.0\%} &\revision{} &\revision{0.0\%} &\revision{}\\

\multicolumn{2}{l}{Task Duration (s)} & 92.3 & 78.0  &  211.1 & 154.9  & \revision{324.8} & \revision{160.5} &\revision{650.9} &\revision{341.8} &  \revision{1072.2} & \revision{337.9} &\revision{134.9} &\revision{20.0} &\revision{584.9} &\revision{214.7} &\revision{257.2} &\revision{145.3}\\

\multicolumn{2}{l}{\revision{\# Applications}} & \revision{2.2} & \revision{0.0}  &  \revision{2.2} & \revision{0.7}  & \revision{2.2} & \revision{1.0} &\revision{2.2}&\revision{0.9} & \revision{3.4} & \revision{1.4} &\revision{2.08} &\revision{0.9} &\revision{0.7} &\revision{0.6} &\revision{0.8} &\revision{1.0}  \\

\multicolumn{2}{l}{\# Events} & 92.5 & 104.6 &  237.3 & 224.0  &\revision{110.0} &\revision{278.9}& \revision{281.7}&\revision{305.9}& \revision{335.6} & \revision{347.7} &\revision{146.78} &\revision{165.5} &\revision{67.4} &\revision{102.1} &\revision{77.7} &\revision{88.8}\\

\multicolumn{2}{l}{\# Mouse Actions}& 51.9 & 66.3 & 0.0 & 0.0 &\revision{37.6} &\revision{18.6} &-& -&  \revision{64.6} &
\revision{23.6}  &\revision{83.3} &\revision{30.9} &\revision{-} &\revision{-} &\revision{67.6} &\revision{81.8}\\

\multicolumn{2}{l}{\hspace*{\cuaindent}\# Mouse up} & 18.0 & 15.8 & 0.0 & 0.0  & \revision{16.7}&\revision{8.8} & -& -& \revision{26.8} & \revision{10.6} &\revision{40.4} &\revision{16.4} &\revision{-} &\revision{-} &\revision{30.7} &\revision{38.8} \\

\multicolumn{2}{l}{\hspace*{\cuaindent}\# Mouse click} & 19.0 & 17.41  &  0.0 & 0.0  &\revision{18.0} &\revision{9.3} &- &- &  \revision{31.4} & \revision{11.7} &\revision{41.6} &\revision{15.5} &\revision{-} &\revision{-} &\revision{36.8} &\revision{44.2} \\

\multicolumn{2}{l}{\hspace*{\cuaindent}\# Scroll} &13.5 &53.8  &  0.0 & 0.0  & \revision{6.3} & \revision{5.9} &- &-  & \revision{0.0} & \revision{0.5} &\revision{0.4} &\revision{1.9} &\revision{-} &\revision{-}&\revision{0.0} &\revision{0.0} \\

\multicolumn{2}{l}{\hspace*{\cuaindent}\# Drag-and-drop} & 1.2 & 2.5 &  0.0 & 0.0  & \revision{0.0}&\revision{0.4} &- &-  &  \revision{0.0} & \revision{0.5} &\revision{0.7} &\revision{5.5} &\revision{-} &\revision{-} &\revision{0.0} &\revision{0.0} \\

\multicolumn{2}{l}{\# Keyboard Actions}& 21.0 & 32.7  &  179.4 & 163.4   & \revision{67.1} & \revision{258.7} & \revision{210.6} & \revision{230.9} & \revision{241.8} & \revision{292.5} &\revision{55.2} &\revision{173.1} &\revision{16.6} &\revision{39.1} &\revision{4.45} &\revision{19.8} \\

\multicolumn{2}{l}{\hspace*{\cuaindent}\# Character Keys} & 14.6 & 23.6  & 36.1 & 41.5  &\revision{44.6} &\revision{146.1} & \revision{125.6}& \revision{145.5}& \revision{150.0} &
\revision{180.8} &\revision{46.1} &\revision{154.2} &\revision{1.1} &\revision{3.9} &\revision{1.6} &\revision{12.9}  \\

\multicolumn{2}{l}{\hspace*{\cuaindent}\# Arrow keys} & 5.5 & 12.9  &  129.6 & 143.3  & \revision{22.5}& \revision{113.7}&\revision{77.9} &\revision{115.3} &  \revision{83.6} & \revision{113.7} &\revision{8.8} &\revision{20.4} &\revision{2.0} &\revision{12.3} &\revision{0.0} &\revision{0.0} \\

\multicolumn{2}{l}{\hspace*{\cuaindent}\# Hotkeys} &  0.8 & 2.3  &  13.6 & 25.4  & \revision{0.0}& \revision{0.0}& \revision{7.0}& \revision{5.3}& \revision{8.1} & \revision{4.1} &\revision{0.3} &\revision{2.5} &\revision{13.4} &\revision{33.2} &\revision{2.7} &\revision{15.3} \\

\multicolumn{2}{l}{\hspace*{2\cuaindent}\# Ctrl-based} & 0.7 & 2.2  &   10.0 & 22.1  & \revision{0.0}& \revision{0.0}& \revision{4.1}&\revision{3.3}& \revision{2.5} &\revision{3.5} & \revision{0.3} &\revision{2.5} &\revision{2.0} &\revision{12.4} &\revision{1.6} &\revision{12.6} \\

\multicolumn{2}{l}{\hspace*{2\cuaindent}\# Alt-based} & 0.0 & 1.5  &  2.9 & 10.6 & \revision{0.0}& \revision{0.0}& \revision{2.5}& \revision{2.9}&  \revision{3.2} & \revision{3.0} &\revision{3.2} &\revision{3.0} &\revision{12.9} &\revision{33.2} &\revision{0.0} &\revision{0.0} \\

\multicolumn{2}{l}{\hspace*{2\cuaindent}\# Win-based}& 0.0 & 0.1 &  0.9 & 4.7  & \revision{0.0}& \revision{0.0}& \revision{0.5}& \revision{0.7} &  \revision{1.3} & \revision{1.4} &\revision{0.0} &\revision{0.0} &\revision{0.0} &\revision{0.0} &\revision{1.1} &\revision{8.9} \\

\multicolumn{2}{l}{\hspace*{2\cuaindent}\# Shift-based}& 0.0 & 0.1  &  1.4 & 7.3  & \revision{0.0}& \revision{0.0}& \revision{0.3}& \revision{0.7}&  \revision{0.2} & \revision{0.6} &\revision{0.0} &\revision{0.0} &\revision{0.4} &\revision{3.2} &\revision{0.0} &\revision{0.0}\\
\bottomrule
\end{tabular}
\caption{\revision{Per-task metrics for participants (SUs, BLVUs) and CUA conditions for Claude Sonnet 4.5 and Qwen3-VL-32B-Instruct models. Default-CUA, SR-CUA, and Magnifier-CUA are evaluated at 50, 100, and 100 steps respectively.} Values are mean $(\mu)$ and standard deviation $(\sigma)$. }
\label{tab:task-metrics-cua}
\end{table*}

\section{Analysis of CUA Performance and Behavior Across Assistive Technology Conditions}
Our goal is to analyze how state-of-the-art CUAs perform everyday desktop and web tasks under conditions that mirror both sighted use (default) and AT use (screen reader and magnifier) and their readiness to assist users, BLVUs in particular. We ask the following research questions,
\begin{enumerate}
    \item[\textbf{RQ1.}] How does success rate and completion time vary for CUAs under default and AT conditions?
    \item[\textbf{RQ2.}] How do action-space limits (keyboard-only, no mouse) and input-space limits (restricted magnified viewport) change agent behavior?
    \item[\textbf{RQ3.}] How closely do agent traces under each condition match SUs and BLVUs?
\end{enumerate}

\subsection{CUA Conditions}
\revision{Claude Sonnet 4.5 is the best general purpose model currently for computer use as listed on OSWorld-Verified ~\cite{OSWorld-Website} leaderboard. It successfully completes 58.1\% and 62.9\% of the tasks on the OSWorld benchmark in 50 and 100 steps, respectively. Among the open source general models, Qwen3-VL series is also among the top with a 41.4\% success rate at 100 steps. Therefore, we evaluate Anthropic's \texttt{claude-sonnet-4-5-20250929} and Alibaba Cloud's \texttt{qwen3-vl-32b-instruct} models on our \system{} dataset under three conditions that mirror common usage patterns for sighted and AT (screen reader and magnifier) users. We cap the CUAs at 50 steps for the default condition and 100 steps for the AT conditions to match the OSWorld setting. These numbers also approximately mirror the observed SU–BLVU action ratio in our dataset while preventing accidental cost blow-ups.} Each CUA is given the same task context and instructions as the participants to match task conditions. We define the CUA’s \textbf{input space} (what it can see on screen) and \textbf{action space} (what actions are allowed) for each of the three conditions,

\begin{enumerate}
    \item \textbf{Default-CUA.} Uses both keyboard and mouse and views the full desktop at standard scaling. The CUA follows an incremental loop: outline a 1–3 step plan, request a screenshot as the first tool action, execute one action, then request the next screenshot (see full prompt in ~\ref{subsec: appendixa1}).
    \item \textbf{Screen-Reader-CUA.} Operates with a keyboard-only action space and disables mouse actions to mirror screen-reader use. The SR-CUA \revision{still} views the full desktop \revision{context that is available to the default-CUA} while the NVDA screen reader runs. We prompt the SR-CUA to assist a blind user and to complete all operations using keystrokes (see full prompt in ~\ref{subsec: appendixa2}).  
    \item \textbf{Magnifier-CUA.} Uses keyboard and mouse but sees a 150\% magnified viewport. The magnifier limits visible context to the CUA and increases panning and scrolling. We prompt the Magnifier-CUA to assist a low-vision user working within a limited field of view (see full prompt in ~\ref{subsec: appendixa3}). 
\end{enumerate}

\noindent\revision{The approximate cost of running Claude Sonnet 4.5 for the three conditions was \$3060 with 28 hours of execution time. Running Qwen3-VL-32B-Instruct's CUA for default and AT conditions did not incur any cost as we hosted it on our lab servers (two units of NVIDIA RTX 6000 ADA GPU). The execution time was around 16 hours for Qwen3-VL. We automate the data collection of CUA execution for all the tasks. However, we manually evaluated the CUA recordings for error analysis and to determine task success based on applications opened and whether the end state is reached in the visible viewport, which took around 4 hours in total for all models and conditions.} 

\subsection{CUA Behavior Across Conditions (Default, Keyboard-Only, Magnifier)} 

\subsubsection{Task Success Rate} Among the 60 tasks in the \system{} dataset, \revision{Sonnet 4.5's} default-CUA is the only agent with broad coverage. The default-CUA has a significantly higher overall average task success rate of \revision{78.33\%} as compared to SR-CUA \revision{(41.67\%)} and magnifier-CUA \revision{(28.33\%)}. The default-CUA performs successfully on web \& browsing \revision{(83.33\%)}, system operations \revision{(100\%)}, and media \revision{(66.66\%)} tasks but shows lower success on document editing \revision{(66.66\%)}, and particularly workflow tasks \revision{(58.33\%). The SR-CUA (keyboard-only, no mouse) performs best on \revision{system operation tasks (75\%), followed by document editing (58.33\%), web browsing tasks (50\%)}, but at a substantially lower success rate than default-CUA. SR-CUA performance drops sharply for workflow (16.66\%) and media (8.33\%) tasks.} Magnifier-CUA successfully completed one task each in Web \& Browsing and Workflow categories and failed on all the other tasks. \revision{On the other hand, Qwen3-VL's default-CUA achieves a task success rate of 20\% and fails on all tasks under SR-CUA and magnifier-CUA conditions.} \revision{Since Claude's Sonnet 4.5 shows a wide performance gap, we discuss results and observations in the following sections. We report the quantitative analysis of Qwen3-VL model's performance for tasks in \system{} dataset in Table ~\ref{tab:task-metrics-cua}}. 

\subsubsection{Task Completion Time \& Speed.} In addition to higher success rates, the \revision{Sonnet 4.5's} default-CUA also recorded a lower task completion time of \revision{324.87 seconds ($\sigma = 160.57$ seconds)} with an average speed of \revision{0.34 actions per second}. SR-CUA took \revision{2x times longer per task ($\mu = 650.91, \sigma = 341.84$ seconds)} but operated with slightly higher speed \revision{($\mu = 0.32$ actions per second)}. Default-CUA can point-and-click visible controls, while SR-CUA must step through long Tab/Arrow sequences to reach the same targets, requiring longer time. The magnifier-CUA also had a much longer average task completion time \revision{($\mu = 1072.20, \sigma = 337.90$ seconds)} as compared to the default-CUA. Magnifier-CUA executed actions at a speed of \revision{0.28 actions per second}, lesser than SR-CUA. 
 
\subsubsection{Interaction Methods.} On an average, the number of mouse and keyboard actions combined per task is the highest with magnifier-CUA \revision{(305.51 actions)}, followed by  SR-CUA \revision{(keyboard-only, 210.61 actions)} and then default-CUA \revision{(104.8 actions) for Sonnet 4.5}. 
For the same tasks, the SR-CUA performs significantly more number of arrow actions per total task actions \revision{(51.9\%)} than default-CUA \revision{(21.46\%)} or magnifier-CUA \revision{(27.30\%)} indicating extra navigation steps and retries (e.g., cycling between panes, re-focusing input fields). Default-CUA reaches the goal with fewer steps with typically short sequences of mouse clicks. Default-CUA never used drag-and-drop and never invoked any of the hotkeys (e.g., \texttt{Ctrl+L/F/V}) during any of the tasks. It instead relied on pointing to menu items and toolbar buttons. Magnifier-CUA did attempt drag-and-drop action once but failed. Magnifier-CUA also shows similar hotkey interactions behavior as SR-CUA despite being able to use mouse actions. Magnifier-CUA used \texttt{Win+R} hotkey activations \revision{($\mu = 1.36, \sigma = 1.47$)} more frequently compared to SR-CUA \revision{($\mu = 0.5, \sigma = 0.72$)} to open applications needed for the tasks using Run program.  
\revision{We observed that Sonnet 4.5 often completed tasks from the command prompt itself, instead of navigating to and opening the specified folder where the task resource existed and then taking relevant actions for task completion}.

\subsubsection{Failure Modes.}\label{cua-failure}
A recurring failure across tasks for all three agents was completing every intermediate step but omitting the final confirming action (e.g., exporting to PDF without clicking ``Save'', or changing a setting without clicking ``Apply''/``Done''). This pattern suggests weak end-of-task validation and missing explicit success checks. Because the environment started each run with all applications closed, SR-CUA often struggled to locate the required documents. On many tasks, both SR-CUA and Magnifier-CUA eventually found the document but stopped there, implicitly treating ``locate file'' as the goal rather than proceeding with the required operation. Magnifier-CUA frequently used character entry in File Explorer to jump to paths.
Default-CUA often failed when the next action depended on non-visible state or controls buried in sub-menus (e.g., selecting a theme inside PowerPoint’s Design gallery, changing a setting in a second-level Settings pane, or revealing a media player menu hidden by an overlay). In these cases it clicked near the target but struggled to reach the specific nested control, or acted with the wrong selection. \revision{Qwen3-VL performed extremely poorly under the SR-CUA condition, yielding a 0\% success rate. The primary source of failure was the agent’s repeated use of \texttt{Alt+Tab} to cycle through application windows in an attempt to locate the target application, which was never successfully identified. Qwen’s Magnifier-CUA also achieved a 0\% success rate with the model consistently producing incorrect coordinates: rather than clicking within the magnified viewport, it generated coordinates relative to the full-screen viewport. This mismatch led to systematic misclicks and task failures.} 

\subsection{Comparing CUA Interactions to Sighted and Blind and Low-Vision Users Workflows}
\subsubsection{Sighted Users vs. Default-CUA} 
Default-CUA is much less successful than SUs \revision{(78.33\% vs. 99.1\%)} and takes more than \revision{thrice} as long per task on average \revision{(default-CUA: $\mu = 324.87.7, \sigma=160.57$ seconds vs. SUs: $\mu = 92.3, \sigma= 78.0$ seconds)}. Default-CUA executes approximately 1.4x more total mouse and keyboard actions combined \revision{($\mu=104.83$)} compared to SUs ($\mu=72.9$) per task. The input mix also diverges: SUs show mouse-dominant behavior (52.89 mouse actions vs 21.0 keyboard actions per task) with routine scrolling \revision{(13.5 per task)} and some drag-and-drop \revision{(1.2 per task)}, whereas default-CUA performs much fewer mouse actions \revision{(37.6 per task)} and scrolling \revision{(2.8 per task)}, and almost never uses drag-and-drop. SUs occasionally used \texttt{Ctrl} based hotkeys \revision{($\mu = 0.7, \sigma = 2.2$ actions per task)} to copy/cut/paste text/files, but default-CUA uses none. \revision{Default-CUA tends to type text manually (e.g., long URLs), making character keystrokes nearly half of total actions (42.5\%), whereas SUs usually copy and paste wherever possible.} Thus, the default-CUA only partially resembles sighted behavior in that it points and clicks visible controls, but it avoids efficient moves that SUs use such as drag-and-drop, scrolling to reveal context, and hotkeys and instead uses longer, \revision{more keyboard-centric paths}. The task success rate of default-CUA  is similar to BLVUs \revtwo{The performance gap between default-CUA and BLVUs is comparable for} web \& browsing \revision{(default-CUA: 83.33\%, BLVUs: 92.70\%), system operations (default-CUA: 100\%, BLVUs: 93.75\%), document editing (default-CUA: 66.6\%, BLVUs: 58.3\%), and media (default-CUA: 83.33\%, BLVUs: 83.33\%) tasks, but the gap increasingly widens for workflow (default-CUA: 33.33\%, BLVUs: 65.62\%) tasks}.    

\subsubsection{Blind and Low-Vision Users vs. Screen-Reader-CUA}
SR-CUA diverges sharply from workflows of BLVUs who use screen readers for assisting them. BLVUs succeed on most tasks (86.9\% task success) and complete them in 211.1 seconds ($\sigma=154.9$) on average, while SR-CUA succeeds on only \revision{41.67\% and takes 650.9 seconds ($\sigma=341.8$)}. Compared to BLVUs, SR-CUAs also generate similar number of total events \revision{(1.18x)}, produce fewer arrow-key actions \revision{(0.60x)}, and substantially lower hotkeys \revision{(0.51x)} but higher character keys \revision{(3.45x)} per task. \revision{We observed that for tasks involving compressing a folder, changing a file format, navigating to certain URL on the web, Sonnet 4.5 executed long text commands to complete the tasks, differing from the interaction style of BLVUs who first navigate to a specified location and then explore options for task completion.} We also observed that when an incorrect action occurs, BLVUs tried to repair in-place by re-selecting the intended item, or re-doing the most recent step for confirmation, and then continued from the current state. In contrast, the SR-CUA rarely perform minimal repair, they often replayed long navigation sequences or restarted the flow to re-establish context, which increased both the number of actions and the total completion time. While BLVUs look for explicit confirmations (“Saved”, “Submitted”, “Added to cart”) for ensuring task completion, SR-CUA often completed the intermediate steps but omitted the final confirmation (not pressing Save/Submit), and stopped. In addition, BLVUs regularly confirm which window or pane is active (visual highlight spoken focus) before acting. However, we observed that SR-CUA lacked robust focus tracking, so after a dialog or overlay, they often continued issuing commands to the wrong application context.

\section{Discussion}
Our study aims to characterize the accessibility gap of CUAs by building a dataset of BLVUs and SUs performing 60 everyday tasks, comprising over 40+ hours of interaction traces and more than 158k+ events. Analysis of this dataset reveals distinct interaction patterns between SUs and BLVUs, as well as variability in strategies. For example, SUs may rely on context menus, shortcuts, or drag-and-drop actions, while BLVUs primarily use sequential navigation and keyboard shortcuts. Our evaluation of \revision{Sonnet 4.5's CUAs } on these tasks shows that CUAs operating with screen readers perform poorly, completing only \revision{41.67\%} of tasks, whereas the default-CUA achieves a \revision{78.33\%} success rate. These findings, along with our dataset construction process, highlight both the limitations of current CUAs under different assistive technology conditions and the challenges involved in collecting the A11y-CUA dataset. Ultimately, our work moves toward developing CUAs that are more accessibility-aware. We elaborate on these insights below.

\subsection{Limitations of CUAs under AT Conditions}
Across default-CUA and SR-CUA, we observed a consistent pattern: agents could complete straightforward, single-application tasks but became brittle in longer workflows, echoing the findings of prior works, such as ~\cite{xie2024osworld, song2025coact}.
For example, all three successfully completed the task of bookmarking the Facebook homepage. In contrast, multi-app tasks often stalled after finishing steps in one application without reliably handing off to the next. For instance, all three CUAs failed in a task that required them to make a list of popular books from Wikipedia in an Excel sheet. The default-CUA opens the right webpage on Chrome, but fails to copy-paste data into Excel. 
Even though SR-CUA ran with NVDA on and could see focus regions, it could not use the screen reader feedback or access the rich accessibility tree that BLVUs work with. Lacking those cues, it frequently mistimed actions around dialogs and status changes. Providing SR-CUA with a list of task-relevant hotkeys led to a small improvement in success (16.66\%), but variability remained high (we discuss these results in A.5).

The Magnifier-CUA had a perception gap in that magnification narrowed its viewport, so controls were often just off-screen, prompting frequent re-centering maneuvers (e.g., repeated \texttt{Win+R} to open applications). Even when the correct File Explorer window and folder were visible, it re-typed the path into the address bar, failed to commit the navigation, and quit, indicating an action gap. We also observed a cognitive gap: when stuck on a zoomed-in region, it failed to infer where the desired target lay relative to the current view and thus didn’t choose an effective pan direction.

These cases point to three gaps. The \textbf{perception gap} arises when agents cannot access the same signals that users perceive (e.g., screen reader announcements, ARIA state changes, or content off-viewport under magnification). The \textbf{cognitive gap} appears as weak tracking of task state and goal completion across steps and applications. The \textbf{action gap} shows up as under-use of robust input methods (hotkeys), fragile drag-and-drop, and failing to execute reliable “finish moves” in the right context. Future work should run further ablations to isolate causes: add an AT-synchronized perception feed (screen-reader output and accessibility tree input), switch between hotkey-first and scroll-first plans, vary magnification and SR speech rate, and separately test perception vs. execution by replaying identical plans with and without AT signals. Measuring the fraction of actions spent on recovery (retries, backtracks, re-focusing) could also clarify where each CUA loses time and why.

\subsection{Challenges in Collecting the \system{} Dataset}
Our study procedure highlights the trade-off between maintaining participant privacy and capturing participants’ real-world accessibility practices. We chose a controlled setup (i.e., participants remotely controlled our laptop) to minimize any compromise on participants' personal information, to standardize application configurations, and to reduce the burden of installing our custom recorder on participants’ personal computers. However, this design choice surfaces challenges in replicating participants’ authentic accessibility setups and added study.

All participants were given time to configure their screen readers in our computer environment, and they confirmed that the setup was generally similar to their home configurations. However, some personal customizations, such as personalized hotkeys or braille display settings, were not fully transferred to our computer. In a few cases ($N$ = 2 out of 8), participants initially attempted to use their custom shortcuts, but our system did not respond as expected, leading to errors or extra steps. They then switched to more traditional ways and abandoned the custom shortcuts. Although we considered importing configurations via built-in screen reader export features, compatibility issues and participants’ limited knowledge of their own setups made the process not always possible. Despite these challenges, such issues were rare long-tail cases; most participants did not encounter them and all of them were still interact with computers using common hotkeys as shown in our result in Section~\ref{BLVU-interaction}. We highlight these edge cases to underscore the importance of smoother screen reader configuration processes for future data collection. Future work should explore privacy-preserving approaches that can effectively replicate participants’ personal environments. One possible direction is to design sandbox systems that act as secure environment and can seamlessly replicate screen reader or accessibility settings from the participants computer. 

Our data collection used closed-source Windows applications (e.g., MS Office, JAWS), which may introduce additional costs and limit accessibility for those who would like to replicate our setup. In addition, our study focused on close-ended tasks, each with a specific end state. We specifically defined these end goals to simplify analysis of task success rates. Future work could extend this by examining differences in interaction patterns between SUs and BLVUs, as well as evaluating agent performance on open-source applications and more open-ended tasks (e.g., travel planning, image editing).

\subsection{Towards Accessibility-Aware CUA}
Our results reinforce long-standing knowledge in accessibility and HCI: people approach the same goal with different strategies, so one fixed interaction style will not serve everyone. Ability-based design argues that systems should adapt to users’ abilities and strategies, rather than asking users to adapt to systems, motivating personalization in CUAs ~\cite{wobbrock2011ability}. 
\revision{We think about CUAs along a spectrum defined by how closely they simulate BLVUs’ assistive interactions.
At the low-simulation end, we expect a CUA to behave like an efficient autonomous operator that attempts to complete tasks with minimal user oversight. This is useful for faster task execution, but today’s CUAs are slower, inaccurate, and not reliable enough yet for fully hands-off use. Thus, it may be worthwhile to build a shared understanding of the task state to enable BLVUs to take over and recover from mistakes effectively. In the middle spectrum, we imagine a CUA that simulates the general population of BLVU, replaying common navigation and error-recovery patterns to stress-test whether interfaces remain usable under typical BLVU workflows and to probe how different CUA policies might alter those workflows. Finally, at the high-simulation end, a CUA approximates the strategies of a specific BLVU, supporting personalized computer-use instruction or tutorials that build on that user’s habits while suggesting more efficient alternatives.}
To put this into practice, we discuss the potential of CUAs for user simulation and as collaborative assistants. 

\subsubsection{Simulation and Personas} 
The dataset shows that one-style agents cannot serve everyone and that interaction strategies matter. \system{}'s interaction traces could be used to simulate how different people operate computer tasks, or to build personas that CUAs could follow or learn from. This aligns with prior efforts that simulate users to study usability ~\cite{park2022social}, and with ``generative agents'' that reproduce human-like traces over time~\cite{park2024generative}. For a specific user, we could summarize their interaction style (e.g., action mix, tempo, navigation strategy, AT settings) from \system{} logs and create a profile. At run time, an agent could: (i) \textit{monitor} in the background and surface gentle prompts when it sees known roadblocks; (ii) \textit{generate tutorials} that mirror the user’s own shortcuts and screen-reader flows and (iii) \textit{cold-start} by using collaborative-filtering or nearest-neighbor matching to pick a similar profile from our dataset. Beyond one-to-one profiles, \system{} could enable \textit{persona sets}. We could group traces by strategies (walking via Tab/Arrow, chunk-jumping via hotkeys, ribbon/OS routes), AT configurations, and task types, then test agents across these groups to estimate robustness. This is consistent with our goal of clustering users for large-scale usability tests and relates to broader simulations that vary user traits to evaluate how systems respond.

Simulating user interactions could also be useful to identify both usability and accessibility issues such as broken or inefficient focus order and keyboard traps and off-screen/low-contrast elements that magnifier users miss and estimate their impact. It could also provide time signals (extra steps, pauses, re-tries) that flag slow execution speed and could provide few-shot examples for training accessibility-aware CUAs. As a next step, we will quantify the fraction of actions spent in error recovery for both SUs and BLVUs to identify where interfaces and agents impose the greatest overhead.

Simulations, while beneficial (e.g., supporting accessibility evaluation, and low-cost prototyping), they also raise important ethical concerns. Disability simulations have been critiqued for reinforcing stereotypes~\cite{Bennett2019ThePO}, trivializing lived experiences~\cite{Cossovich2023CodesigningNK}, misrepresenting disability~\cite{Morris2019AIAA}, and excluding disabled people from the design process~\cite{NarioRedmond2017CripFA}. We acknowledge these valid critiques and agree that simulating disabled experiences should be approached with caution~\cite{andrew2022accessible}, given the associated risks such as errors, misinformation, and over-reliance. We would like to highlight that our work is not to suggest to replace human participants in the design process, but to understand their potential to support early-stage design when direct involvement is not feasible.

\subsubsection{Collaborative Assistants} 
All BLVUs confirmed that many of \system{} tasks were familiar and realistic which they often performed (e.g., moving files, web search, editing documents). BLVUs emphasized that CUA assistance should augment rather than replacing their workflows. They expressed a preference for CUAs that: (1) offer gentle, just-in-time suggestions such as next steps, shortcuts, or safer alternatives; (2) take over routine sub-steps when invited and provide easy undo; and (3) narrate what they are doing in clear and screen-reader-friendly language, similar to prior work \cite{peng2025morae, kodandaram2024enabling}.

\subsubsection{Tutorials}
Personalized CUAs could be used to generate customized step-by-step tutorials and walk-along guidance.  
Prior work shows people learn complex software faster with in-situ, scaffolded instruction~\cite{kelleher2005stencils}. These systems combine short videos, overlays, and step cues to reduce confusion during execution. 
With a few-shot learning of the user interaction, CUA can perhaps be used to infer the user's knowledge and skills of interacting with computers. After a successful run, the CUA could summarize what was learned (shortcuts used, errors avoided) and propose spaced-practice tasks according to the user's needs. 

\subsubsection{\revision{Accessibility-Aware CUA Learning}} \revision{Interaction traces in \system{}’s can also guide CUA learning. Because each action is aligned with screen context and task outcome, the logs support accessibility-aware reward functions that penalize long error-recovery paths or repeated backtracking and reward shorter, stable paths that match BLVUs’ strategies. Common navigation patterns (e.g., step-by-step “walking” vs. larger “chunk-jumps”) and recurring breakdown points highlight where CUAs should intervene by asking for confirmation or offering safer fallback actions when operating in BLVU workflows. We outline a minimal set of checks for CUA accessibility that can serve as a reusable benchmark: (i) compare agent success and path length against human traces under different assistive-technology settings, (ii) measure how much of the agent’s behavior is spent in error recovery (backtracking, retries, undo) versus steady progress, and (iii) verify that the agent preserves key assistive-technology state (e.g., screen reader mode, zoom). We recommend that future CUA accessibility evaluations report, for each task and condition, the AT configuration, task success, mean number of actions, elapsed time, and counts of perception, cognitive, and action errors.}

\subsection{Limitations of \system{} Dataset}
Our study was done on Windows OS and involved the closed-source applications (e.g., MS Office) because they remain the most widely used and screen reader–accessible platforms~\cite{webaim2025screen}.
We did not include other operating systems, such as macOS, where accessibility features and user behaviors may differ.  
Future research could broaden the dataset by incorporating multiple operating systems and applications to capture a broader range of accessibility configurations and usage patterns.
\revision{We also emphasize that our goal was to evaluate how CUAs operate when using AT. The tasks selected for this study were not excessively complex, but they were intentionally crafted to reflect realistic, nontrivial workflows commonly encountered by AT users. Notably, even with these tasks, human participants achieved success rates below 90\%, suggesting that CUAs, given their current capabilities, would likely face even greater difficulty. As models continue to advance and gain the capacity to handle more sophisticated tasks, future work can extend this evaluation to cover increasingly complex tasks.}

\section{Conclusion} 

We present \system{} dataset to characterize the accessibility gap in CUAs by grounding evaluation in real human traces across 60 everyday desktop and web tasks. We compare SU and BLVU workflows on the same tasks to reveal between-group differences and rich within-group diversity. Interaction traces show distinct strategies rather than a single style. SUs work mouse-first with routine scrolling and occasional shortcuts. BLVUs are keyboard-centric, with roughly 72\% of keystrokes used for navigation, and their logs show a consistent “verify-before-commit” pattern. While SUs shift between toolbar, context-menu, and copy-paste paths, BLVUs alternate between walking, chunk-jumping, and ribbon routes.
We also demonstrate that current agents underperform humans under default conditions, and their performance worsens under assistive technology constraints, such as using only a keyboard or a magnified viewport for computer use. We identify recurrent gaps in these CUAs: weak recovery when plans fail, end-state checks and inefficient navigation that compounds in tasks involving multiple applications. \system{} dataset provides synchronized video, audio, and OS input logs with success criteria to benchmark and improve such agents, promoting collaborative and accessible CUAs for everyone. 

% % % Acknowledgement
\begin{acks}
We thank our participants for their time and valuable contributions to our data collection. Rosiana Natalie was partially supported by the Michigan Data Science Fellowship at the University of Michigan. Brandon Kim was supported by the AccessComputing REU program. This research was also supported in part by a Google Cloud Platform Credit Award.
\end{acks}

% References
\bibliographystyle{ACM-Reference-Format}
\bibliography{references}

%%% -*-BibTeX-*-
%%% Do NOT edit. File created by BibTeX with style
%%% ACM-Reference-Format-Journals [18-Jan-2012].

\begin{thebibliography}{72}

%%% ====================================================================
%%% NOTE TO THE USER: you can override these defaults by providing
%%% customized versions of any of these macros before the \bibliography
%%% command.  Each of them MUST provide its own final punctuation,
%%% except for \shownote{} and \showURL{}.  The latter two
%%% do not use final punctuation, in order to avoid confusing it with
%%% the Web address.
%%%
%%% To suppress output of a particular field, define its macro to expand
%%% to an empty string, or better, \unskip, like this:
%%%
%%% \newcommand{\showURL}[1]{\unskip}   % LaTeX syntax
%%%
%%% \def \showURL #1{\unskip}           % plain TeX syntax
%%%
%%% ====================================================================

\ifx \showCODEN    \undefined \def \showCODEN     #1{\unskip}     \fi
\ifx \showISBNx    \undefined \def \showISBNx     #1{\unskip}     \fi
\ifx \showISBNxiii \undefined \def \showISBNxiii  #1{\unskip}     \fi
\ifx \showISSN     \undefined \def \showISSN      #1{\unskip}     \fi
\ifx \showLCCN     \undefined \def \showLCCN      #1{\unskip}     \fi
\ifx \shownote     \undefined \def \shownote      #1{#1}          \fi
\ifx \showarticletitle \undefined \def \showarticletitle #1{#1}   \fi
\ifx \showURL      \undefined \def \showURL       {\relax}        \fi
% The following commands are used for tagged output and should be
% invisible to TeX
\providecommand\bibfield[2]{#2}
\providecommand\bibinfo[2]{#2}
\providecommand\natexlab[1]{#1}
\providecommand\showeprint[2][]{arXiv:#2}

\bibitem[Com(2025a)]%
        {Computer_use_tool}
 \bibinfo{year}{2025}\natexlab{a}.
\newblock \bibinfo{title}{Anthropic Computer-Use-Tool}.
\newblock
\urldef\tempurl%
\url{https://docs.anthropic.com/en/docs/agents-and-tools/tool-use/computer-use-tool}
\showURL{%
\tempurl}


\bibitem[Com(2025b)]%
        {ComputerUsingAgent}
 \bibinfo{year}{2025}\natexlab{b}.
\newblock \bibinfo{title}{Computer-Using Agent}.
\newblock
\urldef\tempurl%
\url{https://openai.com/index/computer-using-agent/}
\showURL{%
\tempurl}


\bibitem[Int(2025)]%
        {Introducing_Operator}
 \bibinfo{year}{2025}\natexlab{}.
\newblock \bibinfo{title}{Introducing Operator}.
\newblock
\urldef\tempurl%
\url{https://openai.com/index/introducing-operator/}
\showURL{%
\tempurl}


\bibitem[OBS(2025)]%
        {OBS}
 \bibinfo{year}{2025}\natexlab{}.
\newblock \bibinfo{title}{OBS Studio}.
\newblock
\urldef\tempurl%
\url{https://obsproject.com/}
\showURL{%
\tempurl}


\bibitem[OSW(2025)]%
        {OSWorld-Website}
 \bibinfo{year}{2025}\natexlab{}.
\newblock \bibinfo{title}{OSWorld: Benchmarking Multimodal Agents for Open-Ended Tasks in Real Computer Environments}.
\newblock
\urldef\tempurl%
\url{https://os-world.github.io/}
\showURL{%
\tempurl}


\bibitem[Pro(2025)]%
        {Project_Astra}
 \bibinfo{year}{2025}\natexlab{}.
\newblock \bibinfo{title}{Project Astra}.
\newblock
\urldef\tempurl%
\url{https://deepmind.google/models/project-astra/}
\showURL{%
\tempurl}


\bibitem[Andrew and Tigwell(2022)]%
        {andrew2022accessible}
\bibfield{author}{\bibinfo{person}{Sarah Andrew} {and} \bibinfo{person}{Garreth~W Tigwell}.} \bibinfo{year}{2022}\natexlab{}.
\newblock \showarticletitle{Accessible design is mediated by job support structures and knowledge gained through design career pathways}.
\newblock \bibinfo{journal}{\emph{Proceedings of the ACM on Human-Computer Interaction}} \bibinfo{volume}{6}, \bibinfo{number}{CSCW2} (\bibinfo{year}{2022}), \bibinfo{pages}{1--24}.
\newblock


\bibitem[Anu~Bharath et~al\mbox{.}(2017)]%
        {anu2017performance}
\bibfield{author}{\bibinfo{person}{Pabba Anu~Bharath}, \bibinfo{person}{Charudatta Jadhav}, \bibinfo{person}{Shashank Ahire}, \bibinfo{person}{Manjiri Joshi}, \bibinfo{person}{Rini Ahirwar}, {and} \bibinfo{person}{Anirudha Joshi}.} \bibinfo{year}{2017}\natexlab{}.
\newblock \showarticletitle{Performance of accessible gesture-based indic keyboard}. In \bibinfo{booktitle}{\emph{IFIP Conference on Human-Computer Interaction}}. Springer, \bibinfo{pages}{205--220}.
\newblock


\bibitem[Bennett and Rosner(2019)]%
        {Bennett2019ThePO}
\bibfield{author}{\bibinfo{person}{Cynthia~L. Bennett} {and} \bibinfo{person}{Daniela~Karin Rosner}.} \bibinfo{year}{2019}\natexlab{}.
\newblock \showarticletitle{The Promise of Empathy: Design, Disability, and Knowing the "Other"}.
\newblock \bibinfo{journal}{\emph{Proceedings of the 2019 CHI Conference on Human Factors in Computing Systems}} (\bibinfo{year}{2019}).
\newblock
\urldef\tempurl%
\url{https://api.semanticscholar.org/CorpusID:140445039}
\showURL{%
\tempurl}


\bibitem[Bigham et~al\mbox{.}(2007)]%
        {bigham2007webinsitu}
\bibfield{author}{\bibinfo{person}{Jeffrey~P Bigham}, \bibinfo{person}{Anna~C Cavender}, \bibinfo{person}{Jeremy~T Brudvik}, \bibinfo{person}{Jacob~O Wobbrock}, {and} \bibinfo{person}{Richard~E Ladner}.} \bibinfo{year}{2007}\natexlab{}.
\newblock \showarticletitle{WebinSitu: a comparative analysis of blind and sighted browsing behavior}. In \bibinfo{booktitle}{\emph{Proceedings of the 9th International ACM SIGACCESS Conference on Computers and Accessibility}}. \bibinfo{pages}{51--58}.
\newblock


\bibitem[Borodin et~al\mbox{.}(2010)]%
        {borodin2010more}
\bibfield{author}{\bibinfo{person}{Yevgen Borodin}, \bibinfo{person}{Jeffrey~P Bigham}, \bibinfo{person}{Glenn Dausch}, {and} \bibinfo{person}{IV Ramakrishnan}.} \bibinfo{year}{2010}\natexlab{}.
\newblock \showarticletitle{More than meets the eye: a survey of screen-reader browsing strategies}. In \bibinfo{booktitle}{\emph{Proceedings of the 2010 international cross disciplinary conference on web accessibility (W4A)}}. \bibinfo{pages}{1--10}.
\newblock


\bibitem[Chen et~al\mbox{.}(2024)]%
        {chen2024gui}
\bibfield{author}{\bibinfo{person}{Dongping Chen}, \bibinfo{person}{Yue Huang}, \bibinfo{person}{Siyuan Wu}, \bibinfo{person}{Jingyu Tang}, \bibinfo{person}{Liuyi Chen}, \bibinfo{person}{Yilin Bai}, \bibinfo{person}{Zhigang He}, \bibinfo{person}{Chenlong Wang}, \bibinfo{person}{Huichi Zhou}, \bibinfo{person}{Yiqiang Li}, {et~al\mbox{.}}} \bibinfo{year}{2024}\natexlab{}.
\newblock \showarticletitle{Gui-world: A video benchmark and dataset for multimodal gui-oriented understanding}.
\newblock \bibinfo{journal}{\emph{arXiv preprint arXiv:2406.10819}} (\bibinfo{year}{2024}).
\newblock


\bibitem[Choi et~al\mbox{.}(2024)]%
        {choi2024proxona}
\bibfield{author}{\bibinfo{person}{Yoonseo Choi}, \bibinfo{person}{Eun~Jeong Kang}, \bibinfo{person}{Seulgi Choi}, \bibinfo{person}{Min~Kyung Lee}, {and} \bibinfo{person}{Juho Kim}.} \bibinfo{year}{2024}\natexlab{}.
\newblock \showarticletitle{Proxona: Leveraging LLM-Driven Personas to Enhance Creators' Understanding of Their Audience}.
\newblock \bibinfo{journal}{\emph{arXiv preprint arXiv:2408.10937}} (\bibinfo{year}{2024}).
\newblock


\bibitem[Cossovich et~al\mbox{.}(2023)]%
        {Cossovich2023CodesigningNK}
\bibfield{author}{\bibinfo{person}{Rodolfo Cossovich}, \bibinfo{person}{Steve Hodges}, \bibinfo{person}{Jin Kang}, {and} \bibinfo{person}{Audrey Girouard}.} \bibinfo{year}{2023}\natexlab{}.
\newblock \showarticletitle{Co-designing new keyboard and mouse solutions with people living with motor impairments}.
\newblock \bibinfo{journal}{\emph{Proceedings of the 25th International ACM SIGACCESS Conference on Computers and Accessibility}} (\bibinfo{year}{2023}).
\newblock
\urldef\tempurl%
\url{https://api.semanticscholar.org/CorpusID:264307091}
\showURL{%
\tempurl}


\bibitem[Deka et~al\mbox{.}(2017)]%
        {deka2017rico}
\bibfield{author}{\bibinfo{person}{Biplab Deka}, \bibinfo{person}{Zifeng Huang}, \bibinfo{person}{Chad Franzen}, \bibinfo{person}{Joshua Hibschman}, \bibinfo{person}{Daniel Afergan}, \bibinfo{person}{Yang Li}, \bibinfo{person}{Jeffrey Nichols}, {and} \bibinfo{person}{Ranjitha Kumar}.} \bibinfo{year}{2017}\natexlab{}.
\newblock \showarticletitle{Rico: A mobile app dataset for building data-driven design applications}. In \bibinfo{booktitle}{\emph{Proceedings of the 30th annual ACM symposium on user interface software and technology}}. \bibinfo{pages}{845--854}.
\newblock


\bibitem[Deng et~al\mbox{.}(2023)]%
        {deng2023mind2web}
\bibfield{author}{\bibinfo{person}{Xiang Deng}, \bibinfo{person}{Yu Gu}, \bibinfo{person}{Boyuan Zheng}, \bibinfo{person}{Shijie Chen}, \bibinfo{person}{Sam Stevens}, \bibinfo{person}{Boshi Wang}, \bibinfo{person}{Huan Sun}, {and} \bibinfo{person}{Yu Su}.} \bibinfo{year}{2023}\natexlab{}.
\newblock \showarticletitle{Mind2web: Towards a generalist agent for the web}.
\newblock \bibinfo{journal}{\emph{Advances in Neural Information Processing Systems}}  \bibinfo{volume}{36} (\bibinfo{year}{2023}), \bibinfo{pages}{28091--28114}.
\newblock


\bibitem[Fakrudeen et~al\mbox{.}(2017)]%
        {fakrudeen2017finger}
\bibfield{author}{\bibinfo{person}{Mohammed Fakrudeen}, \bibinfo{person}{Sufian Yousef}, {and} \bibinfo{person}{Mahdi~H Miraz}.} \bibinfo{year}{2017}\natexlab{}.
\newblock \showarticletitle{Finger Based Technique (FBT): An Innovative System for Improved Usability for the Blind Users' Dynamic Interaction with Mobile Touch Screen Devices}.
\newblock \bibinfo{journal}{\emph{arXiv preprint arXiv:1708.05073}} (\bibinfo{year}{2017}).
\newblock


\bibitem[Findlater and Zhang(2020)]%
        {findlater2020input}
\bibfield{author}{\bibinfo{person}{Leah Findlater} {and} \bibinfo{person}{Lotus Zhang}.} \bibinfo{year}{2020}\natexlab{}.
\newblock \showarticletitle{Input accessibility: A large dataset and summary analysis of age, motor ability and input performance}. In \bibinfo{booktitle}{\emph{Proceedings of the 22nd International ACM SIGACCESS Conference on Computers and Accessibility}}. \bibinfo{pages}{1--6}.
\newblock


\bibitem[Godfrey(2013)]%
        {godfrey2013statistical}
\bibfield{author}{\bibinfo{person}{A~Jonathan~R Godfrey}.} \bibinfo{year}{2013}\natexlab{}.
\newblock \showarticletitle{Statistical software from a blind person's perspective}.
\newblock  (\bibinfo{year}{2013}).
\newblock


\bibitem[H{\"a}m{\"a}l{\"a}inen et~al\mbox{.}(2023)]%
        {hamalainen2023evaluating}
\bibfield{author}{\bibinfo{person}{Perttu H{\"a}m{\"a}l{\"a}inen}, \bibinfo{person}{Mikke Tavast}, {and} \bibinfo{person}{Anton Kunnari}.} \bibinfo{year}{2023}\natexlab{}.
\newblock \showarticletitle{Evaluating large language models in generating synthetic hci research data: a case study}. In \bibinfo{booktitle}{\emph{Proceedings of the 2023 CHI Conference on Human Factors in Computing Systems}}. \bibinfo{pages}{1--19}.
\newblock


\bibitem[He et~al\mbox{.}(2024)]%
        {he2024webvoyagerbuildingendtoendweb}
\bibfield{author}{\bibinfo{person}{Hongliang He}, \bibinfo{person}{Wenlin Yao}, \bibinfo{person}{Kaixin Ma}, \bibinfo{person}{Wenhao Yu}, \bibinfo{person}{Yong Dai}, \bibinfo{person}{Hongming Zhang}, \bibinfo{person}{Zhenzhong Lan}, {and} \bibinfo{person}{Dong Yu}.} \bibinfo{year}{2024}\natexlab{}.
\newblock \bibinfo{title}{WebVoyager: Building an End-to-End Web Agent with Large Multimodal Models}.
\newblock
\showeprint[arxiv]{2401.13919}~[cs.CL]
\urldef\tempurl%
\url{https://arxiv.org/abs/2401.13919}
\showURL{%
\tempurl}


\bibitem[Huq et~al\mbox{.}(2025)]%
        {huq2025cowpilot}
\bibfield{author}{\bibinfo{person}{Faria Huq}, \bibinfo{person}{Zora~Zhiruo Wang}, \bibinfo{person}{Frank~F Xu}, \bibinfo{person}{Tianyue Ou}, \bibinfo{person}{Shuyan Zhou}, \bibinfo{person}{Jeffrey~P Bigham}, {and} \bibinfo{person}{Graham Neubig}.} \bibinfo{year}{2025}\natexlab{}.
\newblock \showarticletitle{CowPilot: A Framework for Autonomous and Human-Agent Collaborative Web Navigation}.
\newblock \bibinfo{journal}{\emph{arXiv preprint arXiv:2501.16609}} (\bibinfo{year}{2025}).
\newblock


\bibitem[Joh et~al\mbox{.}(2022)]%
        {10.1145/3493612.3520454}
\bibfield{author}{\bibinfo{person}{Hwayeon Joh}, \bibinfo{person}{Yun~Jung Lee}, {and} \bibinfo{person}{Uran Oh}.} \bibinfo{year}{2022}\natexlab{}.
\newblock \showarticletitle{Understanding the touchscreen-based nonvisual target acquisition task performance of screen reader users}. In \bibinfo{booktitle}{\emph{Proceedings of the 19th International Web for All Conference}} (Lyon, France) \emph{(\bibinfo{series}{W4A '22})}. \bibinfo{publisher}{Association for Computing Machinery}, \bibinfo{address}{New York, NY, USA}, Article \bibinfo{articleno}{11}, \bibinfo{numpages}{10}~pages.
\newblock
\showISBNx{9781450391702}
\href{https://doi.org/10.1145/3493612.3520454}{doi:\nolinkurl{10.1145/3493612.3520454}}


\bibitem[Jordan et~al\mbox{.}(2024)]%
        {jordan2024information}
\bibfield{author}{\bibinfo{person}{J~Bern Jordan}, \bibinfo{person}{Victoria Van~Hyning}, \bibinfo{person}{Mason~A Jones}, \bibinfo{person}{Rachael Bradley~Montgomery}, \bibinfo{person}{Elizabeth Bottner}, {and} \bibinfo{person}{Evan Tansil}.} \bibinfo{year}{2024}\natexlab{}.
\newblock \showarticletitle{Information Wayfinding of Screen Reader Users: Five Personas to Expand Conceptualizations of User Experiences}. In \bibinfo{booktitle}{\emph{Proceedings of the 26th International ACM SIGACCESS Conference on Computers and Accessibility}}. \bibinfo{pages}{1--7}.
\newblock


\bibitem[Kapoor et~al\mbox{.}(2024)]%
        {kapoor2024omniactdatasetbenchmarkenabling}
\bibfield{author}{\bibinfo{person}{Raghav Kapoor}, \bibinfo{person}{Yash~Parag Butala}, \bibinfo{person}{Melisa Russak}, \bibinfo{person}{Jing~Yu Koh}, \bibinfo{person}{Kiran Kamble}, \bibinfo{person}{Waseem Alshikh}, {and} \bibinfo{person}{Ruslan Salakhutdinov}.} \bibinfo{year}{2024}\natexlab{}.
\newblock \bibinfo{title}{OmniACT: A Dataset and Benchmark for Enabling Multimodal Generalist Autonomous Agents for Desktop and Web}.
\newblock
\showeprint[arxiv]{2402.17553}~[cs.AI]
\urldef\tempurl%
\url{https://arxiv.org/abs/2402.17553}
\showURL{%
\tempurl}


\bibitem[Kelleher and Pausch(2005)]%
        {kelleher2005stencils}
\bibfield{author}{\bibinfo{person}{Caitlin Kelleher} {and} \bibinfo{person}{Randy Pausch}.} \bibinfo{year}{2005}\natexlab{}.
\newblock \showarticletitle{Stencils-based tutorials: design and evaluation}. In \bibinfo{booktitle}{\emph{Proceedings of the SIGCHI conference on Human factors in computing systems}}. \bibinfo{pages}{541--550}.
\newblock


\bibitem[Ko et~al\mbox{.}(2021)]%
        {10.1145/3447526.3472022}
\bibfield{author}{\bibinfo{person}{Yu-Jung Ko}, \bibinfo{person}{Aini Putkonen}, \bibinfo{person}{Ali~Selman Aydin}, \bibinfo{person}{Shirin Feiz}, \bibinfo{person}{Yuheng Wang}, \bibinfo{person}{Vikas Ashok}, \bibinfo{person}{IV Ramakrishnan}, \bibinfo{person}{Antti Oulasvirta}, {and} \bibinfo{person}{Xiaojun Bi}.} \bibinfo{year}{2021}\natexlab{}.
\newblock \showarticletitle{Modeling Gliding-based Target Selection for Blind Touchscreen Users}. In \bibinfo{booktitle}{\emph{Proceedings of the 23rd International Conference on Mobile Human-Computer Interaction}} (Toulouse \& Virtual, France) \emph{(\bibinfo{series}{MobileHCI '21})}. \bibinfo{publisher}{Association for Computing Machinery}, \bibinfo{address}{New York, NY, USA}, Article \bibinfo{articleno}{29}, \bibinfo{numpages}{14}~pages.
\newblock
\showISBNx{9781450383288}
\href{https://doi.org/10.1145/3447526.3472022}{doi:\nolinkurl{10.1145/3447526.3472022}}


\bibitem[Kodandaram et~al\mbox{.}(2023)]%
        {kodandaram2023detecting}
\bibfield{author}{\bibinfo{person}{Satwik~Ram Kodandaram}, \bibinfo{person}{Mohan Sunkara}, \bibinfo{person}{Sampath Jayarathna}, {and} \bibinfo{person}{Vikas Ashok}.} \bibinfo{year}{2023}\natexlab{}.
\newblock \showarticletitle{Detecting deceptive dark-pattern web advertisements for blind screen-reader users}.
\newblock \bibinfo{journal}{\emph{Journal of Imaging}} \bibinfo{volume}{9}, \bibinfo{number}{11} (\bibinfo{year}{2023}), \bibinfo{pages}{239}.
\newblock


\bibitem[Kodandaram et~al\mbox{.}(2024)]%
        {kodandaram2024enabling}
\bibfield{author}{\bibinfo{person}{Satwik~Ram Kodandaram}, \bibinfo{person}{Utku Uckun}, \bibinfo{person}{Xiaojun Bi}, \bibinfo{person}{IV Ramakrishnan}, {and} \bibinfo{person}{Vikas Ashok}.} \bibinfo{year}{2024}\natexlab{}.
\newblock \showarticletitle{Enabling uniform computer interaction experience for blind users through large language models}. In \bibinfo{booktitle}{\emph{Proceedings of the 26th International ACM SIGACCESS Conference on Computers and Accessibility}}. \bibinfo{pages}{1--14}.
\newblock


\bibitem[Koh et~al\mbox{.}(2024)]%
        {koh2024visualwebarena}
\bibfield{author}{\bibinfo{person}{Jing~Yu Koh}, \bibinfo{person}{Robert Lo}, \bibinfo{person}{Lawrence Jang}, \bibinfo{person}{Vikram Duvvur}, \bibinfo{person}{Ming~Chong Lim}, \bibinfo{person}{Po-Yu Huang}, \bibinfo{person}{Graham Neubig}, \bibinfo{person}{Shuyan Zhou}, \bibinfo{person}{Ruslan Salakhutdinov}, {and} \bibinfo{person}{Daniel Fried}.} \bibinfo{year}{2024}\natexlab{}.
\newblock \showarticletitle{Visualwebarena: Evaluating multimodal agents on realistic visual web tasks}.
\newblock \bibinfo{journal}{\emph{arXiv preprint arXiv:2401.13649}} (\bibinfo{year}{2024}).
\newblock


\bibitem[Komninos et~al\mbox{.}(2023)]%
        {komninos2023review}
\bibfield{author}{\bibinfo{person}{Andreas Komninos}, \bibinfo{person}{Vassilios Stefanis}, {and} \bibinfo{person}{John Garofalakis}.} \bibinfo{year}{2023}\natexlab{}.
\newblock \showarticletitle{A review of design and evaluation practices in mobile text entry for visually impaired and blind persons}.
\newblock \bibinfo{journal}{\emph{Multimodal Technologies and Interaction}} \bibinfo{volume}{7}, \bibinfo{number}{2} (\bibinfo{year}{2023}), \bibinfo{pages}{22}.
\newblock


\bibitem[Lazar et~al\mbox{.}(2007)]%
        {Lazar01052007}
\bibfield{author}{\bibinfo{person}{Jonathan Lazar}, \bibinfo{person}{Aaron Allen}, \bibinfo{person}{Jason Kleinman}, {and} \bibinfo{person}{Chris Malarkey}.} \bibinfo{year}{2007}\natexlab{}.
\newblock \showarticletitle{What Frustrates Screen Reader Users on the Web: A Study of 100 Blind Users}.
\newblock \bibinfo{journal}{\emph{International Journal of Human–Computer Interaction}} \bibinfo{volume}{22}, \bibinfo{number}{3} (\bibinfo{year}{2007}), \bibinfo{pages}{247--269}.
\newblock
\showeprint{https://doi.org/10.1080/10447310709336964}
\href{https://doi.org/10.1080/10447310709336964}{doi:\nolinkurl{10.1080/10447310709336964}}


\bibitem[Leporini et~al\mbox{.}(2025)]%
        {10.1145/3733155.3737910}
\bibfield{author}{\bibinfo{person}{Barbara Leporini}, \bibinfo{person}{Marina Buzzi}, {and} \bibinfo{person}{Giuseppe Della~Penna}.} \bibinfo{year}{2025}\natexlab{}.
\newblock \showarticletitle{A Preliminary Evaluation of Generative AI Tools for Blind Users: Usability and Screen Reader Interaction}. In \bibinfo{booktitle}{\emph{Proceedings of the 18th ACM International Conference on PErvasive Technologies Related to Assistive Environments}} \emph{(\bibinfo{series}{PETRA '25})}. \bibinfo{publisher}{Association for Computing Machinery}, \bibinfo{address}{New York, NY, USA}, \bibinfo{pages}{562–568}.
\newblock
\showISBNx{9798400714023}
\href{https://doi.org/10.1145/3733155.3737910}{doi:\nolinkurl{10.1145/3733155.3737910}}


\bibitem[Li and Li(2022)]%
        {li2022spotlight}
\bibfield{author}{\bibinfo{person}{Gang Li} {and} \bibinfo{person}{Yang Li}.} \bibinfo{year}{2022}\natexlab{}.
\newblock \showarticletitle{Spotlight: Mobile ui understanding using vision-language models with a focus}.
\newblock \bibinfo{journal}{\emph{arXiv preprint arXiv:2209.14927}} (\bibinfo{year}{2022}).
\newblock


\bibitem[Liu et~al\mbox{.}(2018)]%
        {liu2018reinforcement}
\bibfield{author}{\bibinfo{person}{Evan~Zheran Liu}, \bibinfo{person}{Kelvin Guu}, \bibinfo{person}{Panupong Pasupat}, \bibinfo{person}{Tianlin Shi}, {and} \bibinfo{person}{Percy Liang}.} \bibinfo{year}{2018}\natexlab{}.
\newblock \showarticletitle{Reinforcement Learning on Web Interfaces using Workflow-Guided Exploration}. In \bibinfo{booktitle}{\emph{International Conference on Learning Representations ({ICLR})}}.
\newblock
\urldef\tempurl%
\url{https://arxiv.org/abs/1802.08802}
\showURL{%
\tempurl}


\bibitem[Liu et~al\mbox{.}(2023)]%
        {liu2023webglm}
\bibfield{author}{\bibinfo{person}{Xiao Liu}, \bibinfo{person}{Hanyu Lai}, \bibinfo{person}{Hao Yu}, \bibinfo{person}{Yifan Xu}, \bibinfo{person}{Aohan Zeng}, \bibinfo{person}{Zhengxiao Du}, \bibinfo{person}{Peng Zhang}, \bibinfo{person}{Yuxiao Dong}, {and} \bibinfo{person}{Jie Tang}.} \bibinfo{year}{2023}\natexlab{}.
\newblock \showarticletitle{WebGLM: towards an efficient web-enhanced question answering system with human preferences}. In \bibinfo{booktitle}{\emph{Proceedings of the 29th ACM SIGKDD conference on knowledge discovery and data mining}}. \bibinfo{pages}{4549--4560}.
\newblock


\bibitem[L{\`u} et~al\mbox{.}(2024)]%
        {lu2024weblinx}
\bibfield{author}{\bibinfo{person}{Xing~Han L{\`u}}, \bibinfo{person}{Zden{\v{e}}k Kasner}, {and} \bibinfo{person}{Siva Reddy}.} \bibinfo{year}{2024}\natexlab{}.
\newblock \showarticletitle{Weblinx: Real-world website navigation with multi-turn dialogue}.
\newblock \bibinfo{journal}{\emph{arXiv preprint arXiv:2402.05930}} (\bibinfo{year}{2024}).
\newblock


\bibitem[Mialon et~al\mbox{.}(2023)]%
        {mialon2023gaia}
\bibfield{author}{\bibinfo{person}{Gr{\'e}goire Mialon}, \bibinfo{person}{Cl{\'e}mentine Fourrier}, \bibinfo{person}{Thomas Wolf}, \bibinfo{person}{Yann LeCun}, {and} \bibinfo{person}{Thomas Scialom}.} \bibinfo{year}{2023}\natexlab{}.
\newblock \showarticletitle{Gaia: a benchmark for general ai assistants}. In \bibinfo{booktitle}{\emph{The Twelfth International Conference on Learning Representations}}.
\newblock


\bibitem[Morris.(2019)]%
        {Morris2019AIAA}
\bibfield{author}{\bibinfo{person}{M.~G.~R. Morris.}} \bibinfo{year}{2019}\natexlab{}.
\newblock \showarticletitle{AI and accessibility}.
\newblock \bibinfo{journal}{\emph{Commun. ACM}}  \bibinfo{volume}{63} (\bibinfo{year}{2019}), \bibinfo{pages}{35 -- 37}.
\newblock
\urldef\tempurl%
\url{https://api.semanticscholar.org/CorpusID:201645229}
\showURL{%
\tempurl}


\bibitem[Nario-Redmond et~al\mbox{.}(2017)]%
        {NarioRedmond2017CripFA}
\bibfield{author}{\bibinfo{person}{Michelle~R Nario-Redmond}, \bibinfo{person}{Dobromir Gospodinov}, {and} \bibinfo{person}{Angela Cobb}.} \bibinfo{year}{2017}\natexlab{}.
\newblock \showarticletitle{Crip for a Day: The Unintended Negative Consequences of Disability Simulations}.
\newblock \bibinfo{journal}{\emph{Rehabilitation Psychology}}  \bibinfo{volume}{62} (\bibinfo{year}{2017}), \bibinfo{pages}{324–333}.
\newblock
\urldef\tempurl%
\url{https://api.semanticscholar.org/CorpusID:2456886}
\showURL{%
\tempurl}


\bibitem[Natalie et~al\mbox{.}(2025)]%
        {natalie2025not}
\bibfield{author}{\bibinfo{person}{Rosiana Natalie}, \bibinfo{person}{Wenqian Xu}, \bibinfo{person}{Ruei-Che Chang}, \bibinfo{person}{Rada Mihalcea}, {and} \bibinfo{person}{Anhong Guo}.} \bibinfo{year}{2025}\natexlab{}.
\newblock \showarticletitle{Not There Yet: Evaluating Vision Language Models in Simulating the Visual Perception of People with Low Vision}.
\newblock \bibinfo{journal}{\emph{arXiv preprint arXiv:2508.10972}} (\bibinfo{year}{2025}).
\newblock


\bibitem[Niu et~al\mbox{.}(2024)]%
        {niu2024screenagent}
\bibfield{author}{\bibinfo{person}{Runliang Niu}, \bibinfo{person}{Jindong Li}, \bibinfo{person}{Shiqi Wang}, \bibinfo{person}{Yali Fu}, \bibinfo{person}{Xiyu Hu}, \bibinfo{person}{Xueyuan Leng}, \bibinfo{person}{He Kong}, \bibinfo{person}{Yi Chang}, {and} \bibinfo{person}{Qi Wang}.} \bibinfo{year}{2024}\natexlab{}.
\newblock \showarticletitle{Screenagent: A vision language model-driven computer control agent}.
\newblock \bibinfo{journal}{\emph{arXiv preprint arXiv:2402.07945}} (\bibinfo{year}{2024}).
\newblock


\bibitem[Pan et~al\mbox{.}(2024)]%
        {pan2024webcanvas}
\bibfield{author}{\bibinfo{person}{Yichen Pan}, \bibinfo{person}{Dehan Kong}, \bibinfo{person}{Sida Zhou}, \bibinfo{person}{Cheng Cui}, \bibinfo{person}{Yifei Leng}, \bibinfo{person}{Bing Jiang}, \bibinfo{person}{Hangyu Liu}, \bibinfo{person}{Yanyi Shang}, \bibinfo{person}{Shuyan Zhou}, \bibinfo{person}{Tongshuang Wu}, {et~al\mbox{.}}} \bibinfo{year}{2024}\natexlab{}.
\newblock \showarticletitle{Webcanvas: Benchmarking web agents in online environments}.
\newblock \bibinfo{journal}{\emph{arXiv preprint arXiv:2406.12373}} (\bibinfo{year}{2024}).
\newblock


\bibitem[Park et~al\mbox{.}(2022)]%
        {park2022social}
\bibfield{author}{\bibinfo{person}{Joon~Sung Park}, \bibinfo{person}{Lindsay Popowski}, \bibinfo{person}{Carrie Cai}, \bibinfo{person}{Meredith~Ringel Morris}, \bibinfo{person}{Percy Liang}, {and} \bibinfo{person}{Michael~S. Bernstein}.} \bibinfo{year}{2022}\natexlab{}.
\newblock \showarticletitle{Social Simulacra: Creating Populated Prototypes for Social Computing Systems}. In \bibinfo{booktitle}{\emph{Proceedings of the 35th Annual ACM Symposium on User Interface Software and Technology}} (Bend, OR, USA) \emph{(\bibinfo{series}{UIST '22})}. \bibinfo{publisher}{Association for Computing Machinery}, \bibinfo{address}{New York, NY, USA}, Article \bibinfo{articleno}{74}, \bibinfo{numpages}{18}~pages.
\newblock
\showISBNx{9781450393201}
\href{https://doi.org/10.1145/3526113.3545616}{doi:\nolinkurl{10.1145/3526113.3545616}}


\bibitem[Park et~al\mbox{.}(2024)]%
        {park2024generative}
\bibfield{author}{\bibinfo{person}{Joon~Sung Park}, \bibinfo{person}{Carolyn~Q Zou}, \bibinfo{person}{Aaron Shaw}, \bibinfo{person}{Benjamin~Mako Hill}, \bibinfo{person}{Carrie Cai}, \bibinfo{person}{Meredith~Ringel Morris}, \bibinfo{person}{Robb Willer}, \bibinfo{person}{Percy Liang}, {and} \bibinfo{person}{Michael~S Bernstein}.} \bibinfo{year}{2024}\natexlab{}.
\newblock \showarticletitle{Generative agent simulations of 1,000 people}.
\newblock \bibinfo{journal}{\emph{arXiv preprint arXiv:2411.10109}} (\bibinfo{year}{2024}).
\newblock


\bibitem[Peng et~al\mbox{.}(2025)]%
        {peng2025morae}
\bibfield{author}{\bibinfo{person}{Yi-Hao Peng}, \bibinfo{person}{Dingzeyu Li}, \bibinfo{person}{Jeffrey~P Bigham}, {and} \bibinfo{person}{Amy Pavel}.} \bibinfo{year}{2025}\natexlab{}.
\newblock \showarticletitle{Morae: Proactively Pausing UI Agents for User Choices}.
\newblock \bibinfo{journal}{\emph{arXiv preprint arXiv:2508.21456}} (\bibinfo{year}{2025}).
\newblock


\bibitem[Rawles et~al\mbox{.}(2023)]%
        {rawles2023androidinthewild}
\bibfield{author}{\bibinfo{person}{Christopher Rawles}, \bibinfo{person}{Alice Li}, \bibinfo{person}{Daniel Rodriguez}, \bibinfo{person}{Oriana Riva}, {and} \bibinfo{person}{Timothy Lillicrap}.} \bibinfo{year}{2023}\natexlab{}.
\newblock \showarticletitle{Androidinthewild: A large-scale dataset for android device control}.
\newblock \bibinfo{journal}{\emph{Advances in Neural Information Processing Systems}}  \bibinfo{volume}{36} (\bibinfo{year}{2023}), \bibinfo{pages}{59708--59728}.
\newblock


\bibitem[sadadow(2025)]%
        {sadadow}
\bibfield{author}{\bibinfo{person}{sadadow}.} \bibinfo{year}{2025}\natexlab{}.
\newblock \bibinfo{title}{Learn how to use Microsoft 365 Copilot}.
\newblock
\urldef\tempurl%
\url{https://learn.microsoft.com/en-us/copilot/}
\showURL{%
\tempurl}


\bibitem[Savva(2017)]%
        {savva2017understanding}
\bibfield{author}{\bibinfo{person}{Andreas Savva}.} \bibinfo{year}{2017}\natexlab{}.
\newblock \emph{\bibinfo{title}{Understanding accessibility problems of blind users on the web}}.
\newblock \bibinfo{thesistype}{Ph.\,D. Dissertation}. \bibinfo{school}{University of York}.
\newblock


\bibitem[Schaadhardt et~al\mbox{.}(2021)]%
        {10.1145/3411764.3445242}
\bibfield{author}{\bibinfo{person}{Anastasia Schaadhardt}, \bibinfo{person}{Alexis Hiniker}, {and} \bibinfo{person}{Jacob~O. Wobbrock}.} \bibinfo{year}{2021}\natexlab{}.
\newblock \showarticletitle{Understanding Blind Screen-Reader Users’ Experiences of Digital Artboards}. In \bibinfo{booktitle}{\emph{Proceedings of the 2021 CHI Conference on Human Factors in Computing Systems}} (Yokohama, Japan) \emph{(\bibinfo{series}{CHI '21})}. \bibinfo{publisher}{Association for Computing Machinery}, \bibinfo{address}{New York, NY, USA}, Article \bibinfo{articleno}{270}, \bibinfo{numpages}{19}~pages.
\newblock
\showISBNx{9781450380966}
\href{https://doi.org/10.1145/3411764.3445242}{doi:\nolinkurl{10.1145/3411764.3445242}}


\bibitem[Sharif et~al\mbox{.}(2021)]%
        {10.1145/3441852.3471202}
\bibfield{author}{\bibinfo{person}{Ather Sharif}, \bibinfo{person}{Sanjana~Shivani Chintalapati}, \bibinfo{person}{Jacob~O. Wobbrock}, {and} \bibinfo{person}{Katharina Reinecke}.} \bibinfo{year}{2021}\natexlab{}.
\newblock \showarticletitle{Understanding Screen-Reader Users’ Experiences with Online Data Visualizations}. In \bibinfo{booktitle}{\emph{Proceedings of the 23rd International ACM SIGACCESS Conference on Computers and Accessibility}} (Virtual Event, USA) \emph{(\bibinfo{series}{ASSETS '21})}. \bibinfo{publisher}{Association for Computing Machinery}, \bibinfo{address}{New York, NY, USA}, Article \bibinfo{articleno}{14}, \bibinfo{numpages}{16}~pages.
\newblock
\showISBNx{9781450383066}
\href{https://doi.org/10.1145/3441852.3471202}{doi:\nolinkurl{10.1145/3441852.3471202}}


\bibitem[Sharif et~al\mbox{.}(2024)]%
        {10.1145/3677846.3677867}
\bibfield{author}{\bibinfo{person}{Ather Sharif}, \bibinfo{person}{Venkatesh Potluri}, \bibinfo{person}{Jazz R.~X. Ang}, \bibinfo{person}{Jacob~O. Wobbrock}, {and} \bibinfo{person}{Jennifer Mankoff}.} \bibinfo{year}{2024}\natexlab{}.
\newblock \showarticletitle{Touchpad Mapper: Exploring Non-Visual Touchpad Interactions for Screen-Reader Users}. In \bibinfo{booktitle}{\emph{Proceedings of the 21st International Web for All Conference}} (Singapore, Singapore) \emph{(\bibinfo{series}{W4A '24})}. \bibinfo{publisher}{Association for Computing Machinery}, \bibinfo{address}{New York, NY, USA}, \bibinfo{pages}{42–44}.
\newblock
\showISBNx{9798400710308}
\href{https://doi.org/10.1145/3677846.3677867}{doi:\nolinkurl{10.1145/3677846.3677867}}


\bibitem[Shi et~al\mbox{.}(2019)]%
        {shi2019vipboard}
\bibfield{author}{\bibinfo{person}{Weinan Shi}, \bibinfo{person}{Chun Yu}, \bibinfo{person}{Shuyi Fan}, \bibinfo{person}{Feng Wang}, \bibinfo{person}{Tong Wang}, \bibinfo{person}{Xin Yi}, \bibinfo{person}{Xiaojun Bi}, {and} \bibinfo{person}{Yuanchun Shi}.} \bibinfo{year}{2019}\natexlab{}.
\newblock \showarticletitle{Vipboard: Improving screen-reader keyboard for visually impaired people with character-level auto correction}. In \bibinfo{booktitle}{\emph{Proceedings of the 2019 CHI Conference on Human Factors in Computing Systems}}. \bibinfo{pages}{1--12}.
\newblock


\bibitem[Shirogane et~al\mbox{.}(2008)]%
        {shirogane2008accessibility}
\bibfield{author}{\bibinfo{person}{Junko Shirogane}, \bibinfo{person}{Takashi Mori}, \bibinfo{person}{Hajime Iwata}, {and} \bibinfo{person}{Yoshiaki Fukazawa}.} \bibinfo{year}{2008}\natexlab{}.
\newblock \showarticletitle{Accessibility evaluation for GUI software using source programs}.
\newblock In \bibinfo{booktitle}{\emph{Knowledge-Based Software Engineering}}. \bibinfo{publisher}{IOS Press}, \bibinfo{pages}{135--144}.
\newblock


\bibitem[Silva et~al\mbox{.}(2024)]%
        {10.1145/3649223}
\bibfield{author}{\bibinfo{person}{Jorge Sassaki~Resende Silva}, \bibinfo{person}{Paula Christina~Figueira Cardoso}, \bibinfo{person}{Raphael~Winckler De~Bettio}, \bibinfo{person}{Daniela~Cardoso Tavares}, \bibinfo{person}{Carlos~Alberto Silva}, \bibinfo{person}{Willian~Massami Watanabe}, {and} \bibinfo{person}{Andr\'{E}~Pimenta Freire}.} \bibinfo{year}{2024}\natexlab{}.
\newblock \showarticletitle{In-Page Navigation Aids for Screen-Reader Users with Automatic Topicalisation and Labelling}.
\newblock \bibinfo{journal}{\emph{ACM Trans. Access. Comput.}} \bibinfo{volume}{17}, \bibinfo{number}{2}, Article \bibinfo{articleno}{12} (\bibinfo{date}{July} \bibinfo{year}{2024}), \bibinfo{numpages}{45}~pages.
\newblock
\showISSN{1936-7228}
\href{https://doi.org/10.1145/3649223}{doi:\nolinkurl{10.1145/3649223}}


\bibitem[Song et~al\mbox{.}(2025)]%
        {song2025coact}
\bibfield{author}{\bibinfo{person}{Linxin Song}, \bibinfo{person}{Yutong Dai}, \bibinfo{person}{Viraj Prabhu}, \bibinfo{person}{Jieyu Zhang}, \bibinfo{person}{Taiwei Shi}, \bibinfo{person}{Li Li}, \bibinfo{person}{Junnan Li}, \bibinfo{person}{Silvio Savarese}, \bibinfo{person}{Zeyuan Chen}, \bibinfo{person}{Jieyu Zhao}, {et~al\mbox{.}}} \bibinfo{year}{2025}\natexlab{}.
\newblock \showarticletitle{Coact-1: Computer-using agents with coding as actions}.
\newblock \bibinfo{journal}{\emph{arXiv preprint arXiv:2508.03923}} (\bibinfo{year}{2025}).
\newblock


\bibitem[Sun et~al\mbox{.}(2022)]%
        {sun2022meta}
\bibfield{author}{\bibinfo{person}{Liangtai Sun}, \bibinfo{person}{Xingyu Chen}, \bibinfo{person}{Lu Chen}, \bibinfo{person}{Tianle Dai}, \bibinfo{person}{Zichen Zhu}, {and} \bibinfo{person}{Kai Yu}.} \bibinfo{year}{2022}\natexlab{}.
\newblock \showarticletitle{Meta-gui: Towards multi-modal conversational agents on mobile gui}.
\newblock \bibinfo{journal}{\emph{arXiv preprint arXiv:2205.11029}} (\bibinfo{year}{2022}).
\newblock


\bibitem[Taeb et~al\mbox{.}(2024)]%
        {taeb2024axnav}
\bibfield{author}{\bibinfo{person}{Maryam Taeb}, \bibinfo{person}{Amanda Swearngin}, \bibinfo{person}{Eldon Schoop}, \bibinfo{person}{Ruijia Cheng}, \bibinfo{person}{Yue Jiang}, {and} \bibinfo{person}{Jeffrey Nichols}.} \bibinfo{year}{2024}\natexlab{}.
\newblock \showarticletitle{Axnav: Replaying accessibility tests from natural language}. In \bibinfo{booktitle}{\emph{Proceedings of the 2024 CHI Conference on Human Factors in Computing Systems}}. \bibinfo{pages}{1--16}.
\newblock


\bibitem[Tian et~al\mbox{.}(2024)]%
        {tian2024mmina}
\bibfield{author}{\bibinfo{person}{Shulin Tian}, \bibinfo{person}{Ziniu Zhang}, \bibinfo{person}{Liangyu Chen}, {and} \bibinfo{person}{Ziwei Liu}.} \bibinfo{year}{2024}\natexlab{}.
\newblock \showarticletitle{Mmina: Benchmarking multihop multimodal internet agents}.
\newblock \bibinfo{journal}{\emph{arXiv preprint arXiv:2404.09992}} (\bibinfo{year}{2024}).
\newblock


\bibitem[Wang and Redmiles(2017)]%
        {10.1145/3131785.3131837}
\bibfield{author}{\bibinfo{person}{Tao Wang} {and} \bibinfo{person}{David Redmiles}.} \bibinfo{year}{2017}\natexlab{}.
\newblock \showarticletitle{Auditory Overview of Web Pages for Screen Reader Users}. In \bibinfo{booktitle}{\emph{Adjunct Proceedings of the 30th Annual ACM Symposium on User Interface Software and Technology}} (Qu\'{e}bec City, QC, Canada) \emph{(\bibinfo{series}{UIST '17 Adjunct})}. \bibinfo{publisher}{Association for Computing Machinery}, \bibinfo{address}{New York, NY, USA}, \bibinfo{pages}{193–195}.
\newblock
\showISBNx{9781450354196}
\href{https://doi.org/10.1145/3131785.3131837}{doi:\nolinkurl{10.1145/3131785.3131837}}


\bibitem[WebAim(2025)]%
        {webaim2025screen}
\bibfield{author}{\bibinfo{person}{WebAim}.} \bibinfo{year}{2025}\natexlab{}.
\newblock \bibinfo{title}{Screen Reader User Survey \#10 Results}.
\newblock
\urldef\tempurl%
\url{https://webaim.org/projects/screenreadersurvey10/}
\showURL{%
\tempurl}


\bibitem[Wei et~al\mbox{.}(2025)]%
        {wei2025browsecomp}
\bibfield{author}{\bibinfo{person}{Jason Wei}, \bibinfo{person}{Zhiqing Sun}, \bibinfo{person}{Spencer Papay}, \bibinfo{person}{Scott McKinney}, \bibinfo{person}{Jeffrey Han}, \bibinfo{person}{Isa Fulford}, \bibinfo{person}{Hyung~Won Chung}, \bibinfo{person}{Alex~Tachard Passos}, \bibinfo{person}{William Fedus}, {and} \bibinfo{person}{Amelia Glaese}.} \bibinfo{year}{2025}\natexlab{}.
\newblock \showarticletitle{Browsecomp: A simple yet challenging benchmark for browsing agents}.
\newblock \bibinfo{journal}{\emph{arXiv preprint arXiv:2504.12516}} (\bibinfo{year}{2025}).
\newblock


\bibitem[Wobbrock et~al\mbox{.}(2011)]%
        {wobbrock2011ability}
\bibfield{author}{\bibinfo{person}{Jacob~O Wobbrock}, \bibinfo{person}{Shaun~K Kane}, \bibinfo{person}{Krzysztof~Z Gajos}, \bibinfo{person}{Susumu Harada}, {and} \bibinfo{person}{Jon Froehlich}.} \bibinfo{year}{2011}\natexlab{}.
\newblock \showarticletitle{Ability-based design: Concept, principles and examples}.
\newblock \bibinfo{journal}{\emph{ACM Transactions on Accessible Computing (TACCESS)}} \bibinfo{volume}{3}, \bibinfo{number}{3} (\bibinfo{year}{2011}), \bibinfo{pages}{1--27}.
\newblock


\bibitem[Xie et~al\mbox{.}(2024)]%
        {xie2024osworld}
\bibfield{author}{\bibinfo{person}{Tianbao Xie}, \bibinfo{person}{Danyang Zhang}, \bibinfo{person}{Jixuan Chen}, \bibinfo{person}{Xiaochuan Li}, \bibinfo{person}{Siheng Zhao}, \bibinfo{person}{Ruisheng Cao}, \bibinfo{person}{Toh~J Hua}, \bibinfo{person}{Zhoujun Cheng}, \bibinfo{person}{Dongchan Shin}, \bibinfo{person}{Fangyu Lei}, {et~al\mbox{.}}} \bibinfo{year}{2024}\natexlab{}.
\newblock \showarticletitle{Osworld: Benchmarking multimodal agents for open-ended tasks in real computer environments}.
\newblock \bibinfo{journal}{\emph{Advances in Neural Information Processing Systems}}  \bibinfo{volume}{37} (\bibinfo{year}{2024}), \bibinfo{pages}{52040--52094}.
\newblock


\bibitem[Yao et~al\mbox{.}(2022)]%
        {yao2022webshop}
\bibfield{author}{\bibinfo{person}{Shunyu Yao}, \bibinfo{person}{Howard Chen}, \bibinfo{person}{John Yang}, {and} \bibinfo{person}{Karthik Narasimhan}.} \bibinfo{year}{2022}\natexlab{}.
\newblock \showarticletitle{Webshop: Towards scalable real-world web interaction with grounded language agents}.
\newblock \bibinfo{journal}{\emph{Advances in Neural Information Processing Systems}}  \bibinfo{volume}{35} (\bibinfo{year}{2022}), \bibinfo{pages}{20744--20757}.
\newblock


\bibitem[Zhang et~al\mbox{.}(2024b)]%
        {zhang2024ufo}
\bibfield{author}{\bibinfo{person}{Chaoyun Zhang}, \bibinfo{person}{Liqun Li}, \bibinfo{person}{Shilin He}, \bibinfo{person}{Xu Zhang}, \bibinfo{person}{Bo Qiao}, \bibinfo{person}{Si Qin}, \bibinfo{person}{Minghua Ma}, \bibinfo{person}{Yu Kang}, \bibinfo{person}{Qingwei Lin}, \bibinfo{person}{Saravan Rajmohan}, {et~al\mbox{.}}} \bibinfo{year}{2024}\natexlab{b}.
\newblock \showarticletitle{Ufo: A ui-focused agent for windows os interaction}.
\newblock \bibinfo{journal}{\emph{arXiv preprint arXiv:2402.07939}} (\bibinfo{year}{2024}).
\newblock


\bibitem[Zhang et~al\mbox{.}(2024a)]%
        {zhang2024accessible}
\bibfield{author}{\bibinfo{person}{Dan Zhang}, \bibinfo{person}{Zhi Li}, \bibinfo{person}{Vikas Ashok}, \bibinfo{person}{William~H Seiple}, \bibinfo{person}{Iv Ramakrishnan}, {and} \bibinfo{person}{Xiaojun Bi}.} \bibinfo{year}{2024}\natexlab{a}.
\newblock \showarticletitle{Accessible gesture typing on smartphones for people with low vision}. In \bibinfo{booktitle}{\emph{Proceedings of the 37th Annual ACM Symposium on User Interface Software and Technology}}. \bibinfo{pages}{1--11}.
\newblock


\bibitem[Zhang et~al\mbox{.}(2024c)]%
        {zhang2024mobile}
\bibfield{author}{\bibinfo{person}{Danyang Zhang}, \bibinfo{person}{Hongshen Xu}, \bibinfo{person}{Zihan Zhao}, \bibinfo{person}{Lu Chen}, \bibinfo{person}{Ruisheng Cao}, {and} \bibinfo{person}{Kai Yu}.} \bibinfo{year}{2024}\natexlab{c}.
\newblock \showarticletitle{Mobile-env: an evaluation platform and benchmark for LLM-GUI interaction}.
\newblock \bibinfo{journal}{\emph{arXiv preprint arXiv:2305.08144}} (\bibinfo{year}{2024}).
\newblock


\bibitem[Zhang et~al\mbox{.}(2019)]%
        {zhang2019text}
\bibfield{author}{\bibinfo{person}{Mingrui~Ray Zhang}, \bibinfo{person}{Shumin Zhai}, {and} \bibinfo{person}{Jacob~O Wobbrock}.} \bibinfo{year}{2019}\natexlab{}.
\newblock \showarticletitle{Text entry throughput: Towards unifying speed and accuracy in a single performance metric}. In \bibinfo{booktitle}{\emph{Proceedings of the 2019 CHI conference on human factors in computing systems}}. \bibinfo{pages}{1--13}.
\newblock


\bibitem[Zheng et~al\mbox{.}(2024a)]%
        {zheng2024gpt}
\bibfield{author}{\bibinfo{person}{Boyuan Zheng}, \bibinfo{person}{Boyu Gou}, \bibinfo{person}{Jihyung Kil}, \bibinfo{person}{Huan Sun}, {and} \bibinfo{person}{Yu Su}.} \bibinfo{year}{2024}\natexlab{a}.
\newblock \showarticletitle{Gpt-4v (ision) is a generalist web agent, if grounded}.
\newblock \bibinfo{journal}{\emph{arXiv preprint arXiv:2401.01614}} (\bibinfo{year}{2024}).
\newblock


\bibitem[Zheng et~al\mbox{.}(2024b)]%
        {zheng2024agentstudio}
\bibfield{author}{\bibinfo{person}{Longtao Zheng}, \bibinfo{person}{Zhiyuan Huang}, \bibinfo{person}{Zhenghai Xue}, \bibinfo{person}{Xinrun Wang}, \bibinfo{person}{Bo An}, {and} \bibinfo{person}{Shuicheng Yan}.} \bibinfo{year}{2024}\natexlab{b}.
\newblock \showarticletitle{Agentstudio: A toolkit for building general virtual agents}.
\newblock \bibinfo{journal}{\emph{arXiv preprint arXiv:2403.17918}} (\bibinfo{year}{2024}).
\newblock


\bibitem[Zhou et~al\mbox{.}(2023)]%
        {zhou2023webarena}
\bibfield{author}{\bibinfo{person}{Shuyan Zhou}, \bibinfo{person}{Frank~F Xu}, \bibinfo{person}{Hao Zhu}, \bibinfo{person}{Xuhui Zhou}, \bibinfo{person}{Robert Lo}, \bibinfo{person}{Abishek Sridhar}, \bibinfo{person}{Xianyi Cheng}, \bibinfo{person}{Tianyue Ou}, \bibinfo{person}{Yonatan Bisk}, \bibinfo{person}{Daniel Fried}, {et~al\mbox{.}}} \bibinfo{year}{2023}\natexlab{}.
\newblock \showarticletitle{Webarena: A realistic web environment for building autonomous agents}.
\newblock \bibinfo{journal}{\emph{arXiv preprint arXiv:2307.13854}} (\bibinfo{year}{2023}).
\newblock


\end{thebibliography}
\appendix

\newpage
\onecolumn
\section{Prompts}

\subsection{Prompt for Default-CUA}
\texttt{
\noindent You are using the computer tool on a real Windows desktop. First, briefly outline a 1–3 step plan, then request a 'screenshot' as your first tool action. Proceed incrementally, requesting a new screenshot after each action.
}

\subsection{Prompt for Screen-Reader-CUA}
\texttt{
\noindent You are controlling a real Windows desktop via KEYBOARD- ONLY for a blind user.
Restricted action space: keystrokes only, no mouse usage
Never use the mouse or scroll wheel: no move, click, drag, or scroll.
\begin{itemize}
    \item After actions that open OS UI, allow ~0.3–0.5s for the UI to appear.
    \item ALWAYS request a fresh `screenshot` before and after each action.
    \item Begin with a concise 1–3 step plan; your first tool action MUST be `screenshot`.
    \item If you ever propose a disallowed action, immediately switch to a keyboard-only path.
    \item Prefer focusing the browser address bar to navigate (no pointing)
\end{itemize}
}

\subsection{Prompt for Magnifier-CUA}
\texttt{You are using the computer tool on a real Windows desktop in LOW-VISION MODE for assisting low-vision users.
The OS display scaling is set to ~150\%, so UI elements are larger and the viewport only shows a portion of the entire OS display. 
Thus, the controls or content required to complete an action may be outside of the viewport. 
To show more of the display, move the viewport by either, 
\begin{itemize}
    \item Moving the pointer to the edge of viewport in the desired movement direction, or
    \item Using keyboard navigation (e.g., Ctrl+L to focus on the address bar; Tab/Shift+Tab to move focus; Ctrl+F to find on page.
\end{itemize} 
One strategy to identify missing content or controls is to perform a grid sweep of the entire OS display by moving the viewport to the top left of the display then moving the viewport over the entire screen to attempt to find the control. }

\newpage
\onecolumn
\section{\system{} Dataset Tasks}
\setlength{\tabcolsep}{3pt}
% assumes \usepackage{array,longtable,booktabs} and \newcolumntype{P}[1]{>{\raggedright\arraybackslash}p{#1}}
\small
\setlength{\LTleft}{0pt}
\setlength{\LTright}{0pt}
\begin{longtable}{@{} P{0.04\textwidth} P{0.26\textwidth} P{0.30\textwidth} P{0.09\textwidth} P{0.23\textwidth} @{}}
\caption{All tasks from our \system{} dataset, which covers five categories of tasks including Browsing \& Web (tasks 1-12), System Operations (tasks 13-24), Document Editing (tasks 25-36), Workflow (tasks 37-48), and Media (tasks 49-60). Every task comes with the Task Context and Instruction that situate the user and specify the required actions. Our tasks have wide range of applications scope, some involve single application, while others require multiple. To determine successful completion, each task is associated with one or more End States that define the expected outcome.}
\label{tab:all_tasks_ids_no_task}\\
\toprule
\textbf{Task ID} & \textbf{Task Context} & \textbf{Task Instruction} & \textbf{Applications} & \textbf{End State} \\
\midrule
\endfirsthead

\toprule
\textbf{Task ID} & \textbf{Task Context} & \textbf{Task Instruction} & \textbf{Applications} & \textbf{End State} \\
\midrule
\endhead

% \midrule
% \multicolumn{5}{r}{\small (continues on next page)}\\
\bottomrule
\endfoot

\bottomrule
\endlastfoot

1 & I am searching for an article on Chrome about "How to do Literature Review". I choose the first article on the search results, but I realize that this article is not helpful, so I return to the search results. & Search for an article on "How to do literature review" on Chrome; Click on the first search result; Return to the search results page & Chrome & The page is showing the original Google search result. \\
2 & I use Facebook regularly so I want easy access. & Bookmark the Facebook homepage on Chrome & Chrome & Facebook appears in the address bar with its "home" screen displayed. The site now appears in "Bookmarks" \\
3 & I'm interested in buying a new laptop and want to learn more about its feature in detail. & Visit apple.com. Go to Macbook Air page on Chrome and find information about the tools Apple Intelligence is built & Chrome & The section showing information about "Apple Intelligence" is open \\
4 & I've been getting into cooking recently, so I've decided to upgrade my frying pan. & Visit walmart.com on Chrome. Search for a frying pan, then click on the first product with a pan size of exactly 10 inches and add it to cart. & Chrome & The page is showing an additional item in the cart. \\
5 & I want to do some shopping on Target and I want to tie my purchase with my account. & Login to target.com on Chrome with username "studyparticipants5@gmail.com" and password "CUAstudy1!" and set the store in 95050 as default. & Chrome & The Target website shows a screen where an account is logged into, and default store set to the one in 95050. \\
6 & I am interested in buying cars and I am exploring car models from different brands, like Subaru, Toyota, and Honda. After looking for some options on Chrome. Then, close the tabs containing Toyota and Honda cars. & Search for Toyota Camry, Subaru Forester, and Honda Civic in separate tabs and open their respective websites on Chrome. Then, close the tabs containing Toyota and Honda cars. & Chrome & One tab remains, which is Subaru Forester page. \\
7 & I am filling out an anonymous feedback form through Google Form. & Go to https://tinyurl.com/facility-survey and fill up the Google Form and submit. & Chrome & The form reponse is submitted. \\
8 & I'm currently watching a lecture on YouTube on "Machine Learning and Generative AI". I want to watch a different video as this one doesn't cover the concept I'm looking to learn. & Visit youtube.com on Chrome and search for "Machine Learning and Generative AI"; Watch the first suggested video for a few seconds. Skip to the next video & Chrome. & A new lecture video is playing. \\
9 & I want to print the lyrics to "Baby Shark" song. & Find the lyrics to "Baby Shark" song on Chrome. Once you find the lyrics, save the webpage as a PDF so I can print it. & Chrome & The PDF file that consist of the lyric exists in the local machine. \\
10 & I want to get up-to-date about some news from France but the reliable news websites are mostly in French. I want to read the English Translation of this. & Visit www.france24.com/fr. Translate the news page in French on this webpage to English using Google Translate. & Chrome & The French news appears in English. \\
11 & I am planning to fly to New York for the Labor Day weekend to see my family. & Visit expedia.com on Chrome. Click on the cheapest round-trip flight from Austin to New York for Sept 2-4. & Chrome & The cheapest flight is displayed on screen. \\
12 & I prefer a privacy-focused search engine and want to use DuckDuckGo by default. & Set DuckDuckGo as the default search engine on Chrome. & Chrome & The DuckDuckGo is become the default search engine. \\
13 & I need an alarm to remind me to badge in to work. & Set an alarm for 8.30 AM that repeats every Monday. & Clock & The alarm clock is set to 8.30 AM; Monday (in Repeats) is checked. \\
14 & I'm trying to organize my computer. I want to move my images to Desktop. & Move image "MyTravelPhoto.jpg" from Documents $>$ Task 14 folder to Desktop & File Explorer & Image "MyTravelPhoto.jpg" is in Desktop. \\
15 & I want to connect my bluetooth mouse to my computer. Currently the bluetooth is off. & Turn the bluetooth on. & Settings & The bluetooth is on. \\
16 & The brightness on my computer is too high. & Reduce the brightness of my computer to 50\% & Settings & The brightness is set to 50\% \\
17 & The wallpaper on my screen is set to the default one and I want to change it to something colorful. & Change the wallpaper to any Solid Color in Settings. & Settings & The wallpaper on the computer is now set to a solid color. \\
18 & I'm browsing on Chrome but it is frozen. & Open Chrome and close the application from Task Manager. & Task Manager, Chrome & The Google chrome is closed (with force exit). \\
19 & I'm flying to Austin next week. I want to check how the weather is and pack accordingly. & Using the Weather app, check the weather in Austin for the following week in Celcius scale. & Weather & The application is open with Austin temperature on screen and then unit is set to Celcius. \\
20 & I want to stay organized will all my assignment documents and resources. & I want to delete a folder "Old Assignment" in Documents $>$ Task 20 folder. Then, create a folder in Documents $>$ Task 20 folder called "Assignments" & File Explorer & A new folder named "Assignments" is created \\
21 & I accidentally deleted a file with the name Task ID 21.jpg. & Recover the file Task ID 21.jpg from Recycle Bin and verify whether it is restored to its previous location in Documents $>$ Task 21 folder. & Recycle Bin & The photo is back to the original folder. \\
22 & I want to do a calculation using the calculator application. & Compute the product of 20 and 483 on Calculator. & Calculator & The calculator app is showing the result of the multiplication of 20 times 483. \\
23 & I joined a new company and my team communicates on Slack. & Install Slack application on my laptop. The installer is in Documents $>$ Task 23 folder. & Chrome, Slack & The slack is installed. \\
24 & I want to share a large folder as an email attachment. & Compress the folder "Final Report" in Documents $>$ Task 24 folder to ZIP format. & File Explorer & The compressed version (in ZIP) of the "Final Report" folder is created. \\
25 & I'm reading an interesting article on solar system in Word. I want to take some notes in a separate document. & Create a new document, add the text "Solar System" in any shade of green, Then, make it a heading and align it to the center of the document. & Word & A blank file in Word is created with a text "Solar System", written in any shade of green, heading, and center aligned. \\
26 & I am tracking all the books I've read in an excel sheet. I have columns for "Book Title", "\# Days to Finish", and "Rating (1-10)". & Make an entry for the book I just completed reading Task ID 26.xlsx in Documents $>$ Task 26 folder. The Book Title is "Atomic Habit" and it took me 10 days to finish reading the book. I really loved the book, I would rate it 9 out of 10. & Excel, File Explorer & A new row entry has been added to the spreadsheet. \\
27 & I have to make a presentation slide to introduce myself to my classmates. & Create a new PowerPoint presentation. In the first slide, add the title "3 Fun Facts About Me". Create a second slide, and add details of my age, favorite color, and favorite animal. I'm 20, I like red, and sea otters are my favorite animals. & PowerPoint, File Explorer & The slide deck has a title slide with the text "About Me" and another slide with title, "3 Fun Facts" and 3 bullet points about favorite color, age, and animal. \\
28 & I want to change my presentation's theme to a colorful one. & Open Powerpoint Slide Task ID 28.pptx in Documents $>$ Task ID 28 folder, and change the presentation theme to "Gallery". & PowerPoint, File Explorer & The theme of slide is now changed to "Gallery". \\
29 & I am tracking all the books I've read in an excel sheet. I want to add a new column, called "Author" to the right of "Book Title". & Open Task ID 29.xlsx in Documents $>$ Task 29 folder. Add a new column next to "Book Title" column and name it "Author." & Excel, File Explorer & A new column is added to the right of "Book Title" named "Author" \\
30 & I am tracking all the books I've read in an excel sheet. I want to make a bar chart for how many days I took to read all the books. & Open Task ID 30.xlsx in Documents $>$ Task 30 folder, create a bar chart where the x-axis is "Book Title" and the y-axis is the "Days to Finish". The title of the chart is "Summer Reading" & Excel, File Explorer & A "Bar Chart" is added with the x-axis titled "Book Title", y-axis titled "Days to Finish", chart titled "Days to finish" \\
31 & I am tracking all the books I've read in an excel sheet. I want to know how many days I took to read a book on average. & Open Task ID 31.xlsx in Documents $>$ Task 31 folder, calculate the average of "\# Days to finish" using mathematical functions available in Excel. & Excel, File Explorer & The average of the ratings for "Rating (1-10)" is reported at the bottom of the column, using the "Average" function \\
32 & I'm reading a word document. I want to suggest changes to the article's title by striking out and highlight the title from the document so that the author is aware that he/she should change the title. & Open Task ID 32.docx in Documents $>$ Task 32 folder, make a strikethrough to the title of the article and highlight the title. & Word, File Explorer & A strike-through is applied to the title of document. The title is also highlighted. \\
33 & I'm working on an introduction PowerPoint for my new job. I placed a photo of myself on the side of the slides but because the photo is vertically oriented, doesn't fit the landscape orientation of the slide. I want to make the slide vertically oriented so that the photo can fit well. & Open Task ID 33.pptx in Documents $>$ Task 33 folder. Orient the slide to vertical so that my photo can fit well. & PowerPoint, File Explorer & The slide is in potrait orientation \\
34 & I am editing my essay report on Word. There is a sentence that I want to revise in one of the sections. & Open Task ID 34.docx in Documents $>$ Task 34 folder. Go to "Knowledge of the Solar System" section. Add a comment to modify the sentence "The solar system has been a topic of study from the beginning of history." to "The solar system is one of the most unique entities in history." & Word, File Explorer & The sentence "The solar system has been a topic of study from the beginning of history." is changed to "The solar system is one of the most unique entities in history." \\
35 & I am writing an article on Word and I want to modify the title format based on the guidelines. Also, I want to add the page number in the document. & Open Task ID 35.docx in Document $>$ Task 35 folder, capitalize the first letter of every word in the title and add the page number & Word, File Explorer & The first letter of each word in the title is in uppercase. Title page is also added. \\
36 & I have to submit my work for an upcoming conference in PDF format, but all the documents are in .docx format currently. & Export the .docx documents in Documents $>$ Task 36 folder to PDF & Word, File Explorer & The desired documents in PDF. \\
37 & I'm giving a presentation tomorrow and I want to move my talking points from a word document to the Notes section of my presentation. & Copy presentation notes from word document located in Task ID 37.docx to my presentation slide Task ID 37.pptx, both in Documents $>$ Task 37 folder. & PowerPoint, Word, File Explorer & The notes from the "Task ID 37.docx" document on the desktop is pasted into the "Notes" section on the powerpoint presentation \\
38 & I want to format my document to match the new font I have installed. & Install the font Organo font.zip, and change the font of the text in Task ID 38.docx to the new font, both in Documents $>$ Task 38 folder. & Word, Font Manager, File Explorer & A .docx saved showing text rendered in the new font (font name visible in the ribbon) \\
39 & I need to convert and export the excel spreadsheet I have to CSV format. I have to have them in Desktop. & Export the Task ID 39.xlsx in Documents $>$ Task 39 folder to .csv and move it to Desktop. & Excel, File Explorer & A new .csv file from the excel sheet is created and located in desktop \\
40 & I'm making a list of popular books from different online sources into an Excel sheet & Visit wikipedia.com and search for the most popular book. Paste in the first two books under the "Between 50 million and 100 million copies" into the Excel Task ID 40.xlsx in Documents $>$ Task 40 folder. & Chrome, Excel, File Explorer & "She: A History of Adventure" and "The Divinci Code" is added into the excel sheet. \\
41 & I want to make a recorded demonstration of a simple calculation using BODMAS rule for my school presentation. & Turn on screen recording using the Snipping Tool. Open calculator app and perform the calculation 26-12/2. Now clear the calculator results and calculate (26-12)/2. Stop the recording on Snipping Tool Insert this video in Task ID 41.pptx in Documents $>$ Task 41 folder. & Snipping Tool, Calculator, PowerPoint, File Explorer & The calculation recording is added to the presentation slide. \\
42 & I want to add a picture of myself to the presentation. & Add the photo Task ID 42.jpg to Task ID 42.pptx, both in Document $>$ Task 42 folder. & PowerPoint, File Explorer & The presentation has an image (Task ID 42.jpg). \\
43 & I want my data visualized in a presentation. & Insert the bar chart visualization from Task ID 43.xlsx into PowerPoint Task ID 43.pptx, both in Documents $>$ Task 43 Folder. & Excel, PowerPoint, File Explorer & The visualization from the excel sheet is copied to Powerpoint slide deck. \\
44 & I am enjoying reading an article in a word document. I want to know more about the author. & Open Task ID 44.docx in Documents $>$ Task 44 folder, locate its author in this article. Search them on Chrome. Go to their personal website. & Word, Chrome, File Explorer & Google Chrome is open with the author's website. \\
45 & I want to know the distance and time needed to cover different sections of the route from The Domain to Austin Airport if I were to take public transport all the way. & Visit maps.google.com. Check for how long it takes to start from The Domain area then reach Austin Airport using public transport. Also get the buses/trains to be taken and time each of them takes then type them into Notepad. & Chrome, Notepad & A notepad file created with the details of different routes. \\
46 & I need to copy all the editable text from the desktop screenshot & Take a screenshot of the desktop window using Snipping Tool and extract all of the text in the image. Paste all the editable text copied to a notepad. & Snipping Tool, Notepad & The notepad has the text copied from Desktop. \\
47 & I want quick access to media folders from Desktop and add each of their paths in Excel & Create new shortcuts to Pictures, Music, and Videos from Desktop and add each of their paths in Task ID 47.xlsx in Documents $>$ Task 47. & File Explorer, Excel & Shortcuts created on desktop; the excel sheet has information on the folder paths. \\
48 & I'm working on creating a personal website. I want to make a list of all the steps for it on my sticky notes so it is easier to access. & Copy all the steps from Task ID 48.docx in Documents $>$ Task 48 folder to a new sticky note and change its color to green. & Chrome, Sticky Notes, File Explorer & The list of steps are in Sticky Notes and the sticky notes color is green. \\
49 & I'm skimming through a video to get an idea about the different concepts it covers. & Open video Task ID 49.mp4 in Documents $>$ Task 49 folder on Media Player. Watch it by skimming through the video and keep fast forwarding the video by 30 seconds until the end of the video. & Media Player, File Explorer & The video is stop at the end of the end of the video. \\
50 & I'm watching a video on Media Player, but I'm on a time crunch so I want to speed through it. & Open video Task ID 50.mp4 in Documents $>$ Task 50 folder on Media Player. Play the video in 1.5x speed. & Media Player, File Explorer & The video is playing at the speed 1.5x \\
51 & I want to make a new playlist to organize my music better. I want to add "Baby Shark" song in my playlist. & Create the playlist "Fun with Kids" on Media Player and add Task ID 51.mp3 in Documents $>$ Task 51 folder to the playlist. & Music, File Explorer & A new playlist is created. Baby Shark song is added to the newly created playlist. \\
52 & The default brightness setting of The Eiffel Tower photo is too high & Open Task ID 52.jpg in Documents $>$ Task 52 folder on Photos app. Turn down the brightness to -1. & Photos, File Explorer & The brightness on the photo of The Eiffel Tower reads -0.1 \\
53 & I want to share a snapshot of the food from the video I took yesterday in a restaurant I visited & Open the video Task ID 53.mp4 in Documents $>$ Task 53 folder on Media Player, and take a screenshot of the frame from the video with food in it. & Media Player, File Explorer & The snapshot of the food is created. \\
54 & I want a mirror version of a picture for layout balance. & Transform the image Task ID 54.jpg in Documents $>$ Task 54 folder on Photos app to make it horizontally flipped. & Photos, File Explorer & The image is flipped horizontally. \\
55 & I'm editing an image to make it the cover photo of a travel album I'm working on. & Open Task ID 55.jpg in Documents $>$ Task 55 folder on Photos app. Edit image to add the text caption "Midnight in Paris". & Photos, File Explorer & The image has additional text caption "Midnight in Paris". \\
56 & I want to watch the visuals of a video without any sound. & Open the video Task ID 56.mp4 in Documents $>$ Task 56 folder on Media Player app. Mute the audio in the video. & Media Player, File Explorer & The video audio is on mute \\
57 & I'm watching a YouTube video, but this video is in English and I want to improve my Spanish ability. & Find "Click This Video To Feed 1 Person" by Mr Beast on YouTube. Change the audio track to Spanish instead. & Chrome & The video is playing in Spanish. \\
58 & I am watching a YouTube video but my network is slow and has an unstable stream & Find "Click This Video To Feed 1 Person" by Mr Beast on YouTube. Change the video quality to 720p. & Chrome & The video is playing in 720p. \\
59 & I want a video I am watching on YouTube to float in a small window while I look for other videos. & Find "Click This Video To Feed 1 Person" by Mr Beast on YouTube. Enable picture-in-picture playback on Youtube. & Chrome & The video is playing picture-in-picture. \\
60 & I have a picture but is too large to share as an email attachment. I need a smaller JPEG. & Open Task ID 60.jpg in Documents $>$ Task 60 folder on Photos app. Reduce the image size to 40\% & Photos, File Explorer & The picture's quality size is reduced to 40\% \\
\end{longtable}

\newpage
\onecolumn
\subsection{Task Outcomes and Timing Across Users and CUA Conditions}

\begin{figure*}[h!]
    \centering
    \includegraphics[width=6in]{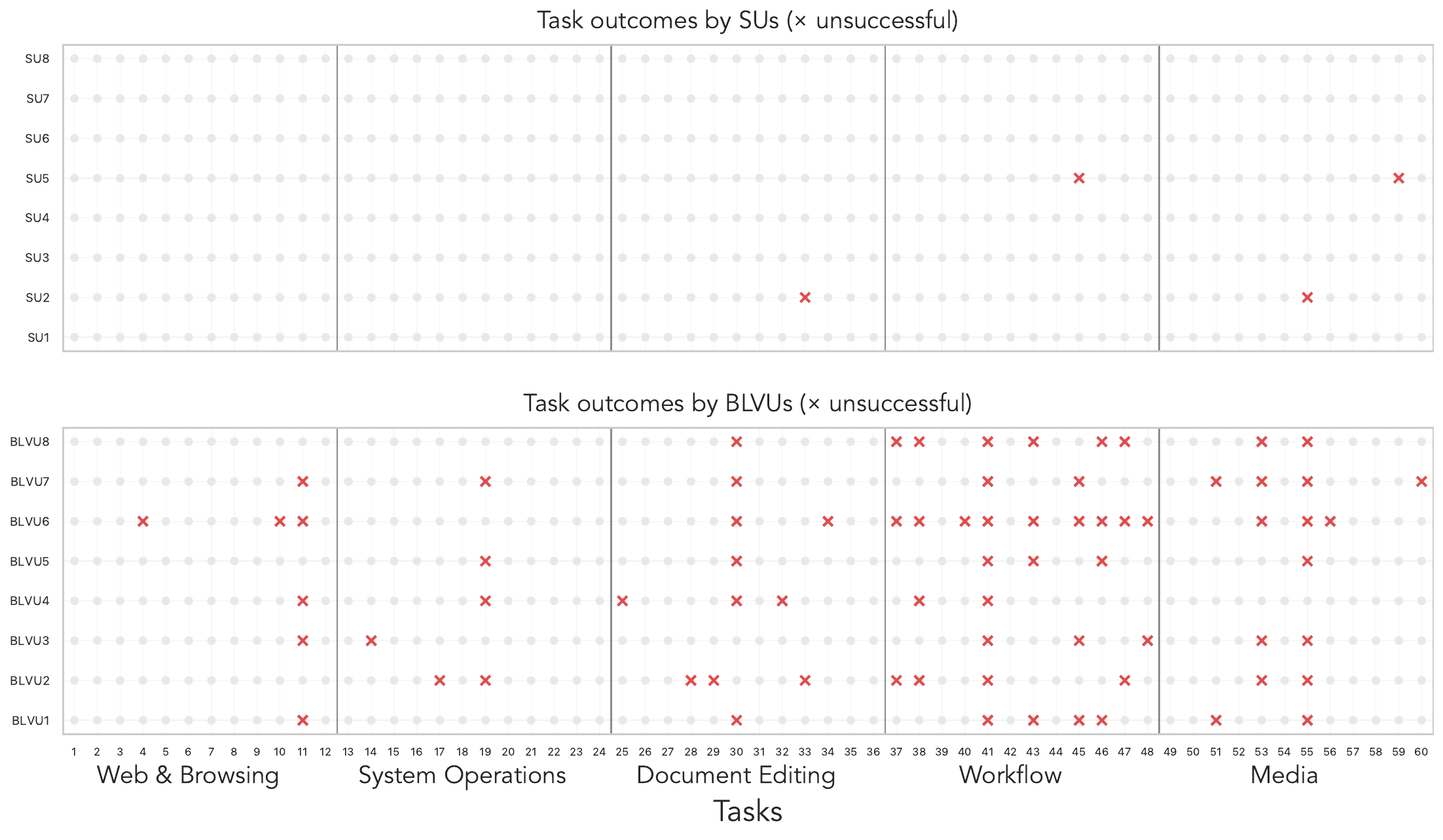}
    \caption{Task outcomes by participant and task. Each row is a participant (N=8 per group); columns show 60 tasks grouped into five categories: web \& browsing, system operations, document editing, workflow, and media. Gray circles indicate successful completions; red × mark unsuccessful attempts. Sighted users (top) show near-ceiling success with only a few isolated failures, whereas BLVUs (bottom) exhibit clustered failures, especially in Workflow and Media tasks, indicating greater difficulty with multi-step coordination and media manipulation.}
    \Description{This figure shows per-participant task outcomes across 60 tasks, split into two panels. The top panel reports sighted users (SUs), and the bottom panel reports blind and low-vision users (BLVUs). Each row is a participant; each column is a task. Symbols encode outcomes: gray circles indicate successful completions and red × marks indicate unsuccessful attempts. Tasks are grouped along the x-axis into five categories—Web & Browsing, System Operations, Document Editing, Workflow, and Media. Sighted users exhibit near-ceiling performance with only a handful of isolated failures. In contrast, BLVUs show many more unsuccessful attempts that cluster within particular regions, especially in Workflow and Media, with additional scattered failures in System Operations and Document Editing. The pattern suggests that multi-step coordination and media-manipulation tasks present greater challenges for BLVUs relative to SUs, and that difficulty varies by participant as well as by task category.}
    \label{fig:task_outcomes}
\end{figure*}

\begin{figure*}[h!]
    \centering
    \includegraphics[width=6in]{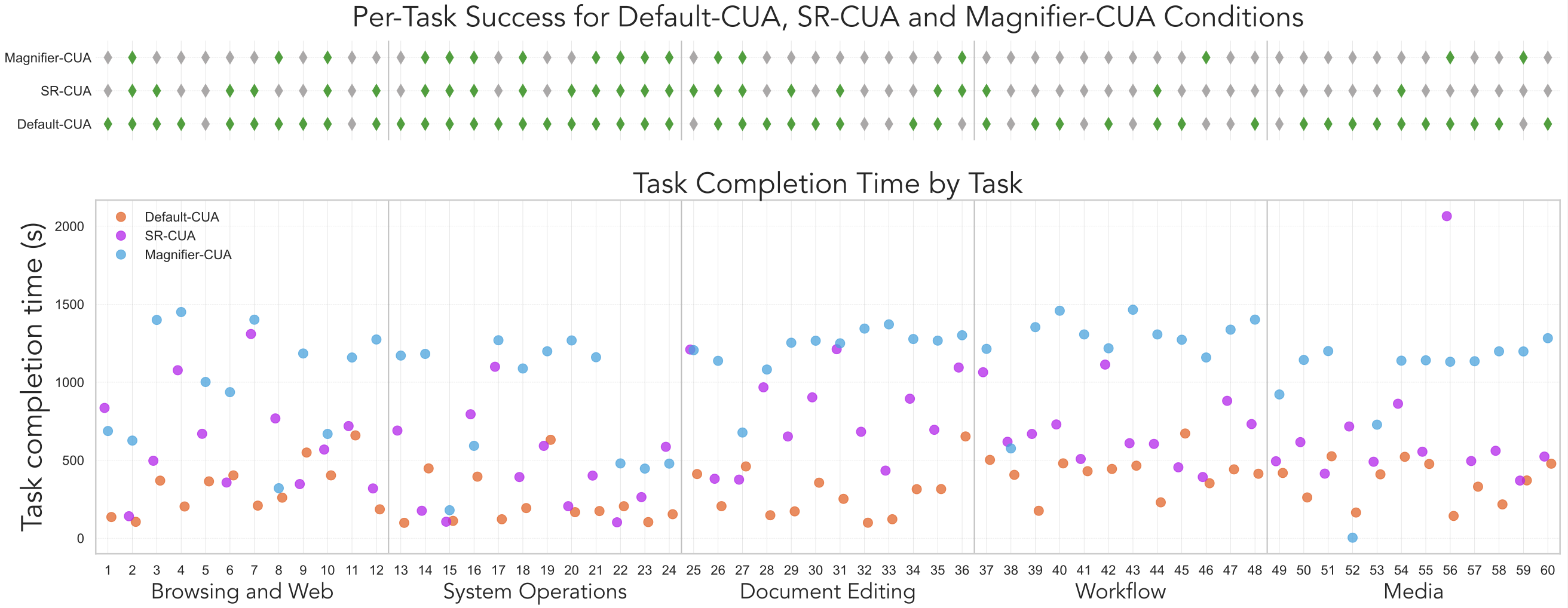}
    \caption{\revision{Per-task success and completion time for 60 tasks under three CUA conditions (Default-CUA, SR-CUA, Magnifier-CUA). Default-CUA succeeds on the largest number of tasks and is generally fastest, SR-CUA completes fewer tasks and takes longer, and magnifier-CUA has the fewest successful tasks with the longest completion times, especially for workflow and media tasks.}}
    \Description{The figure has two stacked panels. The top panel is a per-task success grid with three horizontal rows labeled “Default-CUA,” “SR-CUA,” and “Magnifier-CUA.” Each column is one task (1–60), grouped under Browsing and Web, System Operations, Document Editing, Workflow, and Media. Green diamonds mark successful tasks and gray diamonds mark failures; Default-CUA shows many green diamonds, SR-CUA shows fewer, and Magnifier-CUA has the most gray diamonds, particularly in later Workflow and Media tasks.
    The bottom panel is a scatter plot of task completion time in seconds versus task index for the same 60 tasks. Orange dots (Default-CUA) are clustered at lower times (roughly 100–600s), purple dots (SR-CUA) are higher and more spread out, and blue dots (Magnifier-CUA) are highest, often between 900–1500s with some points above 2000s. Across categories, completion times rise and become more variable from Document Editing through Workflow and Media, illustrating that harder tasks both fail more often and take substantially longer, especially for Magnifier-CUA.}
    \label{fig:per_user_tct}
\end{figure*}

\begin{figure*}
    \centering
    \includegraphics[width=\linewidth]{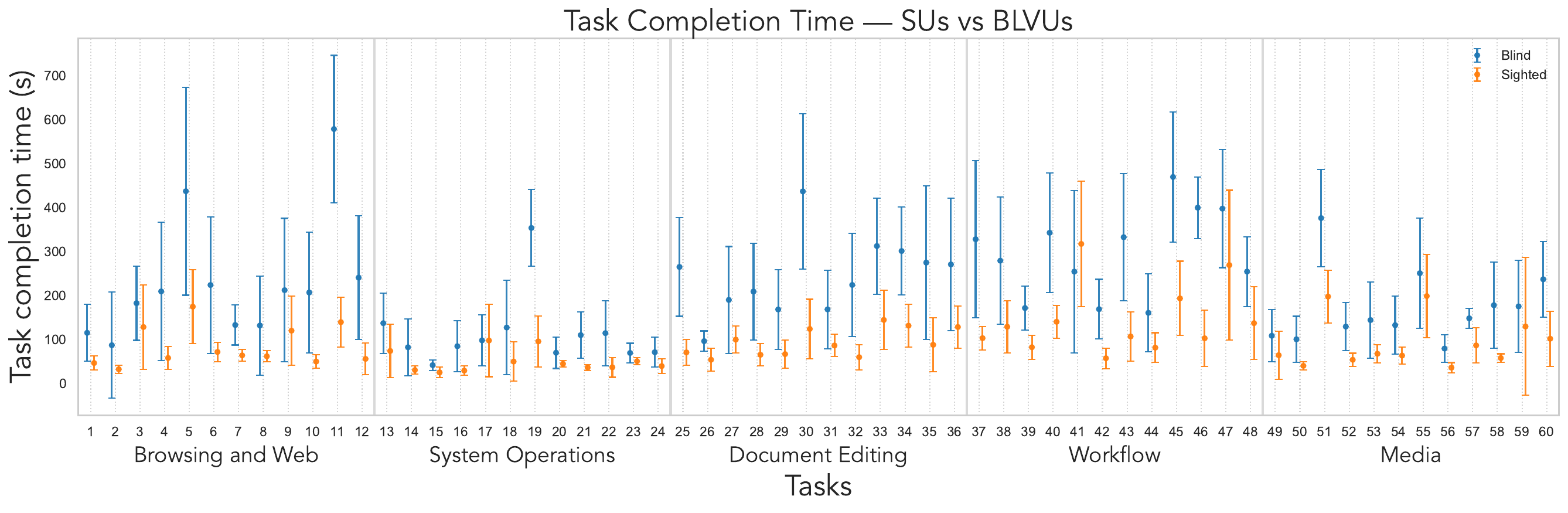}
    \caption{\revision{Mean task completion time for BLVUs and SUs across 60 tasks. Each point is the mean time for a task and vertical bars shows standard deviation. BLVUs generally take longer and show higher variability than SUs.}}
    \Description[]{A wide chart shows task completion time (seconds) on the y-axis and tasks 1–60 on the x-axis. The x-axis is divided into five labeled regions: Browsing and Web (1–12), System Operations (13–24), Document Editing (25–36), Workflow (37–48), and Media (49–60). For each task: A blue dot with a blue error bar marks the mean and standard deviation for BLVUs. An orange dot with an orange error bar marks the mean and standard deviation for SUs. Blue points are usually higher and have taller error bars than orange points, especially in Browsing/Web and Document Editing, indicating longer times and greater variability for BLVUs.}
\end{figure*}

\begin{figure*}
    \centering
    \includegraphics[width=\linewidth]{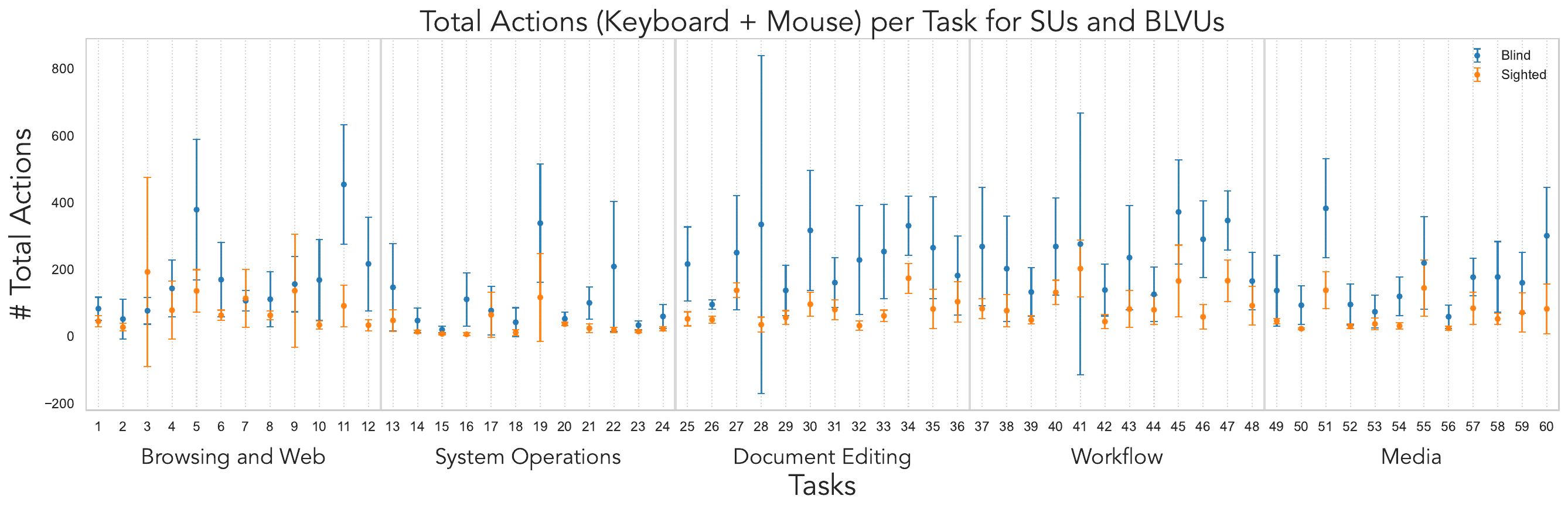}
    \caption{\revision{Mean and standard deviation of total input actions (mouse and keyboard) per task for BLVUs and SUs. Each point shows the average number of actions for a given task, with vertical bars indicating standard deviation across users. Overall, BLVUs use more actions and show greater variability than SUs.}}
    \Description[]{A horizontal sequence of 60 task numbers runs along the x-axis, grouped under labels “Browsing and Web”, “System Operations”, “Document Editing”, “Workflow”, and “Media”. For each task, there are two vertical error bars: a blue bar for BLVUs and an orange bar for SUs. The y-axis shows \#Total Actions, ranging roughly from −200 to 800. Blue markers and bars generally sit higher and are taller than the orange ones, indicating both higher mean numbers of actions and larger variability for BLVUs across almost all tasks.}
\end{figure*}

\newpage
\onecolumn
\subsection{Per-Task Metrics by User Groups and CUA Conditions}
% --------- SU table ---------
\begin{table*}[h!]
\small
\setlength{\tabcolsep}{2pt}
% \begin{tabular}{ll*{9}{cc}}
\begin{tabular}{ll*{9}{c c}}
\toprule
\multicolumn{2}{l}{\multirow{2}{*}{}} &
\multicolumn{2}{c}{\textbf{SU1}} & \multicolumn{2}{c}{\textbf{SU2}} & \multicolumn{2}{c}{\textbf{SU3}} & \multicolumn{2}{c}{\textbf{SU4}} &
\multicolumn{2}{c}{\textbf{SU5}} & \multicolumn{2}{c}{\textbf{SU6}} & \multicolumn{2}{c}{\textbf{SU7}} & \multicolumn{2}{c}{\textbf{SU8}} & \multicolumn{2}{c}{\textbf{Total}} \\
\multicolumn{2}{l}{} &
$\mu$ & $\sigma$ & $\mu$ & $\sigma$ & $\mu$ & $\sigma$ & $\mu$ & $\sigma$ &
$\mu$ & $\sigma$ & $\mu$ & $\sigma$ & $\mu$ & $\sigma$ & $\mu$ & $\sigma$ & $\mu$ & $\sigma$ \\
\midrule
\multicolumn{2}{l}{Task Success Rate (\%)} & 100\% & & 96.6\% & & 100\% & & 100\% & & 96.6\% & & 100\% & & 100\% & & 100\% & & 99.1\% &   \\
\multicolumn{2}{l}{Task Duration (s)} & 58.7 & 40.0 & 66.9 & 51.7 & 103.0 & 81.8 & 92.3 & 56.9 & 151.1 & 130.5 & 95.0 & 62.8 & 101.5 & 76.9 & 70.1 & 47.0 & 92.3 & 78.0\\
\multicolumn{2}{l}{\# Events} & 77.5 & 56.0 & 65.0 & 57.1 & 89.1 & 93.5 & 84.3 & 72.6 & 103.5 & 102.6 & 121.1 & 199.9 & 105.5 & 96.2 & 93.9 & 80.2  & 92.5 & 104.6\\

\multicolumn{2}{l}{\# Mouse Actions} & 46.7 & 39.0 & 32.1 & 30.3 & 51.4 & 54.8 & 47.2 & 43.4 & 55.0 & 53.7 & 84.3 & 137.3 & 51.8 & 49.4 & 46.7 & 50.6 & 51.9 & 66.3\\
\multicolumn{2}{l}{\hspace*{\cuaindent}\# Mouse up} & 18.1 & 14.8 & 14.5 & 11.9 & 18.6 & 18.3 & 15.9 & 12.2 & 20.8 & 18.9 & 19.1 & 18.4 & 20.6 & 16.9 & 16.5 & 12.4 & 18.0 & 15.8\\
\multicolumn{2}{l}{\hspace*{\cuaindent}\# Mouse click} & 19.9 & 17.5 & 14.6 & 12.0 & 20.1 & 18.3 & 16.4 & 12.2 & 22.5 & 19.0 & 19.4 & 18.4 & 22.1 & 16.9 & 17.6 & 12.4 & 19.0 & 17.41\\
\multicolumn{2}{l}{\hspace*{\cuaindent}\# Scroll} & 6.9 & 13.3 & 2.3 & 9.3 & 10.6 & 36.0 & 14.0 & 31.7 & 10.1 & 23.8 & 44.9 & 131.7 & 7.7 & 21.6 & 11.7 & 37.1 &13.5 &53.8\\
\multicolumn{2}{l}{\hspace*{\cuaindent}\# Drag-and-drop} & 1.8 & 3.9 & 0.7 & 1.8 & 2.0 & 3.8 & 0.9 & 1.4 & 1.6 & 2.1 & 0.9 & 2.5 & 1.4 & 2.1 & 0.9 & 1.5 & 1.2 & 2.5\\
\multicolumn{2}{l}{\# Keyboard Actions} & 17.8 & 25.6 & 18.9 & 23.6 & 19.1 & 27.9 & 19.8 & 30.2 & 22.3 & 35.5 & 14.8 & 25.6 & 30.7 & 45.5 & 25.3 & 40.4 & 21.0 & 32.7 \\
\multicolumn{2}{l}{\hspace*{\cuaindent}\# Character Keys} & 13.0 & 21.0 & 14.3 & 19.3 & 14.1 & 21.3 & 14.9 & 23.4 & 11.5 & 16.3 & 10.6 & 19.9 & 20.3 & 31.4 & 18.7 & 31.4 & 14.6 & 23.6\\
\multicolumn{2}{l}{\hspace*{\cuaindent}\# Arrow keys} & 4.1 & 7.2 & 4.1 & 7.9 & 4.5 & 9.8 & 4.1 & 8.5 & 10.1 & 24.9 & 3.9 & 7.5 & 8.2 & 15.8 & 5.6 & 10.3 & 5.5 & 12.9\\
\multicolumn{2}{l}{\hspace*{\cuaindent}\# Hotkeys} & 0.8 & 2.1 & 0.6 & 1.5 & 0.5 & 1.0 & 0.7 & 2.0 & 0.7 & 1.7 & 0.3 & 0.8 & 2.3 & 4.8 & 0.9 & 1.6 &  0.8 & 2.3\\
\multicolumn{2}{l}{\hspace*{2\cuaindent}\# Ctrl-based} & 0.7 & 2.1 & 0.6 & 1.5 & 0.5 & 1.0 & 0.7 & 2.0 & 0.5 & 1.4 & 0.2 & 0.8 & 2.1 & 4.8 & 0.9 & 1.6 & 0.7 & 2.2\\
\multicolumn{2}{l}{\hspace*{2\cuaindent}\# Alt-based} & 0.0 & 0.2 & 0.0 & 0.0 & 0.0 & 0.0 & 0.0 & 0.1 & 0.1 & 0.4 & 0.0 & 0.0 & 0.6 & 4.3 & 0.0 & 0.1 & 0.0 & 1.5 \\
\multicolumn{2}{l}{\hspace*{2\cuaindent}\# Win-based} & 0.0 & 0.0 & 0.0 & 0.0 & 0.0 & 0.0 & 0.0 & 0.0 & 0.0 & 0.1 & 0.0 & 0.1 & 0.1 & 0.3 & 0.0 & 0.1 & 0.0 & 0.1\\
\multicolumn{2}{l}{\hspace*{2\cuaindent}\# Shift-based} & 0.0 & 0.0 & 0.0 & 0.0 & 0.0 & 0.0 & 0.0 & 0.0 & 0.0 & 0.1 & 0.0 & 0.1 & 0.1 & 0.4 & 0.0 & 0.2 & 0.0 & 0.1\\
\bottomrule
\end{tabular}
\caption{Per-task metrics for Sighted Users (SUs). Values shown per user (SU1…SU8) and Total are mean $(\mu)$ and standard deviation $(\sigma)$.}
\end{table*}

% --------- BLVU table ---------
\begin{table*}[h!]
\small
\setlength{\tabcolsep}{2pt}
% \begin{tabular}{ll*{9}{cc}}
\begin{tabular}{ll*{9}{c c}}
\toprule
\multicolumn{2}{l}{\multirow{2}{*}{}} &
\multicolumn{2}{c}{\textbf{BLVU1}} & \multicolumn{2}{c}{\textbf{BLVU2}} & \multicolumn{2}{c}{\textbf{BLVU3}} & \multicolumn{2}{c}{\textbf{BLVU4}} &
\multicolumn{2}{c}{\textbf{BLVU5}} & \multicolumn{2}{c}{\textbf{BLVU6}} & \multicolumn{2}{c}{\textbf{\textbf{BLVU7}}} & \multicolumn{2}{c}{\textbf{BLVU8}} & \multicolumn{2}{c}{\textbf{Total}} \\
\multicolumn{2}{l}{} &
$\mu$ & $\sigma$ & $\mu$ & $\sigma$ & $\mu$ & $\sigma$ & $\mu$ & $\sigma$ &
$\mu$ & $\sigma$ & $\mu$ & $\sigma$ & $\mu$ & $\sigma$ & $\mu$ & $\sigma$ & $\mu$ & $\sigma$ \\

\midrule
\multicolumn{2}{l}{Task Success Rate (\%)} & 88.3\% &  & 81.6\% &  & 88.3\% &  & 88.3\% &  & 90\% &  & 71.6\% & & 86.6\% &  &85\% & & 84.6\% &  \\
\multicolumn{2}{l}{Task Duration (s)} & 197.3 & 139.8 & 192.9 & 132.6 & 208.2 & 164.2 & 133.0 & 114.4 & 177.9 & 138.7 & 265.9 & 160.2 & 237.0 & 172.5 & 277.2 &  164.7 & 211.1 & 154.9\\
\multicolumn{2}{l}{\# Events} & 198.7 & 228.2 & 221.5 & 250.9 & 179.9 & 163.1 & 157.2 & 194.2 & 197.6 & 166.9 & 298.8 & 213.1 & 251.0 & 201.5 & 319.8 & 300.7 &237.3 & 224.0\\

\multicolumn{2}{l}{\# Mouse Actions} & 0.0 & 0.0 & 0.0 & 0.0 & 0.0 & 0.0 & 0.0 & 0.0 & 0.0 & 0.0 & 0.0 & 0.0 & 0.0 & 0.0 & 0.0 & 0.0 & 0.0 & 0.0\\
% \multicolumn{2}{l}{\hspace*{\cuaindent}\# Mouse up} & 0.0 & 0.0 & 0.0 & 0.0 & 0.0 & 0.0 & 0.0 & 0.0 & 0.0 & 0.0 & 0.0 & 0.0 & 0.0 & 0.0 & 0.0 & 0.0 & 0.0 & 0.0 \\
% \multicolumn{2}{l}{\hspace*{\cuaindent}\# Mouse click} & 0.0 & 0.0 & 0.0 & 0.0 & 0.0 & 0.0 & 0.0 & 0.0 & 0.0 & 0.0 & 0.0 & 0.0 & 0.0 & 0.0 & 0.0 & 0.0 & 0.0 & 0.0\\
% \multicolumn{2}{l}{\hspace*{\cuaindent}\# Scroll} & 0.0 & 0.0 & 0.0 & 0.0 & 0.0 & 0.0 & 0.0 & 0.0 & 0.0 & 0.0 & 0.0 & 0.0 & 0.0 & 0.0 & 0.0 & 0.0 & 0.0 & 0.0\\
% \multicolumn{2}{l}{\hspace*{\cuaindent}\# Drag-and-drop} & 0.0 & 0.0 & 0.0 & 0.0 & 0.0 & 0.0 & 0.0 & 0.0 & 0.0 & 0.0 & 0.0 & 0.0 & 0.0 & 0.0 & 0.0 & 0.0& 0.0 & 0.0 \\

\multicolumn{2}{l}{\# Keyboard Actions} & 200.3 & 158.7 & 164.8 & 207.4 & 125.7 & 88.0 & 123.4 & 130.3 & 151.3 & 130.5 & 224.3 & 156.9 & 201.0 & 162.6 & 244.9 & 203.2 & 179.4 & 163.4 \\
\multicolumn{2}{l}{\hspace*{\cuaindent}\# Character Keys} & 42.1 & 51.6 & 39.5 & 37.9 & 37.2 & 27.7 & 22.5 & 27.8 & 34.9 & 37.5 & 56.1 & 60.9 & 29.9 & 30.1 & 27.3 & 39.7 & 36.1 & 41.5\\
\multicolumn{2}{l}{\hspace*{\cuaindent}\# Arrow keys} & 134.5 & 115.0 & 111.9 & 199.3 & 78.1 & 68.8 & 93.3 & 111.2 & 105.1 & 100.6 & 149.9 & 124.4 & 157.0 & 148.1 & 207.3 & 192.0 & 129.6 & 143.3\\
\multicolumn{2}{l}{\hspace*{\cuaindent}\# Hotkeys} & 23.6 & 38.9 & 13.5 & 28.4 & 10.4 & 21.8 & 7.6 & 10.7 & 11.3 & 14.1 & 18.4 & 31.0 & 14.0 & 23.6 & 10.3 & 20.8 & 13.6 & 25.4\\
\multicolumn{2}{l}{\hspace*{2\cuaindent}\# Ctrl-based} & 20.8 & 36.6 & 10.0 & 26.3 & 5.5 & 8.3 & 4.8 & 6.4 & 9.2 & 13.1 & 16.9 & 30.1 & 10.9 & 21.4 & 2.5 & 4.4 & 10.0 & 22.1\\
\multicolumn{2}{l}{\hspace*{2\cuaindent}\# Alt-based} & 2.0 & 4.4 & 1.9 & 4.3 & 5.1 & 18.6 & 2.7 & 6.8 & 1.4 & 3.0 & 0.3 & 0.7 & 2.9 & 5.9 & 6.9 & 20.5 & 2.9 & 10.6\\
\multicolumn{2}{l}{\hspace*{2\cuaindent}\# Win-based} &  1.3 & 3.4 & 1.7 & 8.2 & 0.5 & 1.7 & 0.2 & 0.7 & 0.9 & 0.9 & 1.4 & 9.8 & 0.3 & 0.8 & 1.1 & 1.3 & 0.9 & 4.7\\
\multicolumn{2}{l}{\hspace*{2\cuaindent}\# Shift-based} & 2.6 & 8.4 & 0.5 & 1.4 & 0.5 & 1.7 & 0.2 & 0.6 & 0.7 & 2.4 & 3.2 & 12.8 & 1.0 & 2.6 & 3.1 & 13.1 & 1.4 & 7.3 \\
\bottomrule
\end{tabular}
\caption{Per-task metrics for Blind and Low-Vision Users (BLVUs). Values shown per user (BLVU1…BLVU8) and Total are mean $(\mu)$ and standard deviation $(\sigma)$.}
\end{table*}

\end{document}